\documentclass[graybox,openany]{svmult}

\usepackage{mathptmx}       
\usepackage{helvet}         
\usepackage{courier}        
\usepackage{type1cm}        
\usepackage{makeidx}         
\usepackage{graphicx}        
\usepackage{multicol}        
\usepackage[bottom]{footmisc}

\usepackage{amsmath,mathrsfs}
\usepackage{color}
\usepackage{fancyhdr}

\newcommand{\si}{\sigma}

\newcommand{\norm}{{\mathcal N}}
\newcommand{\be}{\begin{equation}}
\newcommand{\ee}{\end{equation}}
\newcommand{\like}{{\mathcal{L}}}
\newcommand{\Var}{{\rm Var}}
\newcommand{\xhat}{{\hat{x}}}
\newcommand{\ML}{{\rm ML}}
\newcommand{\St}{{\Sigma_\theta}}
\newcounter{excounter}
\addtocounter{excounter}{1}
\newcommand{\attention}{}

\newcommand{\nc}{\newcommand}
\nc{\params}{\theta}
\nc{\data}{d}
\nc{\dr}{{\rm d}}
\nc{\mdl}{{\mathcal M}}
\nc{\It}{I_{10}}
\nc{\gsim}{\mathrel{\rlap{\lower4pt\hbox{\hskip1pt$\sim$}}\raise1pt\hbox{$>$}}}
\nc{\lsim}{\mathrel{\rlap{\lower4pt\hbox{\hskip1pt$\sim$}}\raise1pt\hbox{$<$}}}
\newcommand{\pml}{\params_{\rm max}}
\newcommand{\lmax}{{\mathcal L}_{\rm max}}
\newcommand{\pzero}{{\phi}}
\newcommand{\pone}{{\psi}}
\newcommand{\chisq}{\chi^2}
\newcommand{\LL}{\like}
\newcommand{\Fpost}{{\mathcal F}}
\newcommand{\Omk}{\Omega_\kappa}

\newcommand{\Om}{\Omega_m}
\newcommand{\OL}{\Omega_\Lambda}

\newcommand{\Cp}{{\mathscr{C}}}
\newcommand\Mpc{\,\mbox{Mpc}}

\newcommand{\pval}{\wp}
\newcommand{\obs}{\text{obs}}
\newcommand{\xhati}{\hat{x}_i} 

\begin{document}

\title*{Bayesian Methods in Cosmology}
\titlerunning{Bayesian Methods}
\author{Roberto Trotta}
\institute{Roberto Trotta \at Imperial College London, Imperial Centre for Inference and Cosmology \& Data Science Institute, Blackett Laboratory, Prince Consort Road, London SW7 2AZ \\ \texttt{www.robertotrotta.com}} 
%
%
\maketitle

\abstract*{Short abstract.}

\abstract{These notes aim at presenting an overview of Bayesian statistics, the underlying concepts and application methodology that will be useful to astronomers seeking to analyse and interpret a wide variety of data about the Universe. The level starts from elementary notions, without assuming any previous knowledge of statistical methods, and then progresses to more advanced, research-level topics.  After an introduction to the importance of statistical inference for the physical sciences, elementary notions of probability theory and inference are introduced and explained. Bayesian methods are then presented, starting from the meaning of Bayes Theorem and its use as inferential engine, including a discussion on priors and posterior distributions. Numerical methods for generating samples from arbitrary posteriors (including Markov Chain Monte Carlo and Nested Sampling) are then covered. The last section deals with the topic of Bayesian model selection and how it is used to assess the performance of models, and contrasts it with the classical p-value approach. A series of exercises of various levels of difficulty are designed to further the understanding of the theoretical material, including fully worked out solutions for most of them.}

\tableofcontents

\section{Introduction}

The purpose of physics is to learn about regularities in the natural phenomena in the world, which we call ``Laws of Physics''. Theoretical models expressed in mathematical form (e.g., Newton's theory of gravitation) have to be validated through experiments or observations of the phenomena they aim to describe (e.g., measurement of the time it takes for an apple to fall). Thus an essential part of physics is the quantitative comparison of its theories (i.e., models, equations, predictions) with observations (i.e., data, measurements). This leads to confirm theories or to refute them. 

Measurements often have uncertainties associated with them. Those could originate in the noise of the measurement instrument, or in the random nature of the process being observed, or in selection effects. Statistics is the tool by which we can extract information about physical quantities from noisy, uncertain and/or incomplete data. Uncertainties however are more general than that. There might be uncertainty in the relationship between quantities in a model (as a consequence of limited information or true intrinsic variability of the objects being studied); uncertainty in the completeness of the model itself; and uncertainty due to unmodelled systematics (to name but a few). 

The purpose of these lectures is to provide an appreciation of the fundamental principles underpinning statistical inference, i.e., the process by which we reconstruct quantities of interest from data, subject to the various sources of uncertainty above. The lectures will also endeavour to provide the conceptual, analytical and numerical tools required to approach and solve some of the most common inference problems in the physical sciences, and in particular in cosmology. References are provided so that the reader can further their understanding of the more advanced topics, at research level and beyond. 

Probability theory, as a branch of mathematics, is concerned with studying the properties of sampling distributions, i.e., probability distributions that describe the relative frequency of occurrence of random phenomena. In this sense, probability theory is ``forward statistics'': given the properties of the underlying distributions, it predicts the outcome of data drawn from such distributions. 

Statistical inference, by contrast, asks the question of what can be learnt about the underlying distributions from the observed data. It therefore is sometimes called ``inverse probability'', in that it seeks to reconstruct the parameters of the distributions out of which the data are believed to have been generated. 

Statistics addresses several relevant questions for physicists:
\begin{enumerate}
\item How can we learn about regularities in the physical world given that any measurement is subject to a degree of randomness?
\item How do we quantify our uncertainty about observed properties in the world?
\item How can we make predictions about the future from past experience and theoretical models?.
\end{enumerate}

Inference and statistics are today at the heart of the scientific process, not merely an optional nuisance. Ernest Rutherford is reported to have said, over a century ago: ``If you need statistics, you did the wrong experiment''. While this might have had some merit at the time, it completely misses the point of what science has become today. All scientific questions at the forefront of research involve increasingly complicated models that try to explain subtle effects in complex, multidimensional data sets. The sheer amount of data available to astrophysicists and cosmologists has increased by orders of magnitudes in the last 20 years. Correspondingly, the sophistication of our statistical analysis tools has to keep up: increasingly, the limiting factor of our knowledge about the Universe is not the amount of data we have, but rather our ability of analyse, interpret and make sense of them.

To paraphrase Rutherford, in 21st Century astrophysics f you do {\em not} need statistics, it's because you are doing the wrong kind of physics! There are (at least) five good reasons why every professional astrophysicist and cosmologist ought to have a solid training in advanced statistical methods:
 \begin{enumerate}
 \item  The complexity of the modelling of both our theories and
 observations will always increase, thus requiring correspondingly more refined statistical
 and data analysis skills. In fact, the scientific return of the next generation of
 surveys will be limited by the level of sophistication and
 efficiency of our inference tools.
 \item The discovery zone for new physics is when a potentially new effect
 is seen at the 2--3$\sigma$ level, i.e., with a nominal statistical significance somewhere in the region of 95\% to 99.7\%. This is when tantalizing suggestions for an effect
 start to accumulate but there is no firm evidence yet. In this potential discovery region a careful application
 of statistics can make the difference between claiming or missing a
 new discovery.
 \item If you are a theoretician, you do not want to
 waste your time trying to explain an effect that is not there in
 the first place. A better appreciation of the interpretation of
 statistical statements might help in identifying robust claims
 from spurious ones.
 \item Limited resources mean that we need to focus our efforts on
 the most promising avenues. Experiment forecast and optimization
 will increasingly become prominent as we need to use all of our
 current knowledge ({\em and} the associated uncertainty) to
 identify the observations and strategies that are likely to give the highest
 scientific return in a given field.
 \item Sometimes we don't have the luxury to be able to gather better or further data. This is the case
 for the many problems associated with cosmic variance limited
 measurements on large scales, for example in the cosmic
 background radiation, where the small number of independent
 directions on the sky makes it impossible to reduce the error
 below a certain floor. 
 \end{enumerate}

\section{Elementary notions}

\subsection{The notion of probability} 

There are two different ways of understanding what probability is. The classical (so-called ``frequentist'') notion of probability is that probabilities are tied to the frequency of outcomes over a long series of trials. Repeatability of an experiment is the key concept. 

The Bayesian outlook\footnote{So-called after  Rev.~Thomas Bayes (1701(?)--1761), who was the first to introduce this idea in a paper published posthumously in 1763, ``An essay towards solving a problem in the doctrine of chances''~\cite{Bayes}.} is that probability expresses a degree of belief in a proposition, based on the available knowledge of the experimenter. Information is the key concept. Bayesian probability theory is more general than frequentist theory, as the former can deal with unique situations that the latter cannot handle (e.g., ``what is the probability that it will rain tomorrow?). 

Let $A,B,C, \dots$ denote propositions (e.g., that a coin toss gives tails). Let $\Omega$ describe the sample space (or state space) of the experiment, i.e., $\Omega$ is a list of all the possible outcomes of the experiment.

\example{If we are tossing a coin, $\Omega = \{T,H\}$, where T denotes ``tails'' and H denotes ``head''. If we are rolling a regular die, $\Omega = \{1,2,3,4,5,6\}$. If we are drawing one ball from an urn containing white and black balls, $\Omega = \{W, B\}$, where W denotes a white ball and B a black ball.} 

\begin{svgraybox}
Frequentist definition of probability: The number of times an event occurs divided by the total number of events in the limit of an infinite series of equiprobable trials. 
\end{svgraybox}

\begin{definition}
The {joint probability} of $A$ and $B$ is the probability of $A$ and $B$ happening together, and is denoted by $P(A,B)$. 
The {conditional probability} of $A$ given $B$ is the probability of $A$ happening given that $B$ has happened, and is denoted by $P(A|B)$.
\end{definition}

The sum rule: 
\be
P(A) + P(\overline{A}) = 1 ,
\ee
where $\overline{A}$ denotes the proposition ``not $A$''.

The product rule:
\be \label{eq:prod1}
P(A,B) = P(A|B) P(B).
\ee
By inverting the order of $A$ and $B$ we obtain that
\be\label{eq:prod2}
P(B,A) = P(B|A)P(A)
\ee
\attention and because $P(A,B) = P(B,A)$, we obtain {Bayes theorem} by equating Eqs.~\eqref{eq:prod1} and \eqref{eq:prod2}:
\be \label{eq:bayesth}
P(A|B) = \frac{P(B|A)P(A)}{P(B)}.
\ee

The marginalisation rule follows from the two rules above and it reads:
\be
P(A) = P(A,B_1) +  P(A,B_2) + \dots = \sum_{i} P(A,B_i) = \sum_i P(A|B_i)P(B_i),
\ee 
where the sum is over all possible outcomes for proposition $B$. 

\begin{definition}
Two propositions (or events) are said to be {independent} if and only if 
\be
P(A,B) = P(A)P(B).
\ee
\end{definition}

\subsection{Random variables, parent distributions and samples}

\begin{definition}
A {random variable} (RV) is a function mapping the sample space $\Omega$ of possible outcomes of a random process to the space of real numbers. 
\end{definition}

\example{When tossing a coin once, the RV $X$ can be defined as}
\be
X = \begin{cases}
0, & \text{if coin lands T}\\
1, & \text{if coin lands H}.
\end{cases}
\ee
When rolling a regular, 6-sided die, the RV $X$ can be defined as
\be \label{eq:RV_die}
X = \begin{cases}
1, & \text{if a 1 is rolled}\\
2, & \text{if a 2 is rolled}\\
3, & \text{if a 3 is rolled}\\
4, & \text{if a 4 is rolled}\\
5, & \text{if a 5 is rolled}\\
6, & \text{if a 6 is rolled}.\\
\end{cases}
\ee
When drawing one ball from an urn containing black and white balls, the RV $X$ can be defined as
\be
X = \begin{cases}
0, & \text{if the ball drawn is white}\\
1, & \text{if the ball drawn is black}.
\end{cases}
\ee

A RV can be discrete (only a countable number of outcomes is possible, such as in coin tossing) or continuous (an uncountable number of outcomes is possible, such as in a temperature measurement). It is mathematically subtle to carry out the passage from a discrete to a continuous RV, although as physicists we won't bother too much with mathematical rigour here. Heuristically, we simply replace summation sums over discrete variables with integrals over continuous variables.  

\begin{definition}
Each RV has an associated {probability distribution} to it. The probability distribution of a discrete RV is called {probability mass function} (pmf), which gives the probability of each outcome: $P(X=x_i) = P_i$ gives the probability of the RV $X$ assuming the value $x_i$. In the following we shall use the shorthand notation $P(x_i)$ to mean $P(X=x_i)$. 
\end{definition}

\example {If} $X$ is the RV of Eq.~\eqref{eq:RV_die}, and the die being tossed is fair, then $P_i = 1/6$ for $i=1, \dots, 6$, where $x_i$ is the outcome ``a the face with $i$ pips comes up''.

The probability distribution associated with a continuous RV is called the {probability density function} (pdf), denoted by $p(X)$. The quantity $p(x)dx$ gives the probabilty that the RV $X$ assumes the value between $x$ and $x+dx$.

The choice of probability distribution to associate to a given random process is dictated by the nature of the random process one is investigating (a few examples are given below). 

For a discrete pmf, the {cumulative probability distribution function} (cdf) is given by
\be
C(x_i) = \sum_{j=1}^{i} P(x_j).
\ee
The cdf gives the probabilty that the RV $X$ takes on a value less than or equal to $x_i$, i.e. $C(x_i) = P(X\leq x_i)$. 

For a continuous pdf, the cdf is given by
\be
P(x) = \int_{-\infty}^{x}p(y)dy,
\ee 
with the same interpretation as above, i.e. it is the probability that the RV $X$ takes a value smaller than $x$.

When we make a measurement, (e.g., the temperature of an object, or we toss a coin and observe which face comes up), nature selects an outcome from the sample space with probability given by the associated pmf or pdf. The selection of the outcome is such that if the measurement was repeated an infinite number of times the relative frequency of each outcome is the same as the the probability associated with each outcome under the pmf or pdf. This is another formulation of the frequentist definition of probability given above.

Outcomes of measurements realized by nature are called samples\footnote{The probability theory notion of sample encountered here is not to be confused with the idea of MCMC (posterior) samples, which we will introduce later in section~\ref{sec:MCMC}.}. They are a series of real (or integer) numbers, $\{\hat{x}_1, \hat{x}_2, \dots, \hat{x}_N \}$. In this notes, I will denote samples (i.e., measured values) with a hat symbol, $\hat{\phantom{a}}$.

Definitions and background material on some of the most important and most commonly-encountered  sampling distributions (the uniform, Poisson, Binomial, exponential and Gaussian distributions) are given in Appendix~\ref{app:background}.

\subsection{The Central Limit Theorem} \label{sec:clt}

The Central Limit Theorem (CLT) is a very important result justifying why the Gaussian distribution is ubiquitous. 

\begin{theorem} 
Simple formulation of the CLT: Let $X_1, X_2, \dots, X_N$ be a collection of independent RV with finite expectation value $\mu$ and finite variance $\si^2$. Then, for $N\rightarrow \infty$, thir sum is Gaussian distributed with mean $N\mu$ and variance $N\si^2$.    
\end{theorem}

Note: it does not matter what the detailed shape of the underlying pdf for the individual RVs is! 

Consequence: whenever a RV arises as the sum of several independent effects (e.g., noise in a temperature measurement), we can be confident that it will be very nearly Gaussian distributed. 

\begin{theorem}
More rigorous (and more general) formulation of the CLT:  Let $X_1, X_2, \dots, X_N$ be a collection of independent RV, each with finite expectation value $\mu_i$ and finite variance $\si^2_i$. Then the variable 
\be
Y = \frac{\sum_{i=1}^{N}X_i - \sum_{i=1}^N{\mu_i}}{\sum_{i=1}^N{\si_i^2}}
\ee
is distributed as a Gaussian with expectation value 0 and unit variance. 
\end{theorem}

\subsection{The likelihood function}

The problem of {inference} can be stated as follows: given a collection of samples, $\{\hat{x}_1, \hat{x}_2, \dots, \hat{x}_N \}$, and a generating random process, what can be said about the properties of the underlying probability distribution? 

\example{You toss a coin 5 times and obtain 1 head. What can be said about the fairness of the coin?} 

\example{With a photon counter you observe 10 photons in a minute. What can be said about the average photon rate from the source?}

\example{You measure the temperature of an object twice with two different instruments, yielding the following measurements: $T=256 \pm 10$ K and $T=260\pm 5$ K. What can be said about the temperature of the object?}

\item Schematically, we have that:
\be
\begin{aligned} 
 \text{pdf - e.g., Gaussian with a given ($\mu, \si$)} & \rightarrow \text{Probability of observation} \\
 \text{Underlying $(\mu, \si)$} & \leftarrow \text{Observed events} 
\end{aligned}
\ee
The connection between the two domains is given by the {likelihood function}. 

\begin{definition}
{Given a pdf or a pmf $p(X | \theta)$, where $X$ represents a random variable and $\theta$ a collection of parameters describing the shape of the pdf\footnote{For example, for a Gaussian $\theta = \{\mu, \si\}$, for a Poisson distribution, $\theta = \lambda$ and for a binomial distribution, $\theta = p$, the probability of success in one trial.} and the observed data $\xhat = \{\hat{x}_1, \hat{x}_2, \dots, \hat{x}_N \}$, the likelihood function $\like$ (or ``likelihood'' for short) is defined as  
\be
\like (\theta) = p(X=\xhat | \theta).
\ee
On the right-hand side of the above equation, the probability (density) of observing the data that have been obtained ($X=\xhat$) is considered {\em as a function of the parameters $\theta$}. A very important -- and often misunderstood! -- point is that the likelihood is {\em not} a pdf in $\theta$. This is why it's called likelihood {\em function}! It is normalised over $X$, but not over $\theta$. }
\end{definition}

\begin{figure}
\begin{center}
\includegraphics[width=0.5\linewidth]{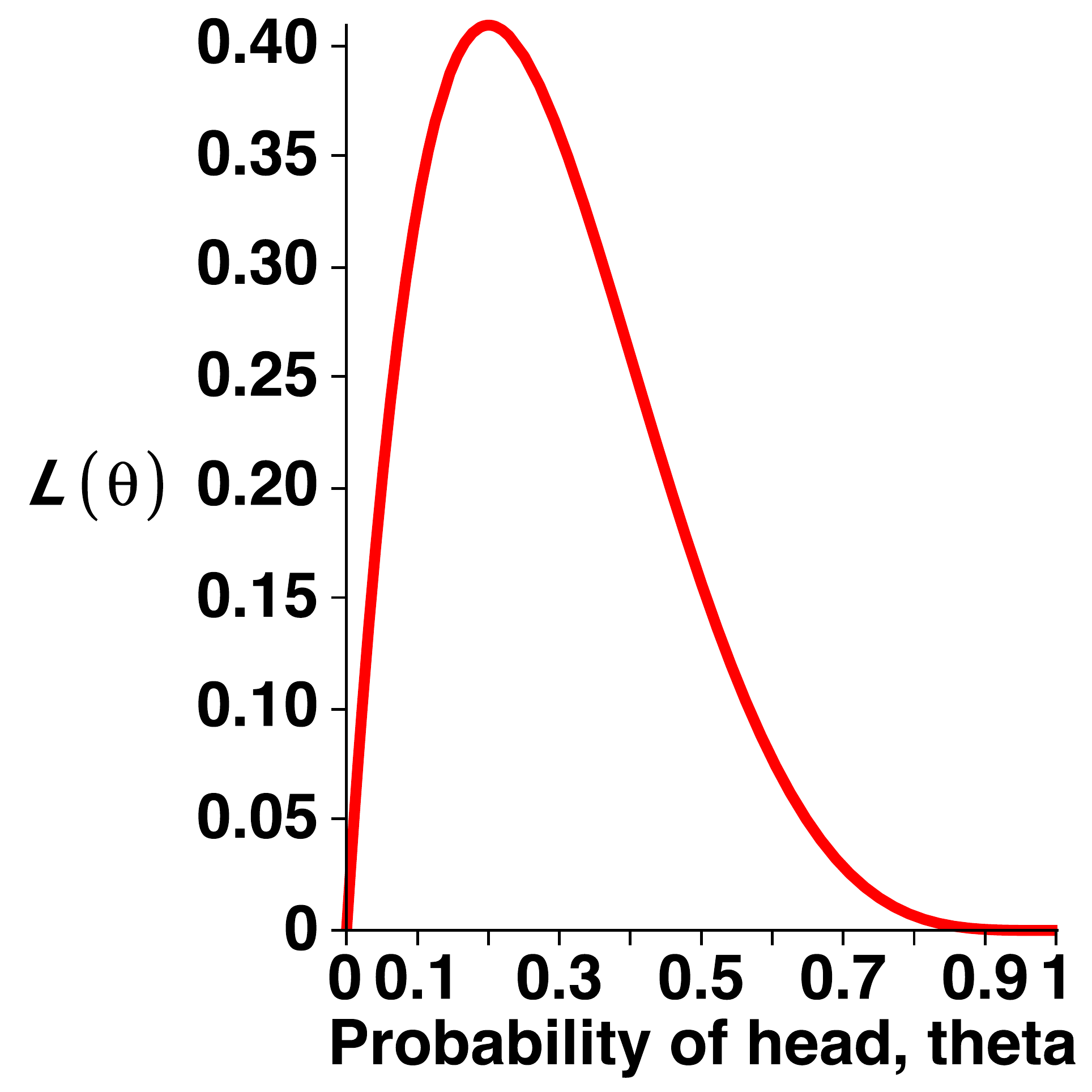} \hfill 
\end{center}
\caption{The likelihood function for the probability of heads ($\theta$) for the coin tossing example, with $N=5, r=1$. }
\label{fig:coinlike}
\end{figure}

\example{In tossing a coin, let $\theta$ be the probability of obtaining heads in one throw. Suppose we make $N=5$ flips and obtain the sequence $\xhat = \{H, T, T, T, T \}$. The likelihood is obtained by taking the binomial, Eq.~\eqref{eq:binomialdef}, and replacing for $r$ the number of heads obtained ($r=1$) in $N=5$ trials, and looking at it {as a function of the parameter we are interested in determining, here $\theta$}. Thus
\be
\like(\theta) = {5\choose1} \theta^1 (1-\theta)^4 = 5 \theta (1-\theta)^4,
\ee
which is plotted as a function of $\theta$ in Fig.~\ref{fig:coinlike}. }

\begin{figure}
\begin{center}
\includegraphics[width=0.5\linewidth]{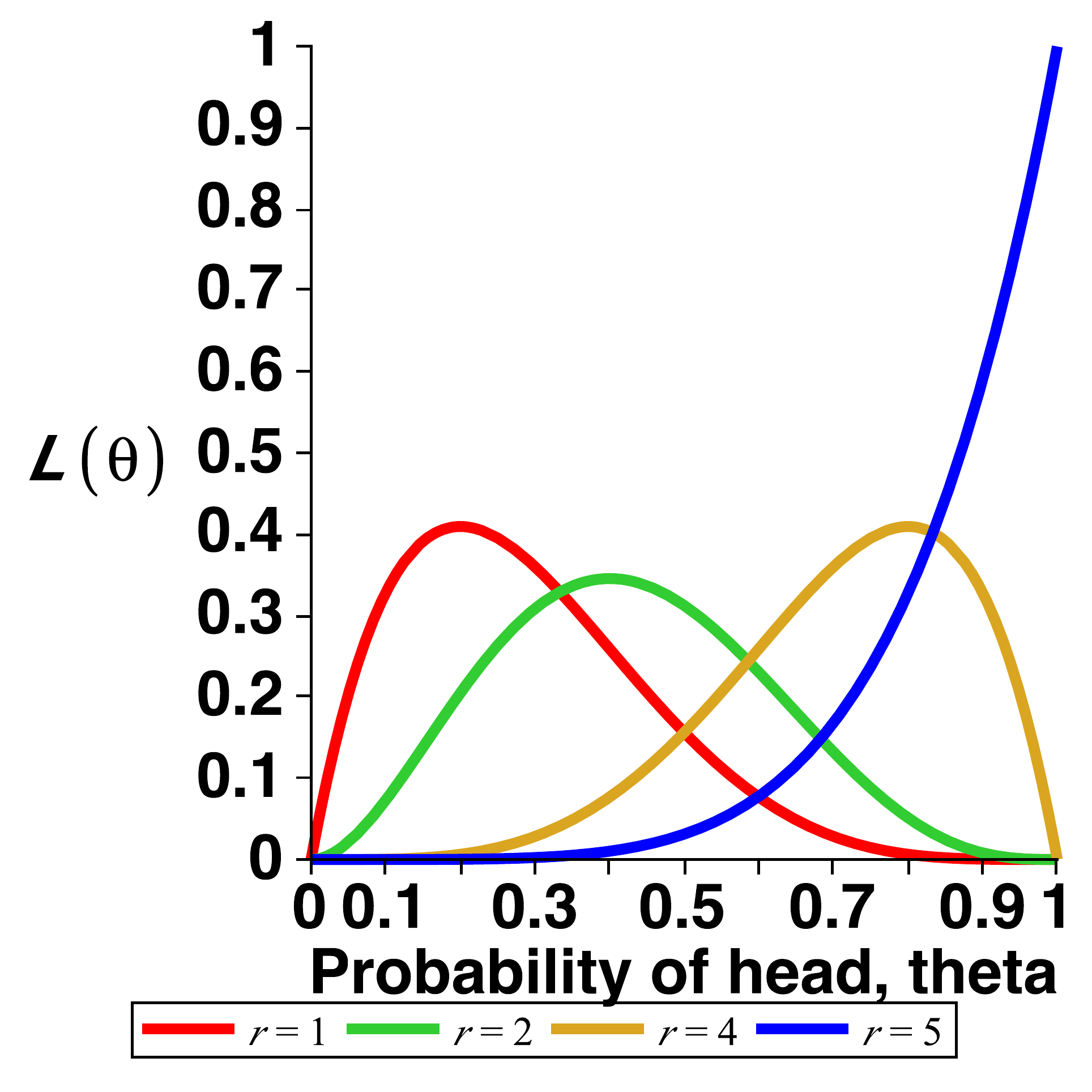} \hfill 
\end{center}
\caption{The likelihood function for the probability of heads ($\theta$) for the coin tossing example, with $n=5$ trials and different values of $r$. }
\label{fig:coinlike2}
\end{figure}

If instead of $r=1$ heads we had obtained a different number of heads in our $N=5$ trials, the likelihood function would have looked as shown in Fig.~\ref{fig:coinlike2} for a few different choices for $r$.

This example leads to the formulation of the Maximum Likelihood Principle: if we are trying to determine the value of $\theta$ given what we have observed (e.g., the sequence of H/T in coin tossing), we should choose the value that maximises the likelihood, because this maximises the probability of obtaining the data that we got. Notice that this is {\em not} necessarily the same as maximising the probability of $\theta$. Doing so requires the use of Bayes theorem, see section~\ref{sec:bayes}.

\subsection{The Maximum Likelihood Principle}

The Maximum Likelihood Principle (MLP): given the likelihood function $\like(\theta)$ and seeking to determine the parameter $\theta$, we should choose the value of $\theta$ in such a way that the value of the likelihood is maximised. 

\begin{definition}
The Maximum Likelihood Estimator (MLE) for $\theta$ is 
\be
\theta_\ML \equiv \max_\theta \like(\theta).
\ee
\end{definition}

It can be shown that the MLE as defined above has the following properties: it is asymptotically unbiased (i.e., $\theta_\ML \rightarrow \theta$ for $N\rightarrow \infty$, i.e., the ML estimate converges to the true value of the parameters for infinitely many data points) and it is asymptotically the minimum variance estimator, i.e. the one with the smallest errors. 

To find the MLE, we maximise the likelihood by requiring its first derivative to be zero and the second derivative to be negative:
\be
\frac{\partial \like(\theta)}{\partial \theta}{\Big\vert_{\theta_\ML}} = 0, \qquad \text{and} \qquad \frac{\partial^2 \like(\theta)}{\partial \theta^2}{\Big\vert_{\theta_\ML}} < 0. 
\ee
 \attention  In practice, it is often more convenient to maximise the logarithm of the likelihood (the ``log-likelihood'') instead. Since log is a monotonic function, maximising the likelihood is the same as maximising the log-likelihood. So one often uses 
\be
\frac{\partial \ln \like(\theta)}{\partial \theta}{\Big\vert_{\theta_\ML}} = 0, \qquad \text{and} \qquad \frac{\partial^2 \ln \like(\theta)}{\partial \theta^2}{\Big\vert_{\theta_\ML}} < 0. 
\ee

\example{MLE of the mean of a Gaussian. Imagine we have $N$ independent measurements of a Gaussian-distributed quantity, and let's denote them by $\{\hat{x}_1, \hat{x}_2, \dots, \hat{x}_N \}$. Here the parameters we are interested in determining are $\mu$ (the mean of the distribution) and $\sigma$ (the standard deviation of the distribution), hence we write $\theta = \{\mu, \si\}$.Then the joint likelihood function is given by
\be \label{eq:Gauss_Nsamples}
\like(\mu, \sigma) = p(\xhat | \mu, \sigma) = \prod_{i=1}^N \frac{1}{\sqrt{2\pi}\si}\exp\left(-\frac{1}{2}\frac{(\hat{x}_i - \mu)^2}{\si^2}\right),
\ee}
Often, the expression above is written as 
\be 
\like = L_0 \exp\left(-\chi^2/2\right)
\ee
where the so-called ``chi-squared'' is defined as 
\be \chi^2 = \sum_{i=1}^N \frac{(\hat{x}_i - \mu)^2}{\si^2}.
\ee

We want to estimate the (true) mean of the Gaussian. The MLE for the mean is obtained by solving 
\be \label{eq:muML}
\frac{\partial \ln \like}{\partial \mu} = 0 \Rightarrow \mu_\ML = \frac{1}{N}\sum_{i=1}^N\hat{x}_i,
\ee
i.e., the MLE for the mean is just the sample mean (i.e., the average of the measurements). 

\example{MLE of the standard deviation of a Gaussian.  If we want to estimate the standard deviation $\sigma$ of the Gaussian, the MLE for $\sigma$ is:
\be \label{eq:sigmaML}
\frac{\partial \ln \like}{\partial \sigma} = 0 \Rightarrow \si^2_\ML = \frac{1}{N}\sum_{i=1}^N(\hat{x}_i - \mu)^2.
\ee
However, the MLE above is ``biased'', i.e. it can be shown that
\be 
E(\sigma^2_\ML) = \left(1-\frac{1}{N}\right)\sigma^2 \neq \sigma^2,
\ee 
where $E(\dot)$ denotes the expectation value. I.e., for finite $N$ the expectation value of the ML estimator is not the same as the true value, $\sigma^2$. In order to obtain an unbiased estimator we replace the factor $1/N$ by $1/(N-1)$. Also, because the true $\mu$ is usually unknown, we replace it in Eq.~\eqref{eq:sigmaML} by the MLE estimator for the mean, $\mu_\ML$. }

Therefore, {the unbiased MLE estimator for the variance} is 
\be \label{eq:sigmaML_unbiased}
 \hat{\si}^2 = \frac{1}{N-1}\sum_{i=1}^N(\hat{x}_i - \mu_\ML)^2.
\ee
In general, you should always use Eq.~\eqref{eq:sigmaML_unbiased} as the ML estimator for the variance, and {not} Eq.~\eqref{eq:sigmaML}.

\example{MLE for the success probability of a binomial distribution. We go back to the coin tossing example, but this time we solve it in all generality. Let's define ``success'' as ``the coin lands heads'' (H). Having observed H heads in a number $N$ of trials, the likelihood function of a binomial is given by Eq.~\eqref{eq:binomialdef}, where the unknown parameter is $\theta$ (the success probability for one trial, i.e., the probability that the coin lands H):
\begin{equation}
\mathcal{L}(\theta) = P(H|\theta,N) = {N\choose{H}} \theta^{H}(1-\theta)^{N-H},
\end{equation} 
The Maximum Likelihood Estimator the success probability is found by maximising the log likelihood: 
\begin{equation}
\begin{split}
\frac{\partial \ln \mathcal{L}(\theta)}{\partial \theta} & =  \frac{\partial}{\partial \theta} \left(\ln {N\choose{H}} + H \ln \theta + (N-H)\ln(1-\theta) \right)  = \frac{H}{\theta} - \frac{N-H}{1-\theta} \stackrel{!}{=} 0 \\ 
& \Leftrightarrow \theta_\text{ML} = \frac{H}{N}.
\end{split}
\end{equation} 
Thus the MLE is simpy given by the observed fraction of heads, which is intuitively obvious. }

\example{MLE for the rate of a Poisson distribution. The likelihood function is given by Eq.~\eqref{eq:poisson}, using the notation $\theta = \lambda$ (i.e., the parameter $\theta$ we are interested in is here the rate $\lambda$):
\begin{equation}
{\mathcal L}(\lambda) = P(n| \lambda) = \frac{(\lambda t)^n}{n!}\exp(-\lambda t),
\end{equation} 
The unknown parameter is the rate $\lambda$, while the data are the observed counts, $n$, in the amount of time $t$.
The Maximum Likelihood Estimate for $\lambda$ is obtained by finding the maximum of the log likelihood as a function of the parameter (here, the rate $\lambda$). Hence we need to find the value of $\lambda$ such that: 
\be 
\frac{\partial \ln P(n|\lambda)}{\partial \lambda} = 0.
\ee
The derivative gives 
\be 
\frac{\partial \ln P(n|\lambda)}{\partial \lambda} = \frac{\partial}{\partial \lambda}\left(n \ln(\lambda t) - \ln n! - \lambda t \right) = 
n \frac{t}{\lambda t} -  t = 0 \Leftrightarrow \lambda_{MLE} = \frac{n}{t}.
\ee
So the maximum likelihood estimator for the rate is the observed average number of counts.} 

We can thus summarise the MLE recipe: \attention 
\begin{enumerate}
\item Write down the likelihood. This depends on the kind of random process you are considering. Identify what is the parameter that you are interested in, $\theta$.
\item Find the ``best fit'' value of the parameter of interest by maximising the likelihood $\like$ as a function of $\theta$. This is your MLE, $\theta_\ML$.
\item Evaluate the uncertainty on $\theta_\ML$, i.e. compute the confidence interval (see next section).
\end{enumerate}

\subsection{Confidence intervals (frequentist)} 

Consider a general likelihood function, $\like(\theta)$ and let us do a Taylor expansion of the log-likelihood $\ln\like$ around its maximum, given by $\theta_\ML$: 
\be
\ln \like(\theta) = \ln \like(\theta_\ML) + \frac{\partial \ln \like(\theta)}{\partial \theta}{\Big\vert_{\theta_\ML}}(\theta-\theta_\ML) + \frac{1}{2} \frac{\partial^2\ln \like(\theta)}{\partial \theta^2}{\Big\vert_{\theta_\ML}}(\theta-\theta_\ML)^2  + \dots
\ee 
The second term on the RHS vanishes (by definition of the Maximum Likelihood value), hence we can approximate the likelihood as
\be \label{eq:likegauss}
\like(\theta) \approx \like(\theta_\ML) \exp\left(-\frac{1}{2}\frac{(\theta-\theta_\ML)^2}{\Sigma_\theta^2}\right) + \dots,
\ee
 \attention  with 
\be \label{eq:Sigma}
\frac{1}{\St^{2}} =  -\frac{\partial^2 \ln \like(\theta)}{\partial \theta^2}{\Big\vert_{\theta_\ML}}.
\ee
A general likelihood function can be approximated to second order as a Gaussian around the ML value, as shown by Eq.~\eqref{eq:likegauss}. Therefore, to the extent that this second order Taylor expansion is sufficiently accurate, the uncertainty around the ML value, $\St$, is approximately given by Eq.~\eqref{eq:Sigma}.

\item \example{Let's go back to the Gaussian problem of Eq.~\eqref{eq:Gauss_Nsamples}. We have seen in Eq.~\eqref{eq:muML} that the sample mean is the MLE for the mean of the Gaussian. We now want to compute the uncertainty on this value. Applying Eq.~\eqref{eq:Sigma} to the likelihood of Eq.~\eqref{eq:Gauss_Nsamples} we obtain  \attention 
\be \label{eq:Gaussian_sigma}
\Sigma_\mu^2 = \sigma^2/N.
\ee
This means that the the uncertainty on our ML estimate for $\mu$ (as expressed by the standard deviation $\Sigma_\mu$) is proportional to $1/\sqrt{N}$, with $N$ being the number of measurements.}

\attention  As the likelihood function can be approximated as a Gaussian (at least around the peak), we can use the results for a Gaussian distribution to approximate the probability content of an interval around the ML estimate for the mean. The interval $[ \mu_\text{min}, \mu_\text{max}]$ is called a $100\alpha$\% {confidence interval} for the mean $\mu$ if $P(\mu_\text{min} < \mu < \mu_\text{max}) = \alpha$. 

\example{For example, the interval $[\mu_\ML - \Sigma_\mu < \mu < \mu_\ML + \Sigma_\mu]$ is a 68.3\% confidence interval for the mean (a so-called ``$1\si$ interval''), while $[\mu_\ML - 2\Sigma_\mu < \mu < \mu_\ML + 2\Sigma_\mu]$ is a 95.4\% confidence interval (a ``$2\si$ interval'').} 

\example{In the temperature measurement example of Eq.~\eqref{eq:T_meas}, the 68.3\% confidence interval for the mean is $198.0 \text{K}< \mu < 201.2 \text{K}$. The 95.4\% confidence interval is $196.4 \text{K}< \mu < 202.8 \text{K}$. 

Generally, the value after the  ``$\pm$'' sign will usually give the 1$\sigma$ (i.e., 68.3\%) region. Sometimes you might find a notation like $50 \pm 1$ (95\% CL), where ``CL'' stands for ``Confidence Level''. In this case, $\pm 1$ encompasses a region of 95\% confidence (rather than 68.3\%), which corresponds to 1.96 $\sigma$ (see Table \ref{tab:gauss_interval}).}

In the multi-dimensional case, additional parameters are eliminated from the likelihood by profiling over them, i.e., maximising over their value. 

\begin{definition}
The profile likelihood for the parameter $\params_1$ (without loss of generality) is defined as 
\be \label{eq:profile}
\like(\params_1) \equiv \max_{\params_2, \dots, \params_N} \like(\params), 
\ee 
where in our case $\like(\params)$ is the full
likelihood function. 
\end{definition}

Thus in the profile likelihood one maximises the
value of the likelihood along the hidden dimensions, rather than
integrating it out as in the marginal posterior (see Eq.~\eqref{eq:marginal_posterior} below). 

The profile likelihood can be directly interpreted as a if it were a genuine likelihood
function, except that it does account for the effect of the
hidden parameters.

Confidence intervals from the
profile likelihood can be obtained via the likelihood ratio test as
follows. 

Classical confidence intervals based on the Neyman
construction are defined as the set of parameter points in which some
real-valued function, or \textit{test statistic}, $t$ evaluated on the
data falls in an acceptance region $W_{\theta} = [t_-, t_+]$.  Likelihood ratios are
often chosen as the test statistic on which frequentist intervals are
based.  When ${\theta}$ is composed of parameters of interest,
$\theta$, and nuisance parameters, $\mbox{$\psi$}$, a common choice of test
statistic is the profile likelihood ratio
\begin{equation}
\lambda(\theta) \equiv \frac{\mathcal{L}(\theta, \hat{\hat{\mbox{$\psi$}}})}{\mathcal{L}(\hat{\theta}, \hat{\mbox{$\psi$}})}.
\label{eq:profile_like}
\end{equation}
where $\hat{\hat{\mbox{$\psi$}}}$ is the conditional maximum likelihood estimate
(MLE) of $\mbox{$\psi$}$ with $\theta$ fixed and $\hat{\theta}, \hat{\mbox{$\psi$}}$ are
the unconditional MLEs.  Under certain regularity conditions\footnote{One important and often-overlooked condition for the validity of Wilks' theorem is that the parameter it is being applied to cannot lie at the boundary of the allowed parameter space. In this case, one ought to employ Chernoff's theorem instead~\cite{Chernoff:1954}. A modern discussion of the regularity conditions necessary for the asymptotic distribution of the likelihood ratio test statistics to be valid can be found in~\cite{Protassov:2002sz}.}, Wilks
showed~\cite{Wilks} that the distribution of $-2\ln\lambda(\theta)$ converges to a
chi-square distribution with a number of degrees of freedom given by
the dimensionality of $\theta$.  

This leads to the following prescription. Starting from the best-fit value in parameter space, an
$\alpha$\% confidence interval encloses all parameter values for which
minus twice the log--likelihood increases less than $\Delta \chisq(\alpha, n)$
from the best fit value. The threshold value depends on $\alpha$ and
on the number $n$ of parameters one is simultaneously considering
(usually $n=1$ or $n=2$), and it is obtained by solving
\be \alpha = \int_0^{\Delta \chisq} \chisq_n (x) d x, \ee 
where $\chisq_n(x)$ is the chi--square distribution for $n$ degrees of
freedom, Eq.~\eqref{eq:chisquare}.

\begin{svgraybox}
One has to be careful with the interpretation of confidence intervals as this is often misunderstood!

Interpretation: if we were to repeat an experiment many times, and each time report the observed $100\alpha\%$ confidence interval, we would be correct $100\alpha\%$ of the time. This means that (ideally) a $100\alpha\%$ confidence intervals contains the true value of the parameter $100\alpha\%$ of the time. 
\end{svgraybox}

In a frequentist sense, it does {not} make sense to talk about ``the probability of $\theta$''. This is because every time the experiment is performed we get a different realization (different samples), hence a different numerical value for the confidence interval. Each time, either the true value of $\theta$ is inside the reported confidence interval (in which case, the probability of $\theta$ being inside is 1) or the true value is outside (in which case its probability of being inside is 0). {Confidence intervals do not give the probability of the parameter!} In order to do that, you need Bayes theorem.


%
\subsection{Exercises} 

These exercises are designed to help you put into practice the above introductory concepts. Please make sure you are familiar with these notions before moving on to the next section. Exercises that are a little more challenging are denoted with a $\dagger$. 
\begin{enumerate}

\item Gaussian 1D problem. The surface temperature on Mars is measured by a probe 10 times, yielding the following data (units of K): 
\begin{equation}
  191.9, 201.6,  206.1,  200.4, 203.2, 201.6, 196.5, 199.5, 194.1, 202.4
\end{equation} 
\begin{enumerate}
\item Assume that each measurement is independently Normally distributed with known variancee $\sigma^2 = 25$ K$^2$. What is the likelihood function for the whole data set?  %
\item Find the Maximum Likelihood Estimate (MLE) for the surface temperature, $T_{\rm ML}$, and express your result to 4 significant figures accuracy. 
\item Determine symmetric confidence intervals at 68.3\%, 95.4\% and 99\% around $T_{\rm ML}$ (4 significant figures accuracy).
\item How many measurements would you need to make if you wanted to have a $1\sigma$ confidence interval around the mean of length  less than 1 K (on each side)?  
\end{enumerate}

\item The surface temperature on Mars is measured by a probe 10 times, yielding the following data (units of K): 
\begin{equation} \label{eq:T_meas}
197.2, 202.4, 201.8, 198.8, 207.6, 191.4, 201.4, 198.2, 195.7, 201.2.
\end{equation} 
\begin{enumerate}
\item Assuming that each measurement is independently Gaussian distributed with known variance $\sigma^2 = 25$ K$^2$, what is the likelihood function for the whole data set? 
\item What is the MLE of the mean, $T_{ML}$? 
\item What is the uncertainty on our MLE for the mean?
\end{enumerate}
\item A laser beam is used to measure the deviation of the distance between the Earth and the Moon from its average value, giving the following data, in units of cm:
\begin{equation}
119, \quad  119, \quad  122, \quad  121, \quad  116.
\end{equation}
\begin{enumerate}
\item Assuming that each measurement above follows an independent Gaussian distribution of known standard deviation $\sigma = 3$ cm, write down the joint likelihood function for $\Delta$, the deviation of the Earth-Moon distance from its average value. 
\item Compute the maximum likelihood estimate for $\Delta$ and its uncertainty, both to 3 significant figures. 
\item How would you report the measurement of $\Delta$ (giving a 1-$\sigma$ confidence interval)? 
\end{enumerate}
\item  
You flip a coin $n=10$ times and you obtain 8 heads. 
\begin{enumerate}
\item What is the likelihood function for this measurement? Identify explicitly what are the data and what is the free parameter you are trying to estimate. \\
\item What is the Maximum Likelihood Estimate for the probability of obtaining heads in one flip, $p$?
\item Approximate the likelihood function as a Gaussian around its peak and derive the 1$\sigma$ confidence interval for $p$. How would you report your result for $p$?
\item With how many $\sigma$ confidence can you exclude the hypothesis that the coin is fair? ({\em Hint: compute the distance between the MLE for $p$ and $p=1/2$ and express the result in number of $\sigma$}).
\item You now flip the coin 1000 times and obtain 800 heads. What is the MLE for $p$ now and what is the 1$\sigma$ confidence interval for $p$? With how many $\sigma$ confidence can you exclude the hypothesis that the coin is fair now?
\end{enumerate}

\item \label{counts} 
An experiment counting particles emitted by a radioactive decay measures $r$ particles per unit time interval. The counts are Poisson distributed. 
\begin{enumerate}
\item If $\lambda$ is the average number of counts per per unit time interval, write down the appropriate probability
  distribution function for $r$. 
\item Now we seek to determine $\lambda$ by repeatedly measuring for $M$ times the number of counts per unit time interval. This series of measurements yields a sequence of counts ${\hat{r} } = \{\hat{r}_1,\,\hat{r}_2,\,\hat{r}_3,\,...,\,\hat{r}_M \}$. Each measurement is assumed to be independent. Derive the joint likelihood function for $\lambda$, $\mathcal{L}(\lambda) = P({\hat{r}} | \lambda)$, given the measured sequence of counts ${\hat{r}}$.
\item Use the Maximum Likelihood Principle applied to the
  the log likelihood $ \ln \mathcal{L}(\lambda)$ to show that
  the Maximum Likelihood estimator for the average rate $\lambda$ is just
  the average of the measured counts, ${\hat{r}}$, i.e.
  \begin{displaymath}
    {\lambda}_{\rm ML}=\frac{1}{M}\sum_{i=1}^{M} \hat{r}_i\, .
  \end{displaymath}
\item By considering the Taylor expansion of $\ln {\cal L}(\lambda)$ to second
  order around ${\lambda}_{\rm ML}$, derive the Gaussian approximation for the likelihood $\mathcal{L}(\lambda)$ around
  the Maximum Likelihood point, and show that it can be written as
    \begin{displaymath}
      {\mathcal L}(\lambda) \approx L_0 \exp\left( - \frac{1}{2}\frac{M}{\lambda_{\rm ML}} (\lambda - {
        \lambda}_{\rm ML})^2\right)\,,
    \end{displaymath}
  where $L_0$ is a normalization constant.
\item    Compare with the equivalent expression for $M$ Gaussian-distributed measurements to show that the variance $\sigma^2$ of the Poisson distribution
  is given by $\sigma^2 = \lambda$. 
\end{enumerate}
\item An astronomer measures the photon flux from a distant star using a very sensitive instrument that counts single photons. After one minute of observation, the instrument has collected $\hat{r}$ photons. One can assume that the photon counts, $\hat{r}$, are distributed according to the Poisson distribution. The astronomer wishes to determine $\lambda$, the emission rate of the source.

\begin{enumerate}
\item What is the likelihood function for the measurement? Identify explicitly what is the unknown parameter and what are the data in the problem.
\item If the true rate is $\lambda = 10$ photons/minute, what is the probability of observing $\hat{r}=15$ photons in one minute? 
\item Find the Maximum Likelihood Estimate for the rate $\lambda$ (i.e., the number of photons per minute). What is the maximum likelihood estimate if the observed number of photons is $\hat{r}=10$?
\item Upon reflection, the astronomer realizes that the photon flux is the superposition of photons coming from the star plus ``background'' photons coming from other faint sources within the field of view of the instrument. The background rate is supposed to be known, and it is given by $\lambda_b$ photons per minute (this can be estimated e.g. by pointing the telescope away from the source and measuring the photon counts there, when the telescope is only picking up background photos). She then points to the star again, measuring $\hat{r}_t$ photons in a time $t_t$. What is her maximum likelihood estimate of the rate $\lambda_s$ from the star in this case? 
{\em Hint:  The total number of photons $\hat{r}_t$ is Poisson distributed with rate $\lambda = \lambda_s + \lambda_b$, where $\lambda_s$ is the rate for the star.}
\item What is the source rate (i.e., the rate for the star) if $\hat{r}_t = 30$, $t_t = 2$ mins, and $\lambda_b = 12$ photons per minute? Is it possible that the measured average rate from the source (i.e., $\hat{r}_t/t_t$) is less than $\lambda_b$?  Discuss what happens in this case and comment on the physicality of this result. 
\end{enumerate}

\item This problem generalizes the Gaussian measurement case to the case where the measurements have different uncertainties among them.

You measure the flux $F$ of photons from a laser source using 4 different instruments and you obtain the following results (units of $10^4$ photons/cm$^2$):
\begin{equation}
34.7 \pm 5.0 , \quad 28.9 \pm 2.0, \quad  27.1 \pm 3.0, \quad  30.6 \pm 4.0. 
\end{equation} 
\begin{enumerate}
\item Write down the likelihood for each measurement, and explain why a Gaussian approximation is justified in this case. 
\item Write down the joint likelihood for the combination of the 4 measurements. 
\item Find the MLE of the photon flux, $F_\text{ML}$, and show that it is given by:
\be
F_\text{ML}  = \sum_i  \frac{\hat{n}_i}{\hat{\sigma}_i^2/\bar{\sigma}^2},
 \ee
 where
 \be 
 \frac{1}{\bar{\sigma}^2} \equiv \sum_i \frac{1}{\hat{\sigma}_i^2}.
 \ee
\item Compute $F_\text{ML}$ from the data above and compare it with the sample mean.
\item Find the $1\sigma$ confidence interval for your MLE for the mean, and show that it is given by:
\be
\left(\sum_i \frac{1}{\hat{\sigma}_i^2}\right)^{-1/2}.
\ee
Evaluate the confidence interval for the above data. How would you summarize your measurement of the flux $F$? 
\end{enumerate} 
\end{enumerate}

\subsection{Solutions to exercises} 
\begin{enumerate}

\item 
\begin{enumerate}
\item The measurements are independent, hence the joint likelihood is the product of the likelihoods for each measurement:
\be 
{\mathcal L}_{\rm tot}(T) = \prod_{i=1}^{10} \frac{1}{\sqrt{2\pi}\sigma} \exp\left(-\frac{1}{2}\frac{(\hat{T}_i - T)^2}{\sigma^2}\right)
\ee
where $\hat{T}_i$ are the data given, $T$ is the temperature we are trying to determine (unknown parameter) and $\sigma = 5$ K.  

\item The MLE for the mean of a Gaussian is given by the mean of the sample, see Eq.~\eqref{eq:muML}, hence 
\be
T_{\rm ML} = \frac{1}{10}\sum_{i=1}^{10} T_i = 199.7 {\rm K}.
\ee
\item The variance of the mean is given by $\sigma^2/N$, see Eq.~\eqref{eq:Gaussian_sigma}. Therefore the standard deviation of our estimate $T_{\rm ML}$ is given by $\Sigma_T = \sigma/\sqrt{N} = 5/\sqrt{10}$ = 1.58 K, which corresponds to the 68.3\% interval: $199.7 \pm 1.6$ K, i.e. the range $[198.1, 201.3]$ K (4 s.f. accuracy). Confidence intervals at 95.4\% and 99\% corresponds to symmetric intervals around the mean of length 2.0 and 2.57 times the standard deviation $\Sigma_T$. Hence the required confidence intervals are $[196.5, 202.9]$ K (95.4\%) and $[195.6, 203.8]$ K (99\%).
\item 
A $1\sigma$ confidence interval lenght 1 K means that the value of $\Sigma_T$ should be 1 K. Using that the standard deviation scales as $1/\sqrt{N}$, we have
\be 
1 = 5/\sqrt{N} \Rightarrow N = 25.
\ee  
You would need $N=25$ measurements to achieve the desired accuracy.
\end{enumerate}

\item 
\begin{enumerate}
\item The measurements are independent, hence the joint likelihood is the product of the likelihoods for each measurement, see Eq.~\eqref{eq:Gauss_Nsamples}:
\be 
\like(T) = \prod_{i=1}^{10} \frac{1}{\sqrt{2\pi}\sigma} \exp\left(-\frac{1}{2}\frac{(\hat{T}_i - T)^2}{\sigma^2}\right)
\ee
\item  the MLE for the mean of a Gaussian is given by the mean of the sample, see Eq.~\eqref{eq:muML}, hence 
\be
T_{\rm ML} = \frac{1}{10}\sum_{i=1}^{10} \hat{T}_i = 199.6 K. \\
\ee
\item The variance of the mean is given by $\Sigma_\mu^2 = \sigma^2/N$, where $\sigma^2 = 25$ K$^2$ and $N=10$. Therefore the standard deviation of our temperature estimate $T_{\rm ML}$ is given by $\Sigma_T = 5/\sqrt{10}$ = 1.6 K. The measurement can thus be summarized as $T = 199.6 \pm 1.6$ K, where the $\pm 1.6$ K gives the range of the $1\sigma$ (or 68.3\%) confidence interval.
\end{enumerate}
%
\item 
\begin{enumerate}
\item The joint Gaussian likelihood function for $\Delta$ is given by
\begin{equation}
P(\Delta | d) \equiv \like(\Delta)= \prod_{i=1}^5 \frac{1}{\sqrt{2\pi} \sigma} \exp\left(-\frac{1}{2} \frac{(\Delta - d_i)^2}{\sigma^2} \right),
\end{equation}
where $\sigma = 3$ cm and $d_i$ are the measurements given in the question. 
 
\item The maximum likelihood estimate for $\Delta$ is found by maximising the log-likelihood function wrt $\Delta$: 
\begin{equation}
\frac{\partial \ln  \like}{\partial \Delta} = - \sum_{i=1}^5 \frac{\Delta - d_i}{\sigma^2} = 0 \rightarrow \Delta_{\rm MLE} = \frac{1}{N} \sum_{i=1}^5 d_i
\end{equation}
The numerical value is $\Delta_{\rm MLE} = 119.4 \text{ cm} \approx 119$ (cm, 3 s.f.).

The uncertainty $\Sigma$ on $\Delta$ is estimated from the inverse curvature of the log likelihood function at the MLE point: 
\begin{equation}
 - \frac{\partial^2 \ln  \like}{\partial \Delta^2} =  \frac{N}{\sigma^2} \rightarrow \Sigma =  \left(- \frac{\partial^2 \ln  \like}{\partial \Delta^2}\right)^{-1/2} = \frac{\sigma}{\sqrt{N}}
 \end{equation}
Numerically this gives $\Sigma = 3/\sqrt{5} = 1.34 \approx  1$ cm. 

\item  The measurement of $\Delta$ would be reported as $\Delta = (119 \pm 1)$ cm.
\end{enumerate}
\item 
\begin{enumerate}
\item The likelihood function is given by
\be
\mathcal{L}(p) = P(r=H|p,n) = {n\choose{H}} p^{H}(1-p)^{n-H},
\ee
where the unknown parameter is $p$ and the data are the number of heads, $H$ (for a fixed number of trials, $n=10$ here). 

\item The Maximum Likelihood Estimator (MLE) for the success probability $p$ is found by maximising the log likelihood:
\be 
\begin{split}
\frac{\partial \ln \mathcal{L}(p)}{\partial p} & =  \frac{\partial}{\partial p} \left(\ln {n\choose{H}} + H \ln p + (n-H)\ln(1-p) \right)  = \frac{H}{p} - \frac{n-H}{1-p} \stackrel{!}{=} 0 \\ 
& \Leftrightarrow p_\text{ML} = \frac{H}{n}.
\end{split}
\ee
Therefore the ML value for $p$ is $p_\text{ML} = 0.8$.

\item We approximate the likelihood function as a Gaussian, with standard deviation given by minus the curvature of the log-likelihood at the peak:
\be \label{eq:gaussapprox}
 \mathcal{L}(p) \approx \mathcal{L}_\text{max} \exp\left(-\frac{1}{2}\frac{(p_\text{ML} - p)^2)}{\Sigma^2} \right),
\ee
where 
\be
\begin{split}
\Sigma^{-2} & = - \frac{\partial^2 \ln\mathcal{L}(p)}{\partial p^2}{\Big\vert_{p=p_\text{ML}}} = - \frac{\partial}{\partial p}\left(\frac{H}{p} - \frac{n-H}{1-p} \right) {\Big\vert_{p=p_\text{ML}}}\\
& = \frac{H-2Hp + p^2 n}{p^2(1-p)^2}{\Big\vert_{p=p_\text{ML}}} = \frac{n}{\frac{H}{n}\left(1-\frac{H}{n}\right)}.
\end{split}
\ee
The $1\sigma$ confidence interval for $p$ is given by $\Sigma = 0.13$. Therefore the result would be reported as $p=0.80\pm 0.13$.

\item Following the hint, the number of $\sigma$ confidence with which the hypothesis that the coin is fair can be ruled out is given by 
\be
\frac{\vert p_\text{ML} - \frac{1}{2}\vert}{\Sigma}= \frac{0.8-0.5}{0.13} = 2.31.
\ee
 Therefore the fairness hypothesis can be ruled out at the $\sim 2.3$ $\sigma$ level. 

\item 
Using above equations, the MLE for the success probability is still $p_\text{ML} = 0.8$, as before. However, the uncertainty is now much reduced, because of the large number of trials. In fact, we get $\Sigma = 0.013$ (notice how the uncertainty has decreased by a factor of $\sqrt{n}$, as expected. I.e., 100 times more trials correspond to a reduction in the uncertainty by a factor of 10). The fairness hypothesis can now be excluded with much higher confidence:s of $p=1/2$, expressed in number of sigmas:
\be
\text{number of sigmas} = \frac{|p_{ML} -  \frac{1}{2}|}{\Sigma} = \frac{0.8 - 0.5}{0.013} = 23.1 \approx 23.
\ee 
This constitutes very strong evidence against the hypothesis that the coin is fair. Notice however that the Gaussian approximation to the likelihood we employed will most probably not be accurate so far into the tails of the likelihood function (i.e., the Taylor expansion on which it is based is a {\em local} expansion around the peak).
\end{enumerate}

\item 
\begin{enumerate}
\item The discrete PMF for the number of counts $r$ of a Poisson process with average rate $\lambda$ is (assuming a unit time, $t=1$ throughout)
    \begin{displaymath}
      P(r) = \frac{\lambda^r}{r!}e^{-\lambda}\, .
    \end{displaymath}
  
\item In this case 
  \begin{displaymath}
    P(\hat{r}_i|\,\lambda) = \frac{\lambda^{\hat{r}_i}}{\hat{r}_i!}e^{-\lambda}\,,
  \end{displaymath}
  for each independent measurement $\hat{r}_i$. So the joint likelihood is given by (as measurements are independent)
  \begin{equation}
    {\mathcal L}(\lambda) = \prod_{i=1}^M P(\hat{r}_i|\,\lambda) = \prod_{i=1}^M \frac{\lambda^{\hat{r}_i}}{\hat{r}_i!}e^{-\lambda}\,.
  \end{equation}

\item The Maximum Likelihood Principle states that the estimator for
  $\lambda$ can be derived by finding the maximum of the likelihood
  function. The maximum is found more easily by considering the log of
  the likelihood
  \begin{displaymath}
   \ln {\cal L}(\lambda) = \sum_{i=1}^M \left[ \hat{r}_i\ln(\lambda)- \ln(\hat{r}_i!) - \lambda \right]\,.
  \end{displaymath}
  with the maximum given by the condition $d{\ln \cal L}/d\lambda = 0$.

  We have
  \begin{eqnarray}
    \frac{d{\ln \cal L}}{d\lambda} &=& \sum_{i=1}^M \left[\frac{\hat{r}_i}{\lambda} -
      1\right]\,\nonumber\\
    &=& \frac{1}{\lambda}\sum_{i=1}^M \hat{r}_i -M\,.\nonumber
  \end{eqnarray}

  So the Maximum Likelihood (ML) estimator for $\lambda$ is 
  \begin{displaymath}
    { \lambda}_{\rm ML} = \frac{1}{M} \sum_{i=1}^M \hat{r}_i\,,
  \end{displaymath}
  which is just the average of the observed counts.

\item The Taylor expansion is
  \begin{displaymath}
    {\ln \cal L}(\lambda) = {\ln \cal L}({ \lambda}_{\rm ML}) + \left.\frac{d{\ln \cal
          L}}{d\lambda}\right|_{\lambda={ \lambda}_{\rm ML}} (\lambda - {
        \lambda}_{\rm ML}) + \frac{1}{2} \left.\frac{d^2{\ln \cal
          L}}{d\lambda^2}\right|_{\lambda={ \lambda}_{\rm ML}} (\lambda - {
        \lambda}_{\rm ML})^2 + \dots \,.
    \end{displaymath}

    By definition the linear term vanishes at the maximum so we just need
    the curvature around the ML point
    \begin{displaymath}
      \frac{d^2{\ln \cal L}}{d\lambda^2} = - \sum_{i=1}^M \frac{\hat{r}_i}{\lambda^2}\,,
    \end{displaymath}
    such that
    \begin{displaymath}
       \left.\frac{d^2{\ln \cal L}}{d\lambda^2}\right|_{\lambda={ \lambda}_{\rm
           ML}} = - \frac{1}{{ \lambda}_{\rm ML}^2}\sum_{i=1}^M
       \hat{r}_i= -\frac{M\lambda_{\rm ML}}{\lambda_{\rm ML}^2}=-\frac{M}{\lambda_{\rm ML}}\,.
    \end{displaymath}

    Putting this into the Taylor expansion gives
    \begin{displaymath}
      {\ln \cal L}(\lambda) = {\ln \cal L}({ \lambda}_{\rm ML}) - \frac{1}{2}\frac{M}{\lambda_{\rm ML}} (\lambda - {
        \lambda}_{\rm ML})^2\,,
    \end{displaymath}
    which gives an approximation of the likelihood function around the
    ML point
    \begin{displaymath}
      {\mathcal L}(\lambda) \approx L_0 \exp\left( - \frac{1}{2}\frac{M}{\lambda_{\rm ML}} (\lambda - {
        \lambda}_{\rm ML})^2\right)\,,
    \end{displaymath}
    (the normalisation constant $L_0$ is irrelevant).

    So the likelihood is approximated by a Gaussian with variance
    \begin{displaymath}
      \Sigma^2  = \frac{{ \lambda}_{\rm ML}}{M}\,.
    \end{displaymath}

\item    
   Comparing this with the standard result for the variance of the mean for the Gaussian case, i.e. 
    \begin{displaymath}
      \Sigma^2 = \frac{\sigma^2}{{M}}\,,
    \end{displaymath}
    where $M$ is the number of measurements and $\sigma$ is the standard deviation of each measurement, we can conclude that
    the variance of the Poisson distribution itself is indeed
    \begin{displaymath}
      \sigma^2 = \lambda\, .
    \end{displaymath}
\end{enumerate}
\item \begin{enumerate}
\item The likelihood function is given by the Poisson distribution evaluated as a function of the parameter, $\lambda$:
\begin{equation}
{\mathcal L}(\hat{r}) = P(\hat{r} | \lambda) = \frac{(\lambda t)^{\hat{r}} }{ \hat{r}!}\exp(-\lambda t),
\end{equation} 
where $t$ is the time of observation in minutes. The unknown parameter is the source strength $\lambda$ (in units of photons/min), while the data are the observed counts, $\hat{r}$.

\item We can compute the requested probability by substituting in the Poisson distribution above the values for $\hat{r}$ and $\lambda$, obtaining:
\be
P(\hat{r}=15 | \lambda = 10, t = 1 \text{ min}) = 0.0347.
\ee

\item 
  The maximum likelihood estimate is obtained by finding the maximum of the log likelihood as a function of the parameter (here, the rate $\lambda$). Hence we need to find the value of $\lambda$ such that: 
\be 
\frac{\partial \ln{\mathcal L}(\hat{r})}{\partial \lambda} = 0.
\ee
The derivative gives 
\be 
\frac{\partial \ln {\mathcal L}(\hat{r})}{\partial \lambda} = \frac{\partial}{\partial \lambda}\left(\hat{r} \ln(\lambda t) - \ln \hat{r}! - \lambda t \right) = 
\hat{r} \frac{t}{\lambda t} -  t = 0 \Leftrightarrow \lambda_{MLE} = \frac{\hat{r}}{t}.
\ee
So the maximum likelihood estimator for the rate is the observed number of counts divided by the time, in agreement with Eq.~\eqref{eq:properties_poisson}. In this case, $t = 1$ min so the MLE for $\lambda$ is 10 photons per minute.

\item  The likelihood function now needs to be modified to account for the fact that the observed counts are the superposition of the background rate and the source rate (the star). According to the hint, the likelihood for the total number counts, $\hat{r}_t$, is Poisson with rate $ \lambda = \lambda_s+\lambda_b$, and thus
\be 
P(\hat{r}_t | \lambda = \lambda_s+\lambda_b)  = \frac{(\lambda t_t)^{\hat{r}_t}}{\hat{r}_t!}\exp(-\lambda t_t).
\ee  
Similarly to what we have done above, the MLE estimate for $\lambda_s$ is found by setting to 0 the derivative of the log likelihood wrt $\lambda_s$: 
\be 
\frac{\partial \ln P(\hat{r}_t|\lambda = \lambda_s+\lambda_b)}{\partial \lambda_s} = \hat{r}_t \frac{t_t}{(\lambda_s + \lambda_b)t_t} - t_t = 0 
\Leftrightarrow \lambda_s = \frac{\hat{r}_t}{t_t} - \lambda_b. 
\ee
So the MLE for the source is given by the observed average total rate ($ \frac{\hat{r}_t}{t_t}$) minus the background rate.

\item 
 Inserting the numerical results, we have that $\lambda_s = 3$. The MLE estimate for $\lambda_s$ gives a negative rate if $\hat{r}_t/t_t < \lambda_b$, which is clearly non-physical. However, this can definitely happen because of downwards fluctuations in the number counts due to the Poisson nature of the signal (even if the background is assumed to be known perfectly). So this is an artefact of the MLE estimator (nothing to do with physics! We {\em know} that the actual physical source rate has to be a non-negative quantity!). The solution is to use Bayes theorem instead. 
\end{enumerate}

\item 
\begin{enumerate}
\item The photon counts follow a Poisson distribution. We know that the MLE for the Poisson distribution is the observed number of counts ($n$) and its standard deviation is $\sqrt{n}$. However, for large $n$ ($\gg 20$) the Poisson distribution is well approximated by a Gaussian of mean $n$ and standard deviation  $\sqrt{n}$. In this case, $n$ is of order $10^5$, hence the standard deviation intrinsic to the Poisson process (the so-called ``shot noise'') is of order $\sqrt{10^5} \approx 3\cdot 10^2$. The quoted experimental uncertainty is much larger than that (of order $10^4$ for each datum), hence we can conclude that the statistical error is dominated by the noise in the detector rather than by the Poisson variance. 

Therefore we can approximate the likelihood for each observation as a Gaussian with mean given by the observed counts $\hat{n}_i$ and standard deviation given by the quoted error, $\hat{\sigma}_i$:
\be
{\mathcal L}_i (F) = \frac{1}{\sqrt{2\pi}\hat{\sigma}_i}\exp\left(-\frac{1}{2} \frac{(F-\hat{n}_i)^2}{\hat{\sigma}_i^2} \right) \quad (i=1,\dots,4).
\ee

\item Since the measurements are independent, the joint likelihood is the product of the 4 terms:
\be
{\mathcal L}(F) = \prod_{i=1}^4 {\mathcal L}_i(F).
\ee

\item 
To estimate the mean of the distribution, we apply the MLE procedure for the mean ($F$), obtaining: 
\be \label{eq:wmean}
\begin{split} 
\frac{\partial \ln {\mathcal L}(F)}{\partial F}  & = - \sum_i  \frac{F-\hat{n}_i}{\hat{\sigma}_i^2} \stackrel{!}{=} 0 \\
 \Leftrightarrow F_\text{ML} & = \sum_i  \frac{\hat{n}_i}{\hat{\sigma}_i^2/\bar{\sigma}^2},
  \end{split}
 \ee
 where
 \be 
 \frac{1}{\bar{\sigma}^2} \equiv \sum_i \frac{1}{\hat{\sigma}_i^2}.
 \ee

 We thus see that the ML estimate for the mean is the mean of the observed counts weighted by the inverse error on each on them (verify that Eq.~\eqref{eq:wmean} reverts to the usual expression for the sample mean for $\hat{\sigma}_i = \hat{\sigma}$ for $(i=1,\dots,4)$, i.e., if all observations have the same error). This automatically gives more weight to observations with a smaller error. 

From the given observations, one thus obtains $F_\text{ML} = 29.2 \times 10^4$ photons/cm$^2$. By comparison the sample mean is $\bar{F} = 30.3  \times 10^4$  photons/cm$^2$. 
 
\item   The inverse variance of the mean is given by the second derivative of the log-likelihood evaluated at the ML estimate:
 \be
 \Sigma^{-2} = -\frac{\partial^2 \ln {\mathcal L}(F)}{\partial F^2}{\Big\vert_{F=F_\text{ML}}} = \sum_i \frac{1}{\hat{\sigma}_i^2}.   
 \ee 
 (again, it is simple to verify that the above formula reverts to the usual $N/\hat{\sigma}^2$ expression if all measurements have the same error). 
 
 Therefore the variance of the mean is given by $\Sigma^2 = 2.16  \times 10^8$  (photons/cm$^2$)$^2$, and the standard deviation is $\Sigma = 1.47  \times 10^4$  photons/cm$^2$. Our measurement can thus be summarized as $F = (29.2 \pm 1.5)  \times 10^4$  photons/cm$^2$.

\end{enumerate}

\end{enumerate}

\section{Bayesian parameter inference} \label{sec:bayes}

In this section we introduce the meaning and practical application of Bayes Theorem, Eq.~\eqref{eq:bayesth}, which encapsulates the notion of {probability as degree of belief}. 

\subsection{Bayes theorem as an inference device} 

As a mathematical result, Bayes Theorem is elementary and uncontroversial. It becomes interesting for the purpose of inference when we replace in Bayes theorem, Eq.~\eqref{eq:bayesth}, $A \rightarrow \theta$ (the parameters) and $B\rightarrow d$ (the observed data, or samples), obtaining  \attention 
\be \label{eq:BT}
P(\theta|d) = \frac{P(d|\theta)P(\theta)}{P(d)}.
\ee
On the LHS, $P(\theta|d)$ is {the posterior probability for $\theta$} (or ``posterior'' for short), and it represents our degree of belief about the value of $\theta$ after we have seen the data $d$. 

On the RHS, $P(d|\theta) = \like(\theta)$ is the likelihood we already encountered. It is the probability of the data given a certain value of the parameters. 

The quantity $P(\theta)$ is {the prior probability distribution} (or ``prior'' for short). It represents our degree of belief in the value of $\theta$ {before} we see the data (hence the name). This is an essential ingredient of Bayesian statistics.  The Bayesian school is divided between ``subjectivists'' (who maintain that the prior is a reflection of the subject state of knowledge of the individual researcher adopting it) and ``objectivists'' (who argue for the use of ``standard'' priors to enforce inter-subjectivity between different researchers). However formulated, the posterior distribution usually converges to a prior-independent regime for sufficiently large data sets.  

In the denominator, $P(d)$ is a normalizing constant (often called ``the evidence'' or ``marginal likelihood''), than ensures that the posterior is normalized to unity:
\be \label{eq:evidence_def}
P(d) = \int d\theta P(d|\theta)P(\theta).
\ee
The evidence is important for Bayesian model selection (see section~\ref{sec:modelselection}). 

\begin{svgraybox}
Interpretation: Bayes theorem relates the posterior probability for $\theta$ (i.e., what we know about the parameter after seeing the data) to the likelihood and the prior (i.e., what we knew about the parameter before we saw the data). It can be thought of as a general rule to update our knowledge about a quantity (here, $\theta$) from the prior to the posterior. \end{svgraybox}

Remember that in general $P(\theta |d) \neq P(d|\theta)$, i.e. the posterior $P(\theta |d)$ and the likelihood $P(d | \theta)$ are two different quantities with different meaning! 

\example{We want to determine if a randomly-chosen person is male (M) or female (F)\footnote{This example is due to Louis Lyons.}. We make one measurement, giving us information on whether the person is pregnant (Y) or not (N). Let's assume we have observed that the person is pregnant, so $d = Y$. 

The likelihood is $P(d=Y | \theta=F) = 0.03$ (i.e., there is a 3\% probability that a randomly selected female is pregnant), but the posterior probability $P(\theta=F | d=Y ) = 1.0$, i.e., if we have observed that the person is pregnant, we are sure she is a woman. This shows that the likelihood and the posterior probability are in general different! 

This is because they mean two different things: the likelihood is the probability of making the observation if we know what the parameter is (in this example, if we know that the person is female); the posterior is the probability of the parameter given that we have made a certain observation (in this case, the probability of a person being female if we know she is pregnant). The two quantities are related by Bayes theorem (prove this in the example given here).}

Bayesian inference works by {updating our state of knowledge} about a parameter (or hypothesis) as new data flow in. The posterior from a previous cycle of observations becomes the prior for the next. 

%
\newcommand{\spazio}{\hspace{0.21\linewidth}}

\subsection{Advantages of the Bayesian approach}

Irrespectively of the philosophical and epistemological views about probability, 
as physicists we might as well take the pragmatic view that the
approach that yields demonstrably superior results ought to be
preferred. In many real--life cases, there are several good
reasons to prefer a Bayesian viewpoint:\\
\begin{enumerate}
 \item Classic frequentist methods are often based on asymptotic properties of
estimators. Only a handful of cases exist that are simple enough
to be amenable to analytic treatment (in physical problems one
most often encounters the Normal and the Poisson distribution).
Often, methods based on such distributions are employed not
because they accurately describe the problem at hand, but because
of the lack of better tools. This can lead to serious mistakes.
Bayesian inference is not concerned by such problems: it can be
shown that {\em application of Bayes' Theorem recovers frequentist
results (in the long run) for cases simple enough where such
results exist}, while remaining applicable to questions that
cannot even be asked in a frequentist context.
 \item Bayesian inference deals effortlessly with {\em nuisance parameters}.
Those are parameters that have an influence on the data but are of
no interest for us. For example, a problem commonly encountered in
astrophysics is the estimation of a signal in the presence of a
background rate The
particles of interest might be photons, neutrinos or cosmic rays.
Measurements of the source $s$ must account for uncertainty in the
background, described by a nuisance parameter $b$. The Bayesian
procedure is straightforward: infer the joint probability of $s$
and $b$ and then integrate over the uninteresting nuisance
parameter $b$ (``marginalization'',
see~Eq.~\eqref{eq:marginalisation_continuous}). Frequentist
methods offer no simple way of dealing with nuisance parameters
(the very name derives from the difficulty of accounting for them
in classical statistics). However neglecting nuisance parameters
or fixing them to their best--fit value can result in a very
serious underestimation of the uncertainty on the parameters of
interest.
 \item In many situations {\em prior information} is highly relevant and
omitting it would result in seriously wrong inferences. The
simplest case is when the parameters of interest have a physical
meaning that restricts their possible values: masses, count rates,
power and light intensity are examples of quantities that must be
positive. Frequentist procedures based only on the likelihood can
give best--fit estimates that are negative, and hence meaningless,
unless special care is taken (for example, constrained likelihood
methods). This often happens in the regime of small counts or low
signal to noise. The use of Bayes' Theorem ensures that relevant
prior information is accounted for in the final inference and that
physically meaningless results are weeded out from the beginning.
 \item Bayesian statistics only deals with the {\em data that were
actually observed}, while frequentist methods focus on the
distribution of possible data that have not been obtained. As a
consequence, {\em frequentist results can depend on what the
experimenter thinks about the probability of data that have not
been observed.} (this is called the ``stopping rule'' problem).
This state of affairs is obviously absurd. Our inferences should
not depend on the probability of what could have happened but
should be conditional on whatever has actually occurred. This is
built into Bayesian methods from the beginning since inferences
are by construction conditional on the observed data.
\end{enumerate}

\begin{figure}
\centering
\includegraphics[width=0.50\linewidth]{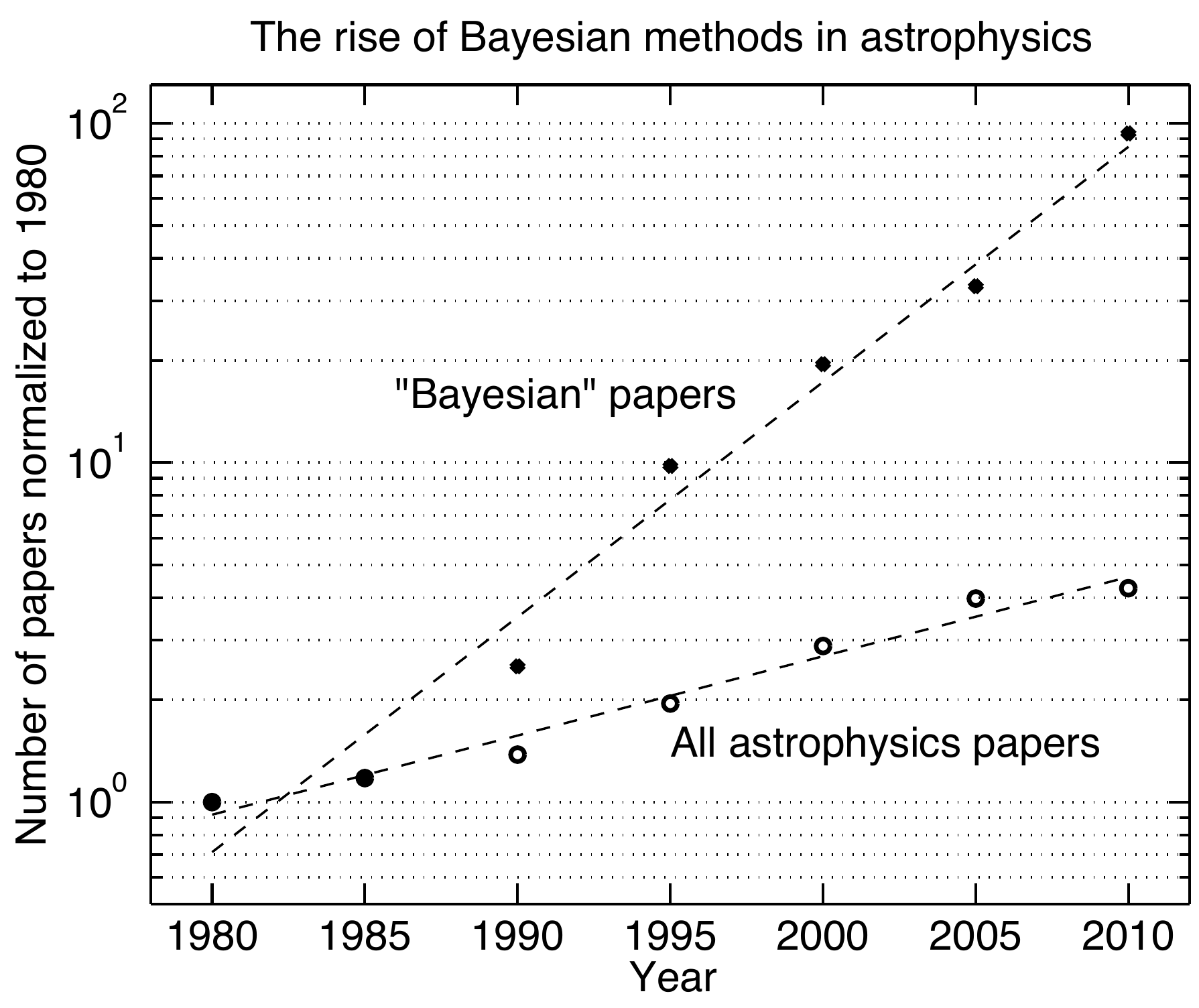}\hfill
\caption{Number of articles in
astronomy and cosmology with ``Bayesian'' in the title, as a
function of publication year (upper data points) and total number of articles (lower data points) as a function of publication year. Numbers are normalized to 1980 levels for each data series. The number of Bayesian papers doubles every 4.3 years, while the total number of papers doubles ``only'' every 12.6 years. At the present rate, by 2060 all papers on the archive will be Bayesian.  (source: NASA/ADS). }
\label{fig:Bayes_papers}
\end{figure}

\item The cosmology and astrophysics communities have been embracing Bayesian mehods since the turning of the Millennium, spurred by the availability of cheap computational power that has ushered in an era of high-performance computing, thus allowing for the first time to deploy the power of Bayesian statistics thanks to numerical implementations (in particular, MCMC and related techniques). The steep increase in the number of Bayesian papers in the astrophysics literature is shown in Fig.~\ref{fig:Bayes_papers}.

\subsection{Considerations and caveats on priors}

\begin{figure}
(a) \spazio (b) \spazio (c) \spazio (d) \\
\centering
\includegraphics[width=0.24\linewidth]{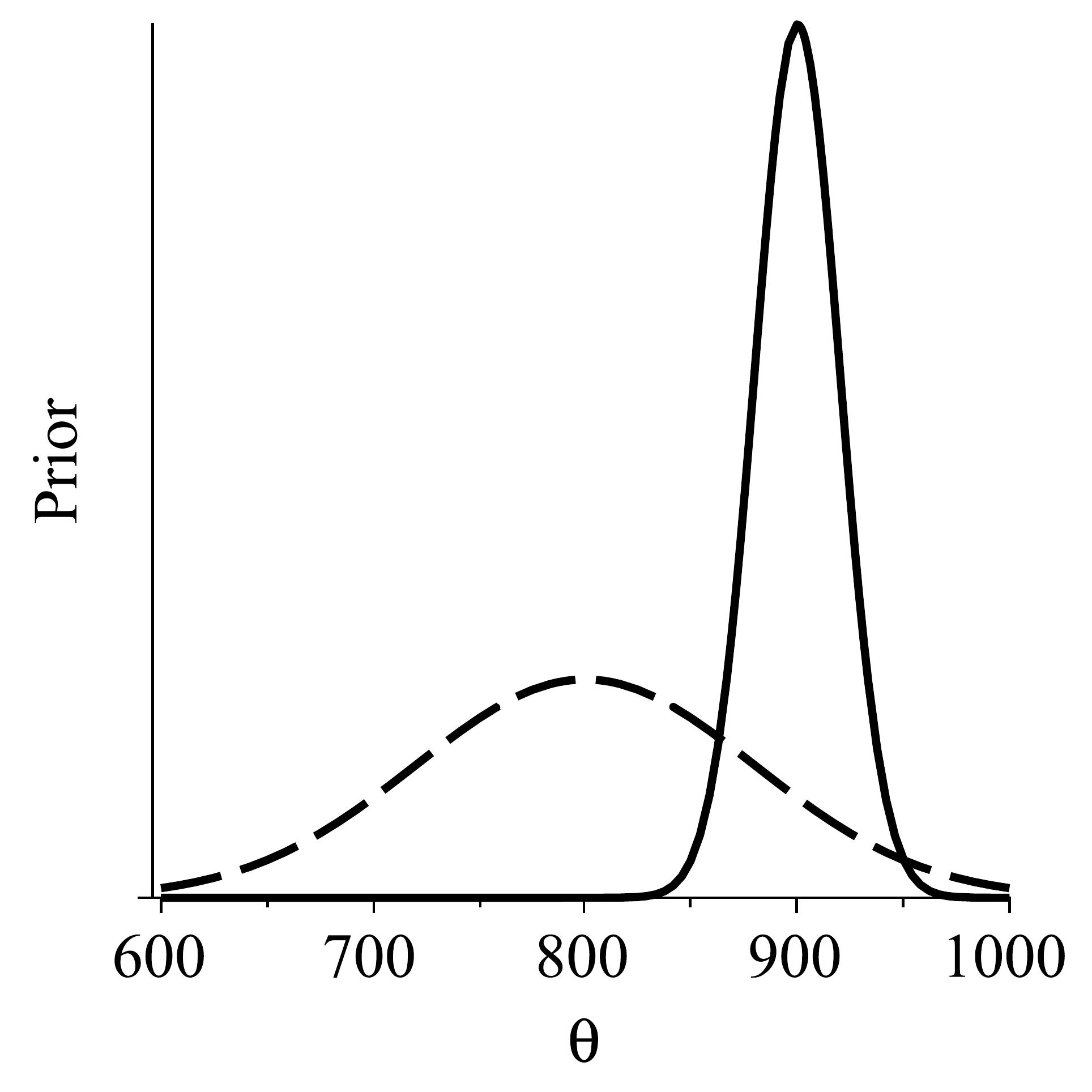}\hfill
\includegraphics[width=0.24\linewidth]{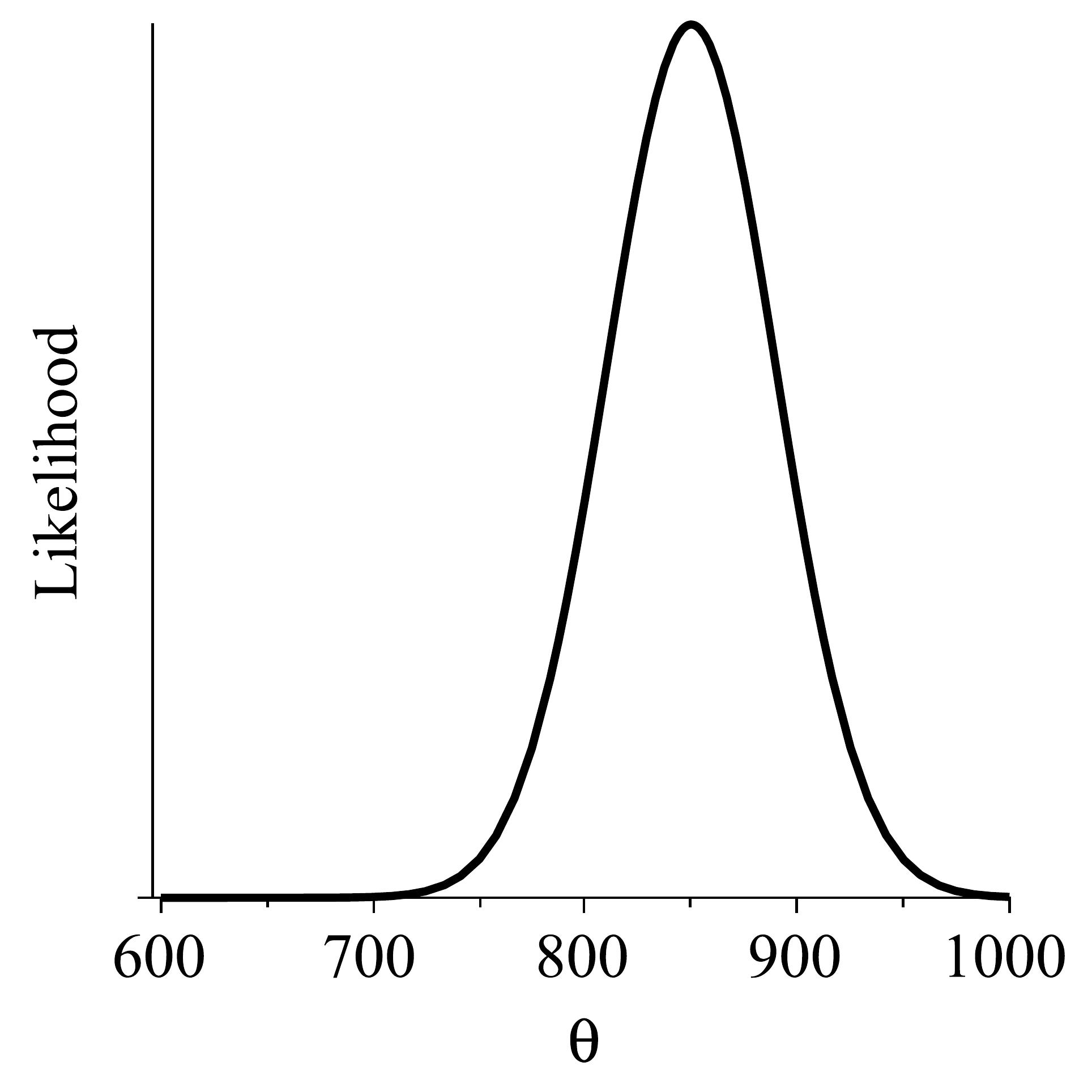}\hfill
\includegraphics[width=0.24\linewidth]{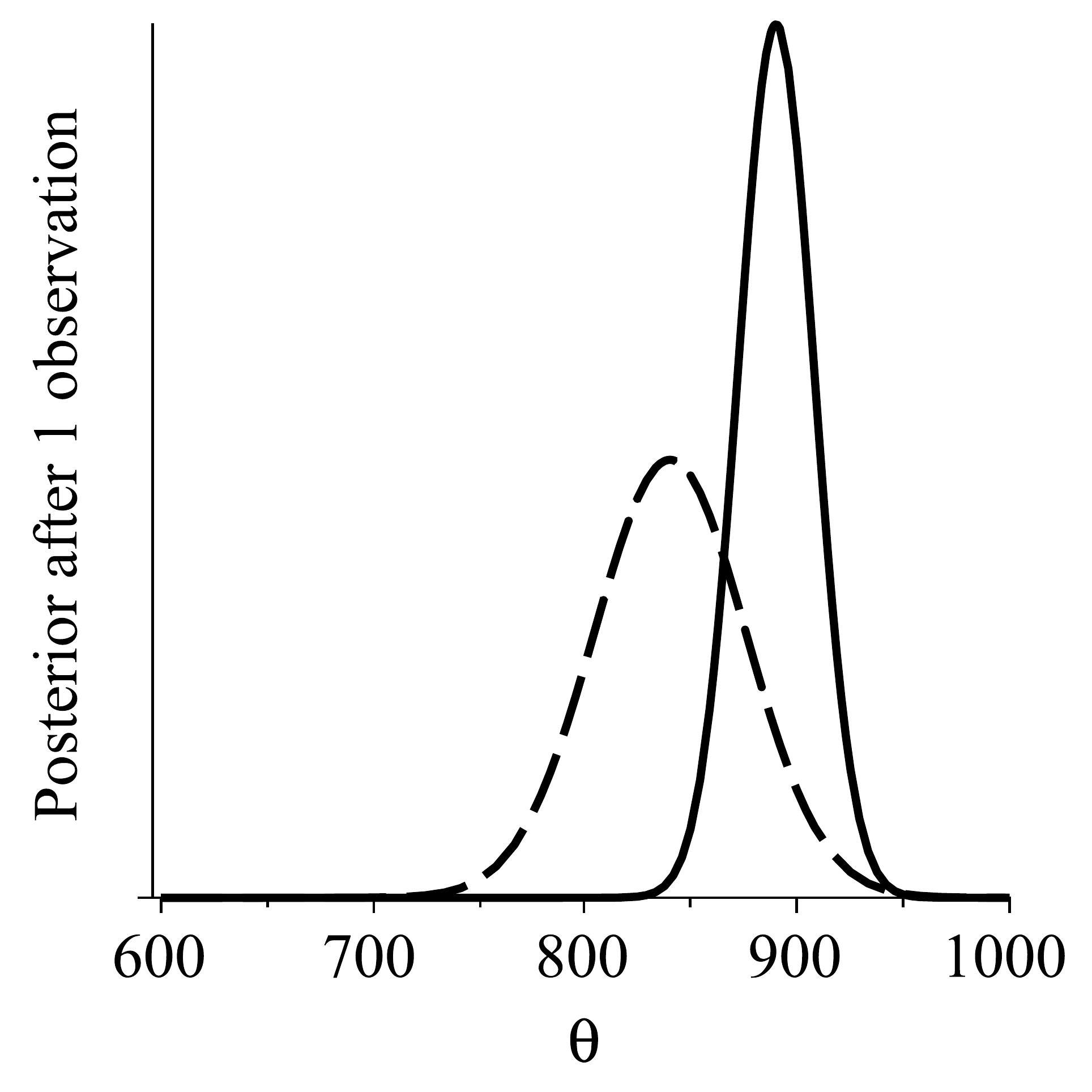}\hfill
\includegraphics[width=0.24\linewidth]{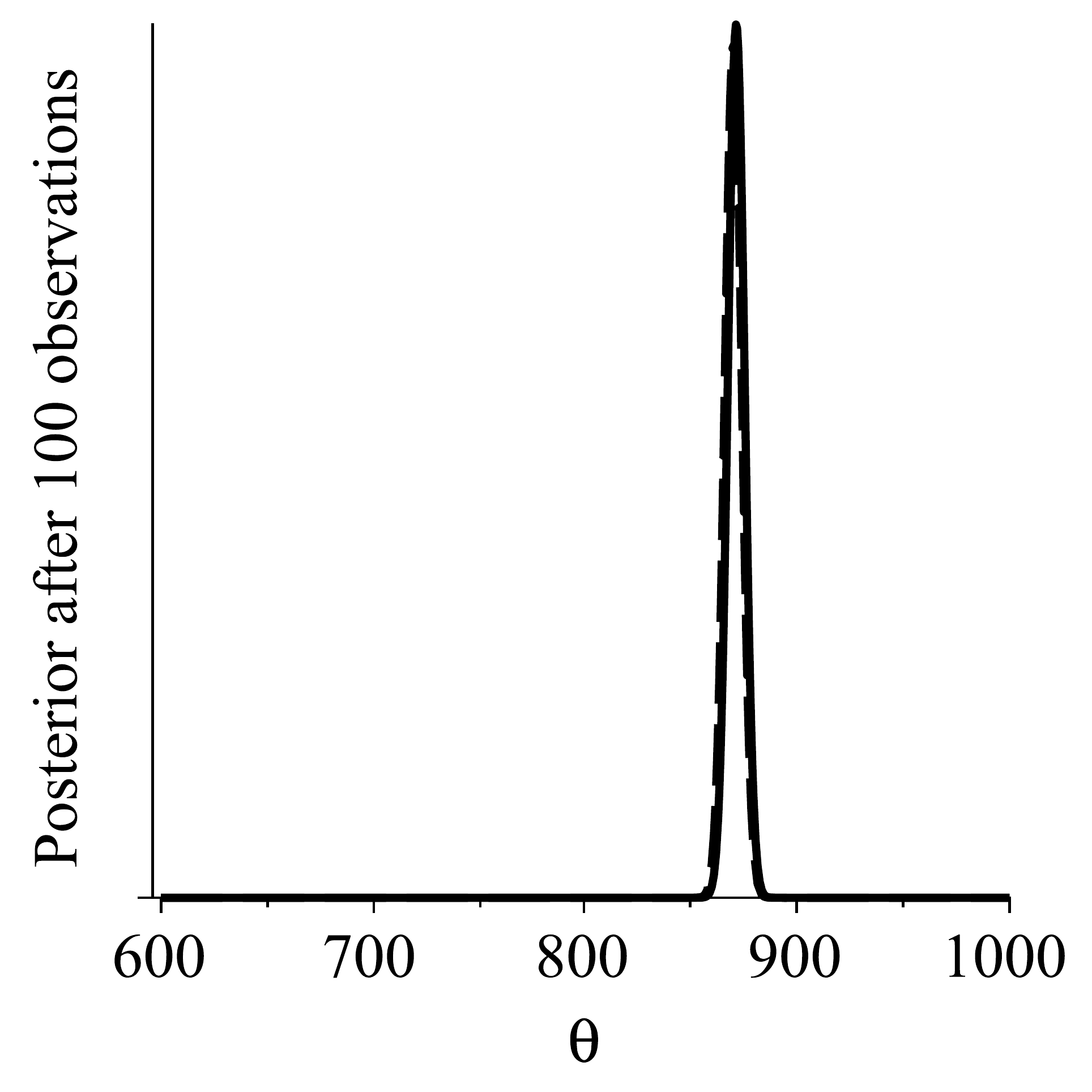}\hfill
\caption{Converging views in Bayesian inference. Two scientists
having different prior believes $p(\theta)$ about the value
of a quantity $\theta$ (panel (a), the two curves representing two different priors) observe
one datum with likelihood $\like(\theta)$ (panel (b)), after
which their posteriors $p(\theta|d)$ (panel (c), obtained via
Bayes Theorem, Eq.~\eqref{eq:bayesth}) represent
their updated states of knowledge on the parameter. This posterior then becomes the prior for the next observation. After
observing 100 data points, the two posteriors have become
essentially indistinguishable (d). }
\label{fig:prior_to_posterior}
\end{figure}

\item Bayesian inference works by updating our state of knowledge about a parameter (or hypothesis) as new data flow in. The posterior from a previous cycle of observations becomes the prior for the next. The price we have to pay is that we have to start somewhere by specifying an initial prior, which is not determined by the theory, but it needs to be given by the user. The prior should represent fairly the state of knowledge of the user about the quantity of interest. Eventually, the posterior will converge to a unique (objective) result even if different scientists start from different priors (provided their priors are non-zero in regions of parameter space where the likelihood is large). See Fig.~\ref{fig:prior_to_posterior} for an illustration. 

\item There is a vast literature about how to select a prior in an appropriate way. Some aspects are fairly obvious: if your parameter $\theta$ describes a quantity that has e.g. to be strictly positive (such as the number of photons in a detector, or an amplitude), then the prior will be 0 for values $\theta < 0$. 

A standard (but by no means harmless, see below) choice is to take a {\em uniform prior} (also called ``flat prior'') on $\theta$, defined as:
\be
P(\theta)  = \left\{
\begin{array}{c l}
  \frac{1}{( \theta_\text{max}- \theta_\text{min})}  & \mbox{for } \theta_\text{min} \leq \theta \leq  \theta_\text{max}  \\
  0 &\mbox{otherwise}
\end{array}
\right.
\ee
With this choice of prior in Bayes theorem, Eq.~\eqref{eq:BT}, the posterior becomes functionally identical to the likelihood up to a proportionality constant:
\be
P(\theta|d) \propto P(d|\theta) = \like(\theta). 
\ee
In this case, all of our previous results about the likelihood carry over (but with a different interpretation). In particular, the probability content of an interval around the mean for the posterior should be interpreted as a statement about our degree of belief in the value of $\theta$ (differently from confidence intervals for the likelihood).
\example{Let's look once more to the temperature estimation problem of Eq.~\eqref{eq:T_meas}. The Bayesian estimation of the temperature proceeds as follows. We first need to specify the likelihood function -- this is the same as before, and it is given by  Eq.~\eqref{eq:T_meas}. If we want to estimate the temperature, we need to compute the posterior probability for $T$, given by (up to a normalization constant)
\be
P(T | d) \propto \like(T)P(T)
\ee
where the likelihood $\like(T)$ is given by  Eq.~\eqref{eq:T_meas}. We also need to specify the prior, $P(T)$. For this particular case, we know that $T>0$ (the temperature in K of an object needs to be positive) and let's assume we know that the temperature cannot exceed 300 K. Therefore we can pick a flat prior of the form 
\be \label{eq:prior_Gauss_example}
P(T)  = \left\{
\begin{array}{c l}
  \frac{1}{300}  & \mbox{for } 0 \text{K } \leq T \leq  300 \text{K}  \\
  0 &\mbox{otherwise.}
\end{array}
\right. 
\ee
The posterior distribution for $T$ then becomes 
\be
P(T| d)  \propto \left\{
\begin{array}{c l}
  \frac{\like(T)}{300}  & \mbox{for } 0 \text{K } \leq T \leq  300 \text{K}  \\
  0 &\mbox{otherwise.}
\end{array}
\right.
\ee
So the posterior is identical to the likelihood (up to a proportionality constant), at least within the range of the flat prior. Hence we can conclude that the posterior is going to be a Gaussian (just like the likelihood) and we can immediately write the 68.3\% posterior range of $T$ as $198.0 \text{K}< \mu < 201.2 \text{K}$. This is {numerically} identical to our results obtained via the MLE. However, in this case the {interpretation} of this interval is that ``after seeing the data, and given our prior as specified in Eq.~\eqref{eq:prior_Gauss_example}, there is 68.3\% probability that the true value of the temperature lies within the range $198.0 \text{K}< \mu < 201.2 \text{K}$''.}

\item Under a change of variable, $\Psi = \Psi(\theta)$, the prior transforms according to:
\be
P(\Psi) = P(\theta) {\Big\vert}\text{det}\left(\frac{\partial\theta}{\partial\Psi}\right){\Big\vert}.
\ee
In particular, a flat prior on $\theta$ is no longer flat in $\Psi$ if the variable transformation is non-linear. 

\begin{figure}
\centerline{\includegraphics[width=0.5\linewidth]{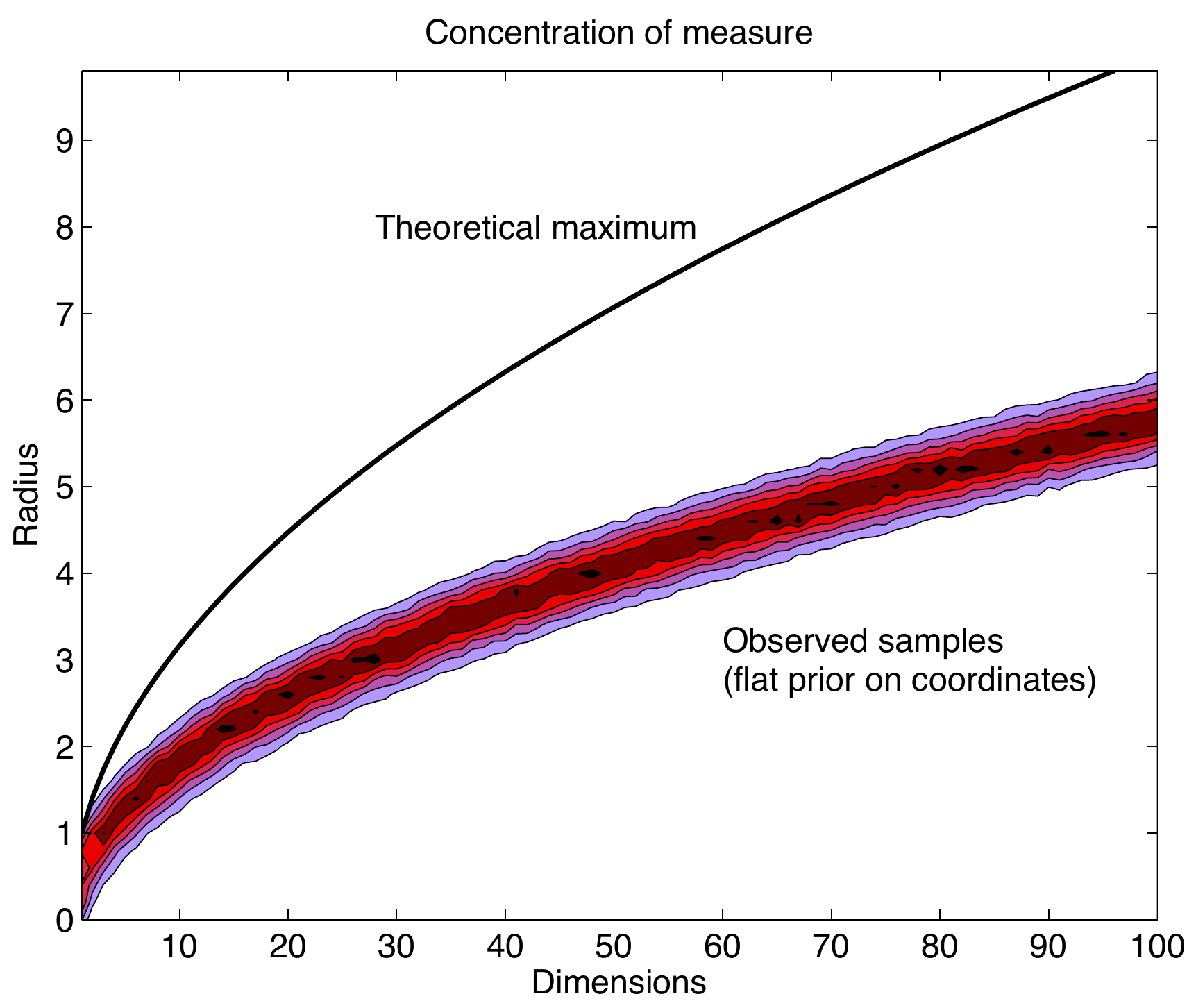}}
\caption{Illustration of the phenomenon of the concentration of measure in parameter spaces with a large number of dimensions. The coloured band represents the density of samples (as a function of the number of dimensions sampled) obtained with a flat prior on the axis coordinates of a D-dimensional hypercube. It can be seen that samples from the prior concentrate in a thin shell of constant variance, leaving most of the parameter space unexplored. The radius of the shell is given in the vertical axis.} \label{fig:concentration}
\end{figure}

It is important to realize that a flat prior is far from harmless, especially in parameter spaces of high dimensionality. This is the so-called ``concentration of measure'' phenomenon. Sampling uniformly (i.e., with a uniform prior) along each dimension $x_i \in [0,1]$ of a $D$-dimensional hypercube leads to the radius $r = \left( \sum_{i=1}^D x_i^2 \right)^{1/2}$ of the samples to concentrate around the value $\langle r \rangle = (D/3)^{1/2}$ with constant variance. As a consequence, all of the samples are found on a thin shell (see Fig.~\ref{fig:concentration} for an illustration). Even worse, in $D$ dimensions the volume of the hypercube is much larger than the volume of the hypersphere, hence most of the volume is in the corners of the hypercube which are not sampled. This means that an MCMC in $D$ dimensions (where $D$ is large) has a prior distribution that is far from being uniformly distributed in the volume of the hypercube -- although any 2-dimensional projection will apparently belie this. 

A sensitivity analysis should always be performed, i.e., change the prior in a reasonable way and assess how robust the ensuing posterior is. Unfortunately, this is seldom done in the astrophysics and cosmology literature.
 
There is a vast body of literature on different types of priors, when to use them and what they are good for. It is a good idea to browse the literature when faced with a new problem, as there is no point in re-inventing the wheel every time. There are essentially two schools of thought: one maintains that priors should be chosen according to subjective degree of belief; the other, that they should be selected according to some formal rule, i.e. priors should be chosen by convention. None of the two approaches is free from difficulties.  To give but some relevant examples:
 \begin{itemize} 
 \item {\em reference priors:} the idea is to define a prior so that the contribution of the data to the posterior is maximised. This is achieved by choosing a prior with maximum entropy. For example, in the case of a Gaussian likelihood this leads to the conclusion that the proper prior for the mean $\mu$ is flat on $\mu$, while for the standard deviation $\sigma$ is should be flat in $\log\sigma$ (with appropriate cutoffs of course).
 \item {\em ignorance priors:} in 1812 Laplace set forth the principle that when nothing else is known priors should be chosen so as to give equal probability to all alternatives (``the principle of indifference''). Unfortunately this is very difficult to do in the case of continuous parameters: part of the reason is that the notion of ``indifference'' is not invariant under non-linear reparameterizations. In some relatively simple cases, ignorance priors can be derived using symmetry or invariance arguments, see~for examples \cite{JaynesBook}. 
 \item {\em conjugate priors:} a prior is said to be conjugate to the likelihood if the resulting posterior is of the same family as the likelihood. The convenience of having conjugate priors is that the likelihood updates the prior to a posterior which is of the same type (i.e., same distributional family). For example, Gaussian distributions are self-conjugate, i.e., a Gaussian prior with a Gaussian likelihood leads to a Gaussian posterior; the conjugate prior to both the Poisson and the exponential likelihood is the Gamma distribution; the conjugate prior to a Binomial likelihood is the Beta distribution. 
 \end{itemize}

\subsection{A general Bayesian solution to inference problems}

The general Bayesian recipe to inferential problems can be summarised as follows: 
\begin{enumerate}
\item Choose a model containing a set of hypotheses in the form of a vector of parameters, $\params$ (e.g., the mass of an extra--solar planet or the abundance of
dark matter in the Universe). 
\item Specify the priors for the parameters. Priors should summarize your state of knowledge about the parameters before you consider the new
data, including an relevant external source of information.
\item Construct the likelihood function for the
measurement, which usually reflects the way the data are obtained (e.g., a measurement with Gaussian noise will be represented
by a Normal distribution, while $\gamma$--ray counts on a detector
will have a Poisson distribution for a likelihood). Nuisance
parameters related to the measurement process might be present in
the likelihood, e.g. the variance of the Gaussian might be unknown
or the background rate in the absence of the source might be
subject to uncertainty. Such nuisance parameters are included in the likelihood (with appropriate prior). If external measurements are available for the nuisance parameters, they can be incorporated either as an informative prior on them, or else as additional likelihood terms. 
\item Obtain the posterior distribution (usually, up to an overall normalisation constant) either by analytical means or, more often, by numerical methods (see below for MCMC and nested sampling algorithms to this effect).
\end{enumerate}

The posterior pdf for one parameter at the time is obtained by {\em marginalization}, i.e., by integrating over the uninteresting parameters. E.g., assume the the vector of parameters is given by $\params = \{\phi, \psi\}$, then the 1D posterior pdf for $\phi$ alone is given by 
 \begin{equation} \label{eq:marginal_posterior}
 p(\phi| \data) \propto \int \like(\phi, \psi) p(\phi, \psi)\dr \psi.
 \end{equation}
The final inference on $\phi$ from the posterior can then be
communicated by plotting $p(\phi| \data)$, with the other components marginalized over.

From an MCMC chain,  one can also obtain the profile
likelihood, Eq.~\eqref{eq:profile}, by maximising the value of the
likelihood in each bin.  The profile likelihood is expected to be prior-independent, as long as the scan has gathered a sufficient number of samples in the favoured region, which is in general a difficult task for multi-dimensional parameter spaces. It is also typically much more expensive to compute as it requires a much larger number of samples than the posterior. 

The profile likelihood and the Bayesian posterior ask two different statistical questions of the data: the latter evaluates which regions of parameter space are most plausible in the light of the measure implied by the prior; the former singles out regions of high quality of fit, independently of their extent in parameter space, thus disregarding the possibility of them being highly fine tuned. The information contained in both is relevant and interesting, and for non-trivial parameter spaces the two different approaches do not necessarily lead to the same conclusions\footnote{In the archetypal case of a Gaussian likelihood and uniform prior, the posterior pdf and the profile likelihood are identical (up to a normalisation constant) and thus the question of which to choose does not arise.}.

\subsection{The Gaussian linear model}

As idealised a case as it is, the Gaussian linear model is a great tool to hone your computational skills and intuition. This is because it can be solved analytically, and any numerical solution can be compared with the exact one. Furthermore, it applies in an approximate way to many cases of interest. Here we solve analytically the general problem in $n$ dimensions. An application to the 2-dimensional case is then given in the Exercises, section ~\ref{sec:bayes_ex}. For a more complete discussion, see~\cite{Kelly07}, where the general case is treated (including errors on the independent variable, general correlations, missing data, upper limits, selection effects and the important subject of Bayesian hierarchical modelling).  

We consider the following {\em linear model}
 \be
 y = F \theta + \epsilon
 \ee
where the dependent variable $y$ is a $d$-dimensional vector of
observations (the {\em data}), $\theta = \{\theta_1, \theta_2, \dots, \theta_n \}$ is a vector of dimension $n$ of unknown
parameters that we wish to determine and $F$ is a $d\times n$ matrix of known
constants which specify the relation between the input variables
$\theta$ and the dependent variables $y$ (so-called ``design matrix'').

In the following, we will specialize to the case where observations $y_i(x)$ are fitted with a linear model of the form $f(x) =
\sum_{j=1}^n \theta_j X^j(x)$. Then the matrix $F$ is given by the basis functions
$X^j$ evaluated at the locations $x_i$ of the observations,
$F_{ij} = X^j(x_i)$. Notice that the model is linear in $\theta_j$, not necessarily in $X^j$, i.e. $X^j$ can very well be a non-linear function of $x$. 

Furthermore, $\epsilon$ is a
$d$-dimensional vector of random variables with zero mean (the
{\em noise}). We assume for simplicity that $\epsilon$ follows a multivariate
Gaussian distribution with uncorrelated covariance matrix $C
\equiv{\rm diag}(\tau_1^2, \tau_2^2, \dots, \tau_d^2)$. The
likelihood function takes the form
 \be p(y | \theta) =
\frac{1}{(2\pi)^{d/2} \prod_j \tau_j}
\exp\left[-\frac{1}{2}(b-A\theta)^t(b-A\theta)\right],
\label{eq:data_like}
 \ee
where we have defined $A_{ij} = F_{ij}/\tau_i$ and $b_i =
y_i/\tau_i$ where $A$ is a $d\times n$ matrix and $b$ is a $d$-dimensional vector. 
This can be re-cast with some simple algebra as 
 \be \label{eq:like_glm}
 p(y | \theta) =
 \LL_0
 \exp\left[-\frac{1}{2}(\theta-\theta_0)^t L (\theta-\theta_0)\right],
 \ee
with the likelihood Fisher matrix $L$ (a $n\times n$ matrix) given by
 \be
 L \equiv A^t A
 \ee
 and a normalization constant
 \be
 \LL_0 \equiv \frac{1}{(2\pi)^{d/2} \prod_j \tau_j}
 \exp\left[-\frac{1}{2}(b - A \theta_0)^t(b-A \theta_0) \right].
 \ee
Here $\theta_0$ denotes the parameter value which maximises the
likelihood (i.e., the maximum likelihood value for $\theta$), given by
 \be \label{maxlike}
 \theta_0 = L^{-1}A^t b.
 \ee

We assume as a prior pdf a multinormal Gaussian distribution with
zero mean and the $n \times n$ dimensional prior Fisher information matrix $P$  (recall that
that the Fisher information matrix is the inverse of the
covariance matrix), i.e.\
 \be \label{eq:gaussprior}
 p(\theta) = \frac{|P|^{1/2}}{(2\pi)^{n/2}}\exp\left[-\frac{1}{2}
 \theta^t P \theta \right] ,
 \ee
where $|P|$ denotes the determinant of the matrix $P$. 

It can be shown that the posterior distribution for $\theta$ is given by multinormal Gaussian with Fisher
information matrix $\Fpost$
    \be \Fpost = L + P
\label{eq:post} \ee
 and mean $\bar{\theta}$ given by
 \be
 \bar{\theta} = \Fpost^{-1}L \theta_0.
 \ee

Finally, the model likelihood (or ``Bayesian evidence'', i.e., the normalizing constant in Bayes theorem) is given by 
 \be \label{eq:modlike}
  \begin{aligned}
    p(y) & = \LL_0 \frac{|\Fpost|^{-1/2}}{|P|^{-1/2}}
  \exp\left[-\frac{1}{2}\theta_0^t(L - L\Fpost^{-1}L )\theta_0
  \right]\\
  & = \LL_0 \frac{|\Fpost|^{-1/2}}{|P|^{-1/2}}
  \exp\left[-\frac{1}{2}(\theta_0^t L \theta_0 - \bar{\theta}^t\Fpost\bar{\theta})
  \right].
  \end{aligned}
   \ee

\subsection{Markov Chain Monte Carlo methods}
\label{sec:MCMC}

\subsubsection{General theory}

The purpose of a Markov chain Monte Carlo algorithm is to
construct a sequence of points (or ``samples'') in parameter space (called ``a
chain''). The crucial property of the chain is that the density of samples is proportional to the posterior pdf. This allows to construct a map of the posterior distribution.

A Markov chain
is defined as a sequence of random variables $\{X^{(0)}, X^{(1)},
\dots, X^{(M-1)}\}$ such that the probability of the $(t+1)$--th
element in the chain only depends on the value of the $t$--th
element. The crucial property of Markov chains is that they can be
shown to converge to a stationary state (i.e., which does not
change with $t$) where successive elements of the chain are
samples from the {\em target distribution}, in our case the
posterior $p(\params|\data)$. 

The generation of the elements of
the chain is probabilistic in nature, and is described by a {\em
transition probability} $T(\params^{(t)},\params^{(t+1)})$, giving
the probability of moving from point $\params^{(t)}$ to
point $\params^{(t+1)}$ in parameter space. A sufficient condition
to obtain a Markov Chain is that the transition probability
satisfy the {\em detailed balance condition}
 \begin{equation} \label{eq:detailed_balance}
 p(\params^{(t)}|\data) T(\params^{(t)},\params^{(t+1)}) =  p(\params^{(t+1)}|\data)
 T(\params^{(t+1)},\params^{(t)}).
 \end{equation}
This is perhaps clearer when recast as follows:
 \begin{equation} \label{eq:detailed_balance2}
\frac{T(\params^{(t)},\params^{(t+1)})}{T(\params^{(t+1)},\params^{(t)})} = \frac{p(\params^{(t+1)}|\data)}{p(\params^{(t)}|\data)},
 \end{equation}
i.e.\. ratio of the transition probabilities is inversely proportional to
the ratio of the posterior probabilities at the two points. 

Once samples from the posterior pdf have been gathered, obtaining Monte Carlo
estimates of expectations for any function of the parameters
becomes a trivial task. The posterior mean is given
by 
 \begin{equation} \label{eq:expectation}
 E[\params] =  \int  P(\params|\data)\params d\params
   \approx \frac{1}{M} \sum_{t=0}^{M-1} \params^{(t)},
 \end{equation}
where the (approximate) equality with the mean of the samples from the MCMC
follows because the samples $\params^{(t)}$ are generated from the
posterior by construction. 

 One can easily obtain the
expectation value of any function of the parameters $f(\params)$
as
 \begin{equation} \label{eq:MC_estimate}
 E[f(\params)] \approx \frac{1}{M}\sum_{t=0}^{M-1}  f(\params^{(t)}).
 \end{equation}
It is usually interesting to summarize the results of the
inference by giving the 1--dimensional {\em marginal probability}
for the $j$--th element of $\params$, $\params_j$, obtained by integrating out all other parameters from the posterior:
 \begin{equation} \label{eq:marginalisation_continuous}
 P(\params_1|\data) = \int  P(\params|\data) d \params_2 \dots d
 \params_n,
 \end{equation}
where $ P(\params_1|\data)$ is the {\em marginal posterior} for the
parameter $\params_1$. While this would usually require an $n-1$-dimensional integration (which can be numerically difficult), it is easily obtained from the Markov chain. Since the elements of
the Markov chains are samples from the full posterior,
$P(\params|\data)$, their density reflects the value of the full
posterior pdf. It is then sufficient to divide the range of
$\params_1$ in a series of bins and {\em count the number of
samples falling within each bin}, simply ignoring the coordinates
values $\params_2, \dots, \params_n$. A 2--dimensional posterior is
defined in an analogous fashion.

 A 1D 2--tail symmetric $\alpha\%$ credible region is given by the interval (for the
parameter of interest) within which fall $\alpha\%$ of the samples, obtained in such a way that a fraction $(1-\alpha)/2$ of the samples lie outside the interval on either side. In the
case of a 1--tail upper (lower) limit, we report the value of the
quantity below (above) which $\alpha\%$ of the sample are to be found. 

Credible regions for a given probability content $\alpha$ can be defined in an infinite number of ways. Two definitions are commonly used. The first is ``symmetric credible interval'' (in 1D) given above. The second definition is that of Highest Posterior Density (HPD) regions. They are obtained by starting from the maximum of the posterior and reducing the level until the desired fraction $\alpha$ of the posterior probability mass is included. Such a definition delimits a region so that every point inside it has by construction a higher posterior density than any point outside it.  For a given probability content $\alpha$, the HPD region is also the shortest interval. For a Normal 1D posterior, the HPD is identical to the symmetric credible region.

\subsubsection{The Metropolis-Hastings algorithm}

The simplest (and widely used) MCMC algorithm is the Metropolis-Hastings algorithm~\cite{Metropolis:1953am,Hastings:1970}: 
 \begin{enumerate}
 \item Start from a random point $\params^{(0)}$,  with
associated posterior probability $p_0 \equiv p(\params^{(0)} |
\data)$.
 \item Propose a candidate point $\params^{(c)}$
 by drawing from the {\em proposal distribution}
$q(\params^{(0)},\params^{(c)})$. The proposal distribution might
be for example a Gaussian of fixed width $\sigma$ centered around
the current point.  For the Metropolis algorithm (as opposed to the more general form due to Hastings), the distribution $q$ satisfies the symmetry
condition, $q({x}, {y}) = q({y}, {x})$.
 \item Evaluate the posterior at the candidate point, $p_c = p(\params^{(c)}
 |\data)$. Accept the candidate point with
probability
 \be \label{eq:acceptance_fct_MH}
 \alpha= \min \left( \frac{p_c q(\params^{(c)}, \params^{(0)})}{p_0 q(\params^{(0)}, \params^{(c)})}, 1 \right).
 \ee
For the Metropolis algorithm (where $q$ is symmetric), this simplifies to 
\be \label{eq:acceptance_fct}
 \alpha = \min \left( \frac{p_c}{p_0}, 1 \right).
 \ee
This accept/reject step can be performed by generating a random
number $u$ from the uniform distribution $[0,1)$ and accepting the
candidate sample if $u < \alpha$, and rejecting it otherwise.
 \item If the candidate point is accepted, add it to the chain and move there.
Otherwise stay at the old point (which is thus counted twice in
the chain). Go back to (ii).
\end{enumerate}
Notice from Eq.~\eqref{eq:acceptance_fct} that whenever the
candidate sample has a larger posterior than the previous one
(i.e., $p_c > p_0$) the candidate is always accepted. Also, in
order to evaluate the acceptance function
\eqref{eq:acceptance_fct} only the unnormalized posterior is
required, as the normalization constant drops out of the ratio. It
is easy to show that the Metropolis algorithm satisfies the
detailed balance condition, Eq.~\eqref{eq:detailed_balance}, with
the transition probability given by $T(\params^{(t)},
\params^{(t+1)}) = q(\params^{(t)},
\params^{(t+1)}) \alpha(\params^{(t)},
\params^{(t+1)})$.

Ref~\cite{GelmanTheorem} shows that an optimal choice of the proposal distribution is such that it leads to an acceptance rate of approximately 25\% (where acceptance rate is the ratio of the number of accepted jumps to the total number of likelihood evaluations). The optimal scale of the proposal distribution is approximately $2.4/\sqrt{d}$ times the scale of the target distribution, where $d$ is the number of dimensions of the parameter space. 

The choice of proposal distribution $q$ is crucial for the efficient exploration of the posterior. If the scale of $q$ is too small compared to the scale of the target distribution, exploration will be poor as the algorithm spends too much time locally. If instead the scale of $q$ is too large, the chain gets stuck as it does not jump very frequently. 

To improve the exploration of the target, it is advisable to run an exploratory MCMC, compute the covariance matrix from the samples, and then re-run with this covariance matrix (perhaps rescaled by a factor $2.4/\sqrt{d}$ as recommended by \cite{GelmanTheorem}) as the covariance of a multivariate Gaussian proposal distribution. This process can be iterated a couple of times. The affine invariant ensemble sampler proposed by~\cite{Goodman10} evolves a series of ``walkers'' rather than just one sampler at the time, and uses the position of the other points in the ensemble to generate a move with greatly reduced auto-correlation length. This also largely dispenses with the need to fine-tune the proposal distribution to match the target density. An algorithm that includes a suitable parallelization of this sampling scheme is described in~\cite{2013PASP..125..306F}, and is implemented in a publicly available Python package, \texttt{emcee}\footnote{Available from: \texttt{http://dan.iel.fm/emcee} (accessed Jan 5th 2017).}. 

\subsubsection{Gibbs sampling}

The Gibbs sampler is a particularly good choice when it is simple (and computationally non-expensive) to sample from the conditional distribution of one of the parameters at the time. It has been shown to work well in a large ($\sim 10^5$) number of dimensions. 

In Gibbs sampling, each of the parameters is updated in turn by drawing the proposal distribution from the univariate conditional distribution of that variable (conditional on all the others). This is best explained in a simple example, where the parameter space is 2-dimensional and $\theta=\{ x, y\}$. In order to obtain the $t$-th sample, one draws
\begin{eqnarray}
x^{(t)} \sim p(x|y=y^{(t-1)}) \\
y^{(t)} \sim p(y|x=x^{(t)}). \\
\end{eqnarray}
Notice that in the second step, when drawing $y$ we condition on a value of $x$ that has been updated to the latest draw of $x$, namely $x^{(t)}$. In the above, $p$ denotes the target distribution, i.e. the posterior density (where we have omitted explicit conditioning on the data for ease of notation). 

In a higher number of dimensions of parameter space, one always draws the $k$-th variable from the conditional distribution $p(\theta_k | \theta_{(-k)})$, where $\theta_{(-k)}$ denotes the vector of variables without the $k$-th variable.

It is perhaps slightly baffling that one can obtain samples from the joint posterior merely from knowledge of the conditional distributions (although this is not generally true). An explanation of why this is the case (under only very mild conditions) can be found in~\cite{Casella}.

The Gibbs sampler can thus be seen as a special case of Metropolis-Hastings, with one-dimensional proposal distributions and an acceptance rate of 1. 

The above can also be generalised to blocks of variables, that are all updated simultaneously conditional on all the others. In the so-called ``blocked Gibbs sampler'' one draws two (or more) variables simultaneously from $p(\theta_{k,j} | \theta_{(-k,j)})$. This can be useful in improving the convergence if the two variables $k,j$ are strongly correlated.  A collapsed Gibbs sampler refers to the case when one of the variables has been marginalised out in one of the sampling steps, i.e.\ one draws from $p(\theta_k | \theta_{(-k,j)})$, where the $j$-th variable has been marginalised from the joint. More sophisticated sampling strategies can also be employed to reduced auto-correlation and improve sampling, see \cite{park:vand:09,park:vand:09} for the Partially Collapsed Gibbs sampler, and~\cite{yu:meng:11} for the ancillarity-sufficiency interweaving strategy.

\subsubsection{Hamiltonian Monte-Carlo}

Hamiltonian Monte Carlo is particularly appealing for physicists, as it is built on the formalism of Hamiltonian dynamics (as the name implies). Only a very sketchy introduction is possible here. Refer to~\cite{NealHMC} for further details. A Python implementation of HMC can be found at: \texttt{mc-stan.org}.

The idea is to augment the vector containing the variable of interest, $q$ (representing position), by another vector of the same dimensionality, $p$ (representing momentum). We then define the potential energy $U(q)$ as the negative log of the unnormalized posterior we wish to sample from, 
\be
U(q) = -\log(\pi(q) \like(q)), 
\ee
where $\pi(q)$ is the prior and $\like(q)$ the likelihood function.  The Hamiltonian of this fictitious system is then given by
\be
H(q,p) = K(p) + U(q)
\ee
where $K(p)$ represents kinetic energy,
\be
K(p) = \sum_i \frac{p_i^2}{2 m_i}.
\ee
Here, the sum runs over the dimensionality of the parameter space, and $m_i$ are ``mass values'' that are chosen for convenience. If we look at the kinetic energy term as the negative log of a probability distribution, then it defines a multivariate Gaussian of 0 mean with variance along each direction given by $m_i^2$. 

From analytical mechanics, we know that physical solutions are obtained by solving the Hamiltonian equations:
\begin{eqnarray}
\frac{d q_i}{dt} = \frac{\partial H}{\partial p_i} \\
\frac{d p_i}{dt} = -\frac{\partial H}{\partial q_i}. 
\end{eqnarray}
Such solutions have the useful properties of preserving energy (i.e., $dH/dt = 0$) and conserving the phase space volume (in virtue of Liouville's theorem). Those properties are crucial in ensuring that the Hamiltonian MC (HMC) algorithm leaves the desired distribution invariant. 

In order to obtain a Markov Chain from the target distribution, the Hamiltonian MC algorithm performs the following steps in each iteration:
\begin{enumerate}
\item resample the momentum variables, $p_i \sim \norm(0, m_i^2)$;
\item obtain a new candidate location  $(q_c, p_c)$ in phase space by evolving the system via approximate Hamiltonian dynamics (e.g. via the leapfrog method);
\item take a Metropolis accept/reject step at the candidate location (this is necessary as in practice numerical approximation schemes mean that the energy of the system is only approximately conserved).
\end{enumerate} 
The Hamiltonian dynamics preserves energy, but it changes the value of both the momentum (in step (1)) and position variables (in step (2)), thus accomplishing a large jump in the parameters of interest, namely $q$. 

The key advantages of HMC is that it produces samples that are much less correlated than ordinary Metropolis-Hastings (in virtue of the large distance travelled via the Hamiltonian dynamics step), and that it scales well with the number of dimensions of the parameter space.  

\subsubsection{Importance sampling}

Importance sampling is a useful technique when we want to sample from a target distribution $p(x)$ (usually the posterior), but we have samples from another distribution $q(x)$ (perhaps because the latter is simpler to sample from). In some applications, $q(x)$ could be the posterior from a certain data set, and we then want to add another data set on top of it, thus obtaining $p(x)$. As long as $p(x)$ is not too dissimilar from $q(x)$, it can be obtained by importance sampling.

The expectation value under $p$ of any function $f(x)$ of the RV $x$ can be written as 
\be
E_p[f(x)] = \int f(x) p(x) dx = \int f(x) q(x) \frac{p(x)}{q(x)} dx = E_q[\frac{p(x)}{q(x)} f(x)].  
\ee
This shows that we can obtain the expectation value under $p$ by computing the expectation value under $q$ but re-weighting the function of interest by the factor $p(x)/q(x)$. 

In terms of the sampling estimate, we can write
\be \label{eq:importance_sampling}
\mu_f \approx \frac{1}{M}  \frac{\sum_{i=0}^M w_i f(x_i)}{\sum_{i=0}^M w_i}
\ee
where $w_i = p(x_i)/q(x_i)$ are the importance sampling weights and $x_i \sim q(x)$. Notice that only the unnormalized values of $p$ and $q$ are necessary in Eq.~\eqref{eq:importance_sampling}, since the normalisation cancels in the ratio.

\subsection{Practical and numerical issues}

It is worth mentioning several important practical issues in working with MCMC
methods.  Poor exploration
of the posterior can lead to serious mistakes in the final
inference if it remains undetected -- especially in high--dimensional
parameter spaces with multi--modal posteriors. It is therefore important {\em
not} to use MCMC techniques as a black box, but to run adequate tests to ensure insofar as possible that the MCMC sampling has converged to a fair representation of the posterior.

\begin{figure}[tb]
\centering
\includegraphics[width=0.5\linewidth]{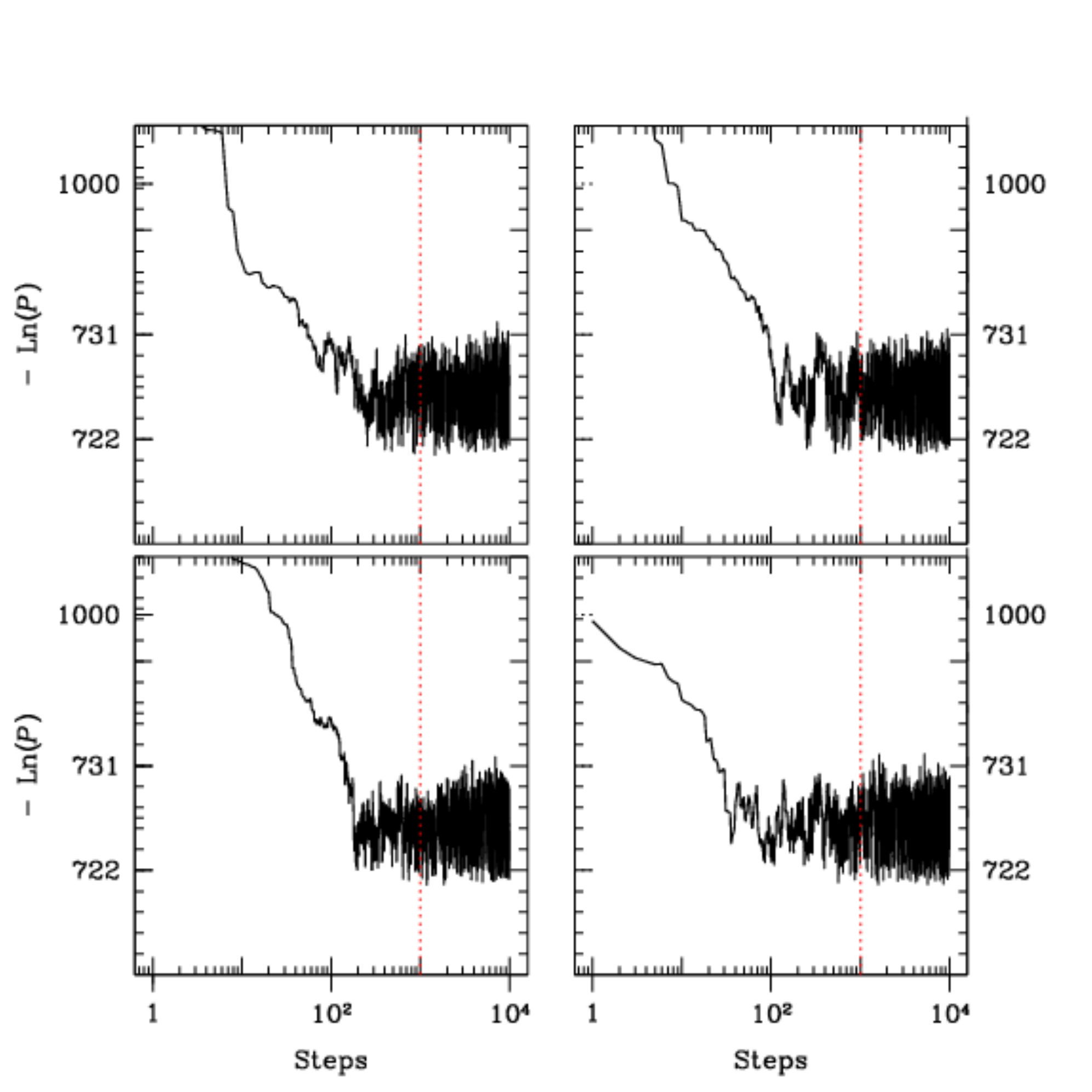}%
\includegraphics[width=0.5\linewidth]{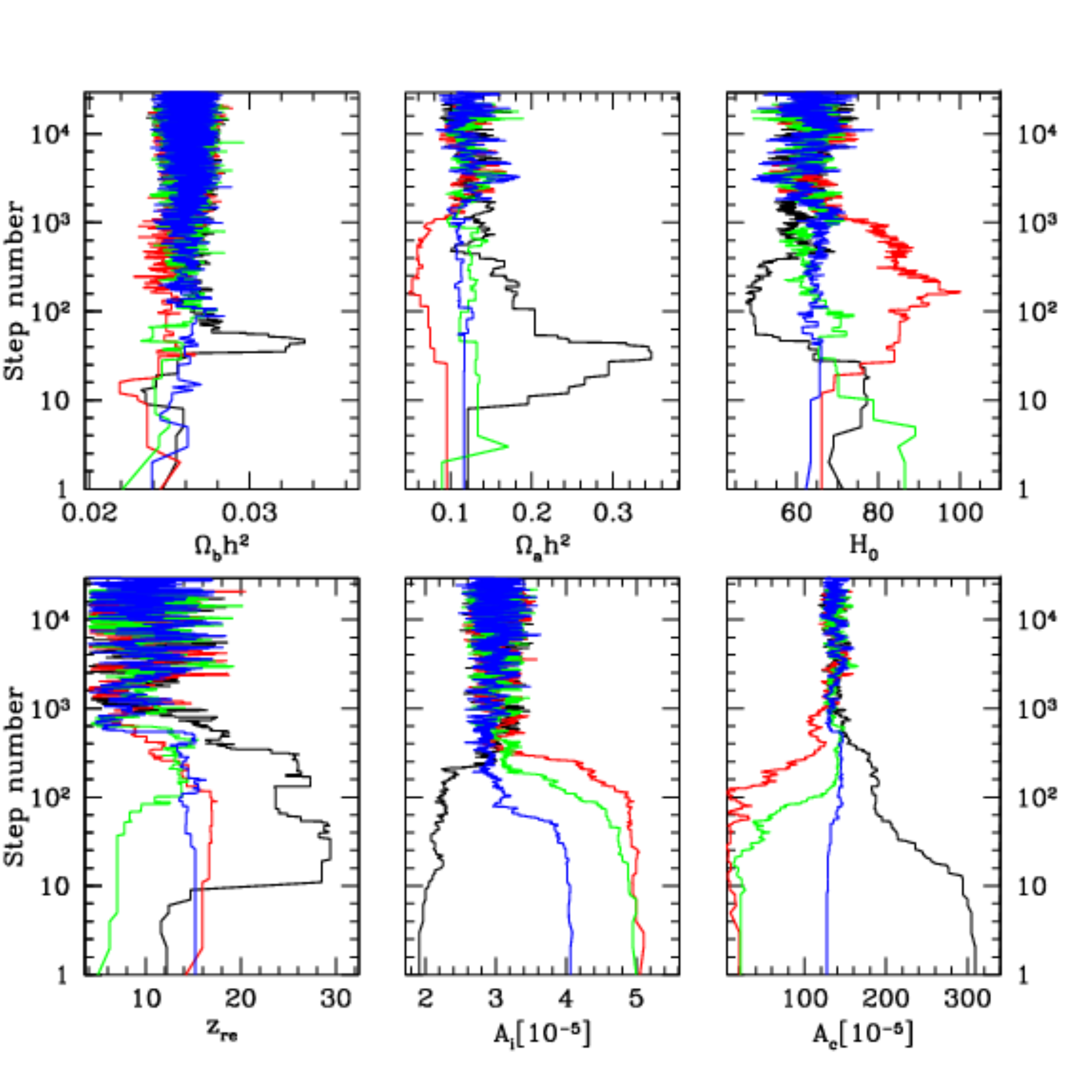}\hfill
\caption[Illustration of the burn-in period for Monte Carlo Markov
chains.]{Illustration of the burn-in period. Left panel: the
logarithm of the log-likelihood, $ - \ln P(\data | \params)$, as a function of the step number for four Monte
Carlo chains. After the burn-in period (dotted, vertical lines),
the value flattens and the chains are sampling from the target
distribution. Right panel: the four chains (in different colors)
are started in different points of a 6-dimensional parameter space
and all converge to the same region after the burn-in. The
vertical axis gives the number of steps.}
\label{fig:chains_evolution}
\end{figure}

Some of the most relevant aspects are: \\
\begin{enumerate}
\item  Initial samples in the chain must be discarded, since the
Markov process is not yet sampling from the equilibrium
distribution (so--called {\em burn--in period}). The length of the burn--in period can
be assessed by looking at the evolution of the posterior
density as a function of the number of steps in the chain. When
the chain is started at a random point in parameter space, the
posterior probability will typically be small and becomes larger
at every step as the chain approaches the region where the fit to
the data is better. Only when the chain has moved in the
neighborhood of the posterior peak the curve of the log posterior
as a function of the step number flattens and the chain begins
sampling from its equilibrium distribution. Samples obtained
before reaching this point must be discarded, see Fig.~\ref{fig:chains_evolution}
 \item A difficult problem is
presented by the assessment of {\em chain convergence}, which aims
at establishing when the MCMC process has gathered enough samples
so that the Monte Carlo estimate~\eqref{eq:MC_estimate} is
sufficiently accurate. Useful diagnostic tools include the Raftery
and Lewis statistics~\cite{Raftery:1995b} and the Gelman and Rubin
criterion~\cite{Gelman:1992b}.
 \item One has to bear in mind that MCMC
is a {\em local algorithm}, which can be trapped around local
maxima of the posterior density, thus missing regions of even
higher posterior altogether. Considerable experimentation is
sometimes required to find an implementation of the MCMC algorithm
that is well suited to the exploration of the parameter space of
interest. Experimenting with different algorithms (each of which
has its own strength and weaknesses) is highly recommended.
 \item Successive samples in a chain are in general correlated. Although this is
not prejudicial for a correct statistical inference, it is often
interesting to obtain {\em independent samples} from the
posterior. This can be achieved by ``thinning'' the chain by an
appropriate factor, i.e. by selecting only one sample every $K$.  The auto-correlation is a good measure of the number of steps required before the chain has ``forgotten'' its previous state. It can be estimated from the MCMC samples as
 \be
\hat{\gamma}(k) = \frac{\sum_{i=0}^{M-k}(\theta_i - \bar{\theta})(\theta_{i+k} - \bar{\theta})}{\sum_{i=0}^{M-k}(\theta_i - \bar{\theta})^2},
\ee
where $k$ is called the lag and $\bar{x}$ is the sample mean (the above equation should be understood component by component if the parameter vector $\theta$ is multi-dimensional). A plot of $\hat{\gamma}$ versus lag $k$ is called ``autocorrelation function'' (ACF) and the value of the lag after which it drops close to 0 provides an estimate of the thinning factor $K$ required to obtain approximate independent samples from the chain. 

A
discussion of samples independence and how to assess it can be
found in~\cite{Dunkley:2004sv}, along with a convergence test
based on the samples' power spectrum.
\end{enumerate}

\subsection{Exercises}

\subsubsection{Bayesian reasoning} 
\begin{enumerate}
\item 
A batch of chemistry undergraduates are screened for a dangerous medical condition called {\em Bacillum Bayesianum} (BB). The incidence of the condition in the population (i.e., the probability that a randomly selected person has the disease) is estimated at about 1\%. If the person has BB, the test returns positive 95\% of the time. There is also a known 5\% rate of false positives, i.e. the test returning positive even if the person is free from BB. One of your friends takes the test and it comes back positive. Here we examine whether your friend should be worried about her health.  
 \begin{enumerate}
 \item Translate the information above in suitably defined conditional probabilities. The two relevant propositions here are whether the test returns positive (denote this with a $+$ symbol) and whether the person is actually sick (denote this with the symbol $BB=1$. Denote the case when the person is healthy as $BB=0$). 
 \item Compute the conditional probability that your friend is sick, knowing that she has tested positive, i.e., find $P(BB=1 | +)$.
 \item Imagine screening the general population for a very rare desease, whose incidence in the population is $10^{-6}$ (i.e., one person in a million has the disease on average, i.e. $P(BB=1) = 10^{-6}$). What should the reliability of the test (i.e., $P(+|BB=1)$) be if we want to make sure that the probability of actually having the disease after testing positive is at least 99\%? Assume first that the false positive rate $P(+| BB=0)$ (i.e, the probability of testing positive while healthy), is 5\% as in part (a). What can you conclude about the feasibility of such a test?\\
 \item Now we write the false positive rate as $P(+|BB=0) = 1 - P(-|BB=0)$. It is reasonable to assume (although this is not true in general) that  $P(- | BB=0) = P(+ | BB=1) $, i.e. the probability of getting a positive result if you have the disease is the same as the probability of getting a negative result if you don't have it.  Find the requested reliability of the test (i.e., $P(+|BB=1)$) so that the probability of actually having the disease after testing positive is at least 99\% in this case. Comment on whether you think a test with this reliability is practically feasible.  \\
 \end{enumerate}

%
\item In a game, you can pick one of three doors, labelled A, B and C. Behind one of the three doors lies a highly desirable price, such as for example a cricket bat. After you have picked one door (e.g., door A) the person who is presenting the game opens one of the remaining 2 doors so as to reveal that there is no prize behind it (e.g., door C might be opened). Notice that the gameshow presenter {\em knows} that the door he opens has no prize behind it. At this point you can either stick with your original choice (door A) or switch to the door which remains closed (door B). At the end, all doors are opened, at which point you will only win if the prize is behind your chosen door. 
\begin{enumerate}
\item Given the above rules (and your full knowledge of them), should you stick with your choice or is it better to switch?
\item In a variation, you are given the choice to randomly pick one of doors B or C and to open it, after you have chosen door A. You pick door C, and upon opening it you discover there is nothing behind it. At this point you are again free to either stick with door A or to switch to door B. Are the probabilities different from the previous scenario? Justify your answers. \\
\end{enumerate}
%
\item In a TV debate, politician $A$ affirms that a certain proposition $S$ is true. You trust politician $A$ to tell the truth with probability 4/5. Politician $B$ then agrees that what politician $A$ has said is indeed true. Your trust in politician $B$ is much weaker, and you estimate that he lies with probability 3/4. 

After you have heard politician $B$, what is the probability that statement $S$ is indeed true? You may assume that you have no other information on the truth of proposition $S$ other than what you heard from politicians $A$ and $B$.

{\em Hint: Start by denoting by $A_T$ the statement ``politician A tells the truth'', and by $B_T$ the statement ``politician B tells the truth''. What you are after is the probability of the statement ``proposition $S$ is true'' after you have heard politician $B$ say so.} 

\item A body has been found on the Baltimore West Side, with no apparent wounds, although it transpires that the deceased, a Mr Fuzzy Dunlop, was a heavy drug user. The detective in charge suggests to close the case and to attribute the death to drugs overdose, rather than murder. 

Knowing that, of all murders in Baltimore, about 30\% of the victims were drug addicts, and that the probability of a dead person having died of overdose is 50\% (without further evidence apart from the body) estimate the probability that the detective's hunch is correct. (For this problem, you may assume that the possible only causes of death are overdose or murder). \\

\subsubsection{Bayesian parameter inference}
\label{sec:bayes_ex}
\item This problem takes you through the steps to derive the posterior distribution for a quantity of interest $\theta$, in the case of a Gaussian prior and Gaussian likelihood, for the 1-dimensional case.

Let us assume that we have made $N$ independent measurements, $\hat{x} = \{\hat{x}_1, \hat{x}_2, \dots, \hat{x}_N \}$ of a quantity of interest $\theta$ (this could be the temperature of an object, the distance of a galaxy, the mass of a planet, etc). We assume that each of the measurements in independently Gaussian distributed with known experimental standard deviation $\sigma$. Let us denote the sample mean by $\bar{x}$, i.e.
\be
\bar{x} = \frac{1}{N} \sum_{i=1}^N \hat{x}_i.
\ee
Before we do the experiment, our state of knowledge about the quantity of interest $\theta$ is described by a Gaussian distribution on $\theta$, centered around 0 (we can always choose the units in such a way that this is the case). Such a prior might come e.g. from a previous experiment we have performed. The new experiment is however much more precise, i.e. $\Sigma \gg \sigma$. Our prior state of knowledge be written in mathematical form as the following Gaussian pdf:
\be
p(\theta) \sim \mathcal{N}(0,\Sigma^2).
\ee
\begin{enumerate}
\item Write down the likelihood function for the measurements and show that it can be recast in the form:
\be
{\mathcal L}(\theta) = L_0 \exp\left(-\frac{1}{2}\frac{(\theta-\bar{x})^2}{\sigma^2/N}\right),
 \ee
where $L_0$ is a constant that does not depend on $\theta$. \\
\item By using Bayes theorem, compute the posterior probability for $\theta$ after the data have been taken into account, i.e. compute $p(\theta | \hat{x})$. Show that it is given by a Gaussian of mean $\bar{x}\frac{\Sigma^2}{\Sigma^2 + \sigma^2/N}$ and variance $\left[\frac{1}{\Sigma^2} + \frac{N}{\sigma^2}\right]^{-1}$. \\
{\em Hint: you may drop the normalization constant from Bayes theorem, as it does not depend on $\theta$.}
\item Show that as $N \rightarrow \infty$ the posterior distribution becomes independent of the prior.
\end{enumerate}
%

\item We already encountered the coin tossing problem, but this time you'll do it in the Bayesian way.  

A coin is tossed $N$ times and heads come up $H$ times. 
 \begin{enumerate}
 \item What is the likelihood function? Identify clearly the parameter, $\theta$, and the data. 
 \item What is a reasonable, non-informative prior on $\theta$? 
 \item Compute the posterior probability for $\theta$. Recall that $\theta$ is the probability that a single flip will give heads. This integral will prove useful:
 \be
 \int_0^1 d\theta \theta^N (1-\theta)^M = \frac{\Gamma(N+1)\Gamma(M+1)}{\Gamma(N+M+2)}.  
 \ee
 \item Determine the posterior mean and standard deviation of $\theta$.
 \item Plot your results as a function of $H$ for $N=10, 100, 1000$. 
 \item $\dagger$  Generalize your prior to the Beta distribution, 
 \be
 p(\theta | \nu_1, \nu_2) = \frac{1}{B(\nu_1, \nu_2)}\theta^{\nu_1-1} (1-\theta)^{\nu_2-1}
 \ee
 where $B(\nu_1, \nu_2) = \Gamma(\nu_1) \Gamma(\nu_2)/\Gamma(\nu_1+\nu_2)$ is the beta function and the ``hyperparameters'' $\nu_1, \nu_2 >0$. Clearly, a uniform prior is given by the choice $(\nu_1, \nu_2) = (1,1)$. Evaluate the dependency of your result to the choice of hyperparameters. 
 \item $\dagger$ What is the probability that the $(N+1)$-th flip will give heads?  
 \end{enumerate}
%
\item Prove Eqs.~\eqref{eq:like_glm}, \eqref{eq:post} and \eqref{eq:modlike} in the notes for the Gaussian linear model given by 
 \be
 y = F \theta + \epsilon.
 \ee
{\em Hint:} recall this standard result for Gaussian integrals:
\be
\int \exp\left[-\frac{1}{2}({x} - {m})^t {\Sigma}^{-1} ({x} - {m})\right] \dr {x} = \sqrt{\text{det}(2\pi{\Sigma}})
\ee
%
\item  Now we specialize to the case $n=2$, i.e.\ we have two parameters of interest, $\theta = \{\theta_1, \theta_2 \}$ and the linear function we want to fit is given by
\be
y = \theta_1 + \theta_2 x. 
\ee
(In the formalism above, the basis vectors are $X^1 = 1, X^2 = x$).

Table~\ref{tab:linear_model} gives an array of $d=10$ measurements $y = \{y_1, y_2, \dots, y_{10}\}$, together with the values of the independent variable $x_i$\footnote{This data set is also provided with this arxiv submission as an ancillary data file called \texttt{LinearModelData.txt}. It can be downloaded from a link below the usual article download links.}. Assume that the uncertainty in the same for all measurements, i.e. $\tau_i = 0.1$ ($i=1,\dots,10$).  You may further assume that measurements are uncorrelated.  The data set is shown in the left panel of Fig.~\ref{fig:exdata}.

\begin{table}
\caption{Data sets for the Gaussian linear model exercise. You may assume that all data points are independently and identically distributed with standard deviation of the noise $\sigma = 0.1$.}
\label{tab:linear_model}       
%
%
\begin{tabular}{p{2cm}p{2.4cm}p{2cm}p{4.9cm}}
\hline\noalign{\smallskip}
$x$ & $y$  \\
\noalign{\smallskip}\svhline\noalign{\smallskip}
0.8308 & 0.9160 \\
0.5853 & 0.7958 \\
0.5497 & 0.8219 \\
0.9172 & 1.3757 \\
0.2858 & 0.4191 \\
0.7572 & 0.9759 \\
0.7537 & 0.9455 \\
0.3804 & 0.3871 \\
0.5678 & 0.7239 \\
0.0759 & 0.0964 \\
\noalign{\smallskip}\hline\noalign{\smallskip}
\end{tabular}
\end{table}

\begin{figure}
\centering
\includegraphics[width = 1.0\linewidth]{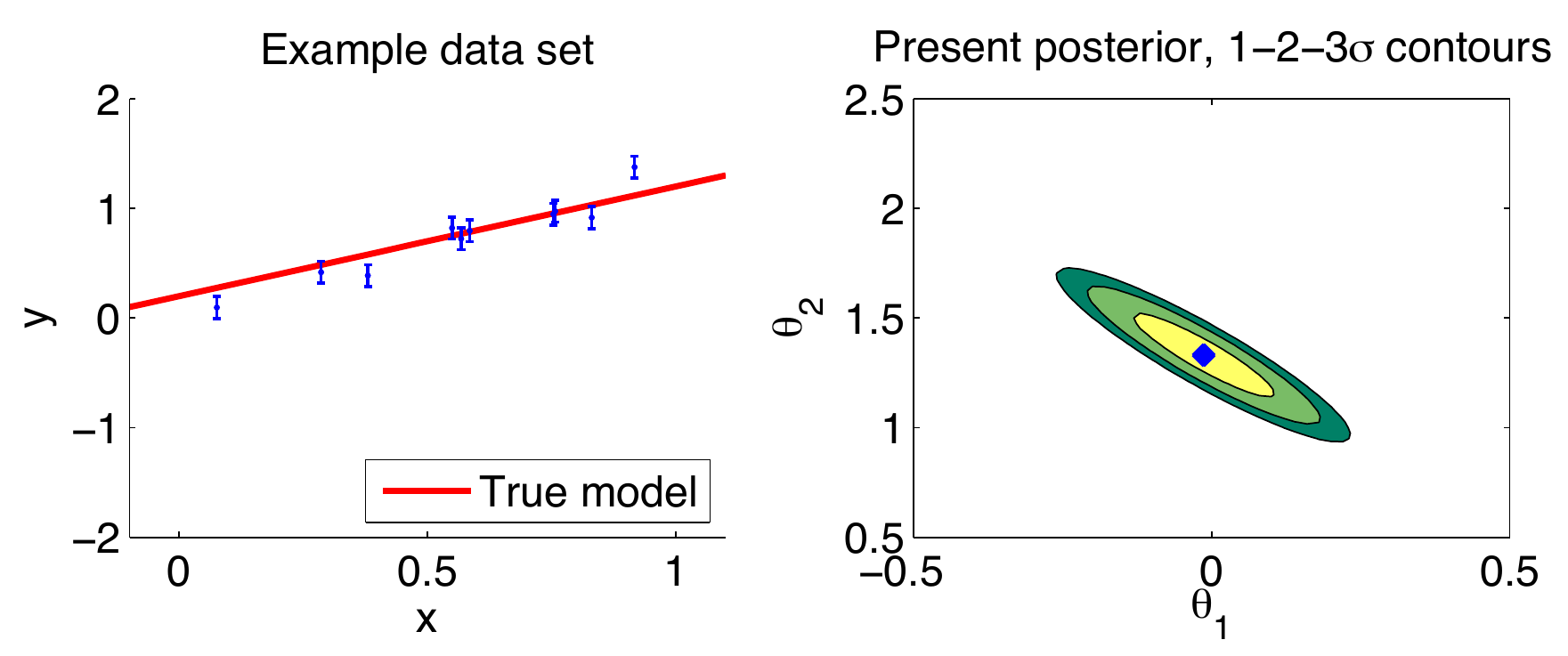}
\caption{Left panel: data set for the Gaussian linear problem. The solid line shows the true value of the linear model from which the data have been generated, subject to Gaussian noise. Right panel: 2D credible intervals from the posterior distribution for the parameters. The the blue diamond is the Maximum Likelihood Estimator, from Eq.~\eqref{maxlike}, whose value for this data set is  $x=  -0.0136, y = 1.3312$.}
\label{fig:exdata}
\end{figure}
  \begin{enumerate}
  \item Assume a Gaussian prior with Fisher matrix $P = \text{diag}\left(10^{-2} ,10^{-2} \right)$ for $\theta$. 
  
Find the posterior distribution for $\theta$ given the data, and plot it in 2 dimensions in the $(\theta_1, \theta_2)$ plane (see right panel of Fig.~\ref{fig:exdata}). 
  
Use the appropriate contour levels to demarcate 1, 2 and 3 sigma joint credible intervals of the posterior.
\item In a language of your choice, write an implementation of the Metropolis-Hastings Markov Chain Monte Carlo algorithm, and use it to obtain samples from the posterior distribution. 
  
Plot {\em equal weight samples}\footnote{Posterior samples obtained via MCMC have a weight associated with them, given by the number of times the samplers has failed to jump from that particular location (i.e., the number of repeat counts of the same coordinate location in parameter space). Producing a scatter plot of such weighted samples would fail to reproduce visually their actual density, for each sample would have a different weight associated to it. Equal weight samples are obtained from the MCMC chain by normalizing all weights to unity (i.e, replacing weights $w_i$ by $u_i \equiv w_i/\text{max}_{i} w_i$, where $i=1,\dots, N$ is the number of samples in the chain) and retaining each sample with probability given by $u_i$.} in the $(\theta_1, \theta_2)$ space, as well as marginalized 1-dimensional posterior distributions for each parameter. 

\item Compare the credible intervals that you obtained from the MCMC with the analytical solution.

\end{enumerate}


\item Supernovae type Ia can be used as standardizable candles to measure distances in the Universe. This series of problems explores the extraction of cosmological information from a simplified SNIa toy model. 

The cosmological parameters we are interested in constraining are 
\be
\Cp = \{\Om, \OL, h \}
\ee 
where $\Om$ is the matter density (in units of the critical energy density) and $\OL$ is the dark energy density, assumed here to be in the form of a cosmological constant, i.e. $w=-1$ at all redshifts.  In the following, we will fix $h=0.72$, where the Hubble constant today is given by $H_0 = 100 h$ km/s/Mpc, since the value of $H_0$ is degenerate with the (unknown) absolute magnitude of the SNIas, $M$.

In an FRW cosmology defined by the parameters $\Cp$, the distance modulus $\mu$ (i.e., the difference between the apparent and absolute magnitudes, $\mu = m-M$) to a SNIa located at redshift $z$ is given by 
\be
\mu(z, \Cp) = 5 \log \left[ \frac{D_L(z,\Om, \OL,h)}{\Mpc} \right] +25,
\ee
where $D_L$ denotes the luminosity distance to the SNIa. Recalling that $D_L = c d_L/H_0$, we can rewrite this as
\be
\mu(z, \Cp) = \eta + 5\log d_L(z,\Om, \OL),
\ee
where
\be \label{eq:defeta}
\eta = -5\log\frac{100h}{c} + 25
\ee
and $c$ is the speed of light in km/s. We have defined the dimensionless luminosity distance 
\be
d_L(z,\Om, \OL) = \frac{(1+z)}{\sqrt{|\Omk|}}\text{sinn} 
\{\sqrt{|\Omk|} \int_0^z \dr z' 
[ (1+z')^3\Om + \OL + (1+z')^2\Omk ]^{-1/2} \} .
\ee
The curvature parameter is given by the constraint equation
\be
\Omk = 1 - \Om - \OL
\ee
and the function 
\be
\text{sinn}(x) = \left\{\begin{array}{ll}
        x & \mbox{for a flat Universe ($\Omk = 0$)};\\
        \sin(x) & \mbox{for a closed Universe ($\Omk < 0$)}; \\
        \sinh(x) & \mbox{for an open Universe ($\Omk > 0$)}.
        \end{array} \right. 
\ee

We now assume that from each SNIa in our sample we have a measurement of its distance modulus with Gaussian noise\footnote{We neglect the important issue of applying the empirical corrections known as Phillip's relations to the observed light curve. This is of fundamental important in order to reduce the scatter of SNIa within useful limits for cosmological distance measurements, but it would introduce a technical complication here without adding to the fundamental scope of this exercise. Furthermore, the correct likelihood function is not of the Gaussian form given here. For a fully Bayesian treatment, see e.g.~\cite{Shariff:2015yoa}.}, i.e., that the likelihood function for each SNIa $i$ ($i=1,\dots,N$) is of the form
\be
\like_i(z_i,\Cp,M) = \frac{1}{\sqrt{2\pi}\sigma_i}\exp\left(-\frac{1}{2}\frac{( \hat{\mu_i}-\mu(z_i, \Cp))^2}{\sigma_i^2}\right).
\ee 
The observed distance modulus is given by $\hat{\mu_i}  = \hat{m}_i - M$, where $\hat{m}_i$ is the observed apparent magnitude and $M$ is the intrinsic magnitude of the SNIa. 
We assume that each SN observation is independent of all the others. 

The provided data file\footnote{The datafile is provided with this arxiv submission as ancillary data file. It can be downloaded from a link below the usual article download links.} (\texttt{SNIa\_SimulatedData}) contains simulated observations from the above simplified model of $N=300$ SNIa. The two columns give the redshift $z_i$ and the observed apparent magnitude $ \hat{m}_i$. The observational error is the same for all SNe, $\sigma_i=\sigma=0.4$ mag  for $i=1,\dots, N$. 

A plot of the data set is shown in the left panel of Fig.~\ref{fig:dataset}. The characteristics of the simulated SNe are designed to mimic currently available datasets (see~\cite{Kowalski:2008ez, Amanullah:2010vv, Kessler:2009ys,Rest:2013bya,Betoule:2014frx}).

\begin{enumerate}
 \item We assume that the intrinsic magnitude\footnote{In reality the SNIas intrinsic magnitude is not the same for all of the objects, but there is an ``intrinsic dispersion'' (even after Phillips' corrections) reflecting perhaps intrinsic variability in the explosion mechanism, or environmental parameters which are currently poorly understood. } is known and fix $M=M_0=-19.3$ and that $h=0.72$. We also assume that the observational error is known, given by the value above. 
 
Using a language of your choice, write a code to carry out an MCMC sampling of the posterior probability for $(\Om, \OL)$ and plot the resulting 68\% and 95\% posterior regions, both in 2D and marginalized to 1D, using uniform priors on $(\Om, \OL)$ (be careful to define them explicitly). 
 
  You should obtain a result similar to the 2D plot shown in the right panel of Fig.~\ref{fig:dataset}. 
 
 \begin{figure}
\begin{center}
\includegraphics[width=0.48\linewidth, angle=0, trim ={100 0 100 50}, clip]{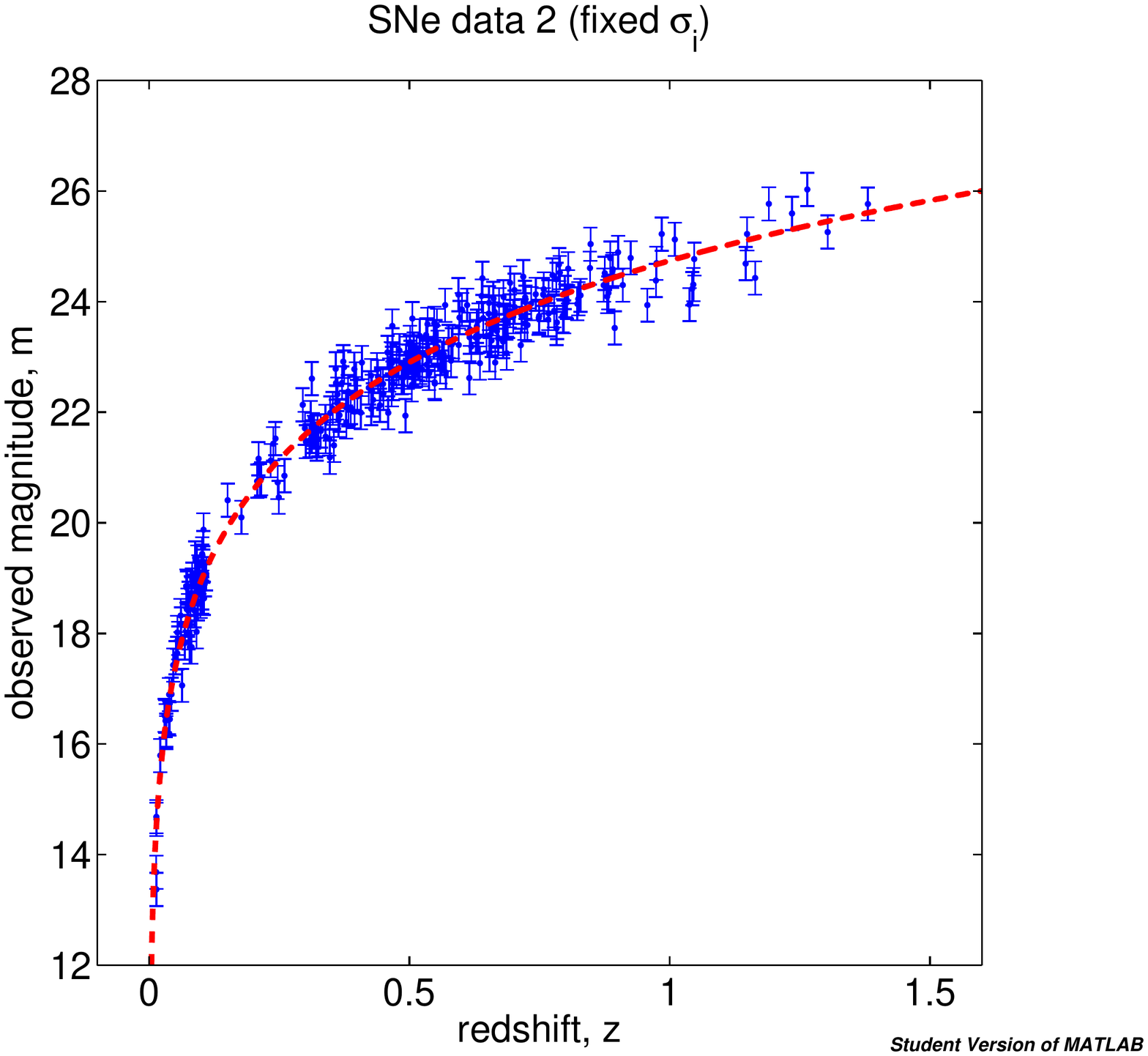} \hfill 
\includegraphics[width=0.48\linewidth, angle=0, trim = {130 0 100 50}, clip]{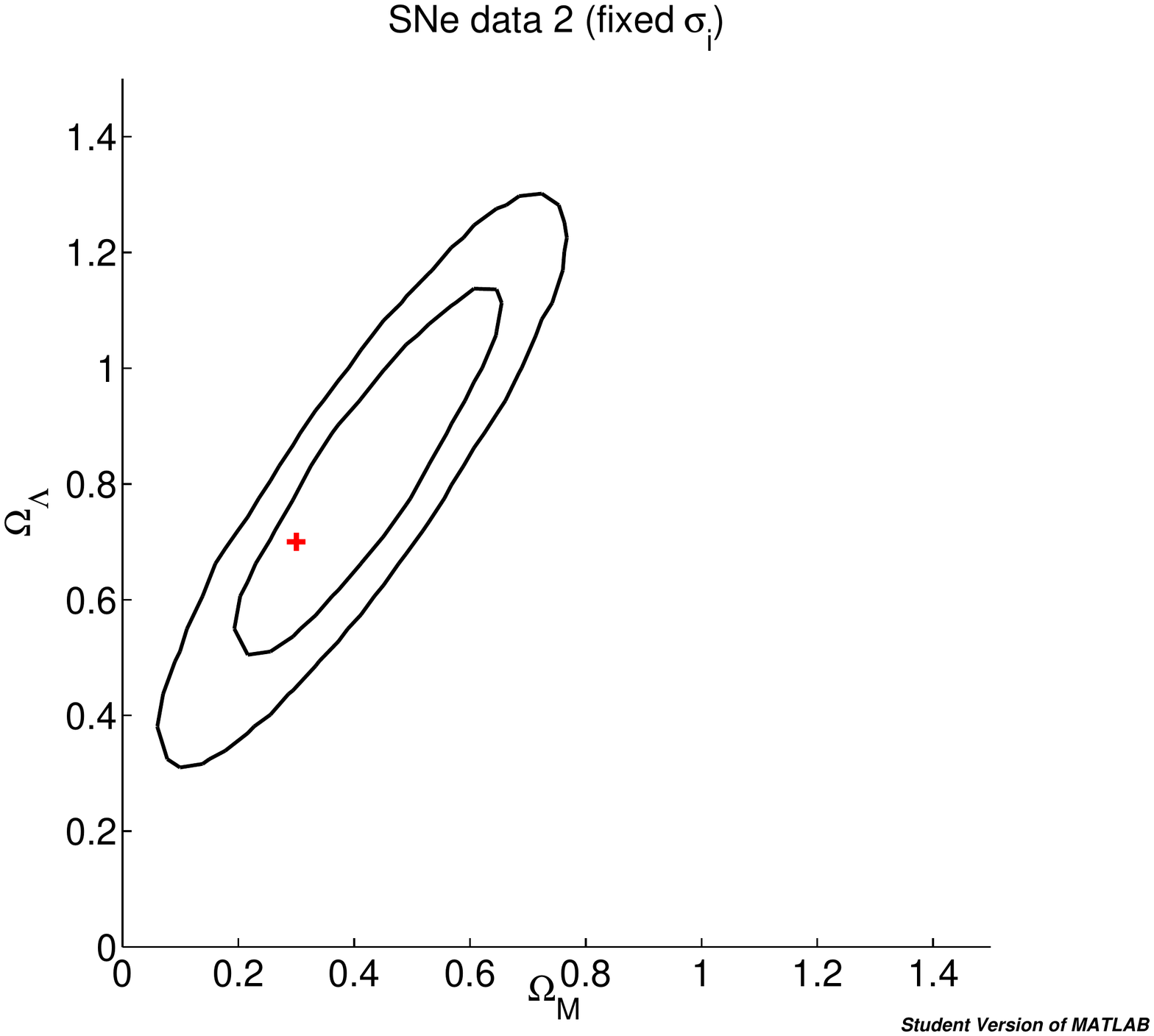}
\end{center}
\caption{Left: Simulated SNIa dataset, \texttt{SNe\_simulated.dat}. The solid line is the true underlying cosmology. Right: constraints on $\Om, \OL$ from this dataset, with contours delimiting 2D joint 68\% and 95\% credible regions (uniform priors on the variables $\Om, \OL$, assuming $M=M_0$ fixed and $h=0.72$). The red cross denotes the true value.}
\label{fig:dataset}
\end{figure}
 
\item $\dagger$ Add the quantity $\sigma$ (the observational error) to the set of unknown parameters and estimate it from the data along with $\Cp$. Notice that since $\sigma$ is a ``scale parameter'', the appropriate (improper) prior is $p(\sigma) \propto 1/\sigma$ (see~\cite{Box:1992} for a justification).

\item The location of the peaks in the CMB power spectrum gives a precise measurement of the angular diameter distance to the last scattering surface, divided by the sound horizon at decoupling. This approximately translates into an effective constraint (see ~\cite{Spergel:2006hy}, Fig.~20) on the following degenerate combination of $\Om$ and $\OL$: 
\be
1.41 \OL + \Om = 1.30 \pm 0.04.
\ee 
Add this constraint (assuming a Gaussian likelihood, with the above mean and standard deviation) to the SNIa likelihood and plot the ensuing combined 2D and 1D limits on $(\Om, \OL)$.

\item The measurement of the baryonic acoustic oscillation scale in the galaxy power spectrum at small redshift gives an effective constraint on the angular diameter distance $D_A$ out to $z \sim 0.6$. This measurement can be summarized as~\cite{Anderson:2013oza}:
\be
D_A(z=0.57) = (1408 \pm 45) \text{ Mpc}.
\ee
Add this constraints (again assuming a Gaussian likelihood) to the above CMB+SNIa limits and plot the resulting combined 2D and 1D limits on $(\Om, \OL)$. \\
{\em Hint:} recall that $D_L(z) = (1+z)^2D_A(z)$.
  
\end{enumerate}
\end{enumerate}

\subsection{Solutions to selected exercises}

\subsubsection{Bayesian reasoning} 
\begin{enumerate}
\item 
 \begin{enumerate}
 \item 
 Let $BB=1$ denote the proposition that your friend has the virus, and $BB=0$ that she does not. We use $+$ ($-$) to denote the test returning a positive (negative) result.
We know from the reliability of the test that
\begin{align}
P(+ | BB=1) & = 0.95 \\
P(+ | BB =0) & = 0.05 \text{ hence } \\ 
P(- | BB =0) & = 0.95.
\end{align}
Given that $1\%$ of the population has the virus, the probability of being one of them (before taking the test) is  $P(BB=1) = 0.01$, while $P(BB=0) = 0.99$. 

 \item The probability of your friend having the virus after she has tested positive is thus
\be
P(BB=1|+) = \frac{P(+|BB=1)P(BB=1)}{P(+)}.
\ee
We can compute the denominator as follows, by combining the marginalization rule with the product rule (a procedure that is sometimes called ``expanding the discourse"): 
\begin{align}
P(+) & = P(+|BB=1)P(BB=1) + P(+| BB=0) P(BB=0) \\ 
& = 0.95 \cdot 0.01 + 0.05 \cdot 0.99 = 0.059.
\end{align}
Therefore the probability that your friend has the virus is much less than 95\%, namely
\be 
P(BB=1|+) = \frac{0.95 \cdot 0.01}{0.059} = 0.16 = 16\%.
\ee

 \item From the above, we have that 
\be \label{eq:solve}
P(BB=1|+) = \frac{P(+|BB=1)P(BB=1)}{P(+|BB=1)P(BB=1) + P(+| BB=0) P(BB=0)}.
\ee
We want to achieve 99\% probability that the person has BB given that they tested positive, i.e., $P(BB=1|+) = 0.99$, and we need to solve the above equation for the reliability, i.e. $P(+|BB=1)$. 

We first assume that $P(+| BB=0) = 0.05$, as in part (a). Since $P(BB=1) = 10^{-6}$, it follows that $P(BB=0) = 1 - P(BB=1)  \approx 1$. Then Eq.~\eqref{eq:solve} becomes 
\be
0.99 = \frac{P(+|BB=1)10^{-6}}{P(+|BB=1)10^{-6} + 0.05 \times 1} \approx \frac{P(+|BB=1)10^{-6}}{0.05}.
\ee
It is clear that this equation has no solution for $P(+|BB=1) \leq 1$. This means that for a 5\% false positive rate and for the given incidence $P(BB=1)$ it is impossible to obtain a test that is 99\% reliable. Therefore in order to achieve 99\% reliability, the false positive rate, $P(+|BB=0)$, has to be reduced, as well.

 \item Let us denote by $x = P(+ | BB=1) = P(- | BB=0)$ the reliability of the test. Then requiring a value of 99\% for  $P(BB=1|+)$ amounts to solving for $x$ the following equation: 
\be 
0.99 = P(BB=1|+) = \frac{x P(BB=1)}{x P(BB=1) + (1-x) P(BB=0)}
\ee
where $P(BB=1) = 10^{-6}$ and $P(BB=0) = 1-10^{-6} \approx 1$. This gives for $x$
\be
x \approx \frac{1}{1+10^{-8}},
\ee which means that the the reliability of the test ought to be in excess of 1 in $10^8$. This is obviously not feasible and hence it is important to screen people before administering the test, i.e., to only test people who already show symptoms of the condition. 

\end{enumerate}

%
\item 
\begin{enumerate}
\item Let's assume that you have chosen door $A$. If the prize is indeed behind that door, than the presenter opens randomly one of $B$ or $C$ (with probability $1/2$). If the prize is behind door $B$, then he must open door $C$ (and viceversa). This means: 

\begin{align} 
P(B \text{ open} | \text{prize behind } A) = \frac{1}{2} & \quad P(C \text{ open} | \text{prize behind } A) = \frac{1}{2}  \\
P(B \text{ open} | \text{prize behind } B) = 0 & \quad P(C \text{ open} | \text{prize behind } B) = 1 \label{eq1}\\
P(B \text{ open} | \text{prize behind } C) = 1 & \quad P(C \text{ open} | \text{prize behind } C) = 0  \label{eq2} 
\end{align}

If the presenter opens door $C$, we obtain the probability that the prize is behind each of the doors by inverting the order of conditioning as follows:

\be \label{PA}
P( \text{prize behind } A | C \text{ open}) = \frac{ P(C \text{ open} | \text{prize behind } A) P(A)}{P( C \text{ open})} 
\ee
and
\be \label{PB}
P( \text{prize behind } B | C \text{ open}) = \frac{ P(C \text{ open} | \text{prize behind } B) P(B)}{P( C \text{ open})} 
\ee
At the beginning of the show, the probability that the prize is behind one of the 3 doors is the same:
\be \label{eq:doorsprior}
P(A) = P(B) = P(C) = \frac{1}{3}.
\ee
We can compute the denominator in Eqs.~\eqref{PA} and \eqref{PB} using again the rules of probability (in particular, the marginalisation rule):
\be
\begin{split} 
P( C \text{ open}) = P( C \text{ open} | \text{prize behind } A) P(A) \\ + P( C \text{ open} | \text{prize behind } B) P(B)  \\ 
+ P( C \text{ open} | \text{prize behind } C) P(C) \\
= \frac{1}{2}\cdot \frac{1}{3} + 1 \cdot \frac{1}{3} + 0 \cdot \frac{1}{3} = \frac{1}{2}.  
\end{split}
\ee
From Eqs.~\eqref{PA} et \eqref{PB} it follows 
\be
P( \text{prize behind } A | C \text{ open}) = \frac{1}{3}, \quad P( \text{prize behind } B | C \text{ open}) = \frac{2}{3}.
\ee
Therefore you should switch in order to increase your probability of winning (from $1/3$ to $2/3$). 

If you are still unconvinced, here is a variant to hone your intuition: there are 1000 doors, and you pick one at the beginning. The presenter then opens 998 doors, revealing that there is no prize behind them (and he knew this when he opened them). At this point you can either switch to the last remaining closed door, or stick with the one you had originally chosen. Which way to go is at this point a no-brainer! 

\item 
In the second scenario, you choose between doors $B$ and $C$ randomly, and therefore the amount of information in the problem changes (in the previous case, the presenter {\em knew} behind which door the prize is). Eqs.~\eqref{eq1} et \eqref{eq2} are modified as follows: 
\begin{align} 
P(B \text{ open} | \text{prize behind } B) = \frac{1}{2} & \quad P(C \text{ open} | \text{prize behind } B) = \frac{1}{2} \\
P(B \text{ open} | \text{prize behind } C) = \frac{1}{2} & \quad P(C \text{ open} | \text{prize behind } C) = \frac{1}{2}  
\end{align}
In this case, the probability of winning is {\em not} modified by you opening a further door at random, and in fact:
 \be
P( \text{prize behind } A | C \text{ open}) = \frac{1}{2}, \quad P( \text{prize behind } B | C \text{ open}) = \frac{1}{2}.
\ee
Another, more formal argument goes as follows. The prior is given by Eq.~\eqref{eq:doorsprior} and
\begin{align}
P(C \text{ open}) & = \sum_{i=A,B,C}P(i \text{ open}|\text{prize behind } i)P(\text{prize behind } i) \\
& =  \frac{1}{2}\frac{1}{3} + \frac{1}{2}\frac{1}{3}+ 0\frac{1}{3} = \frac{1}{3}
\end{align}
This is because 
\be
P(C \text{ open} |  \text{prize behind } A) = P(C\text{ open}|  \text{prize behind } B) = \frac{1}{2}
\ee
but
\be
P(C\text{ open}| \text{prize behind } C) = 0
\ee
since this statement is incompatible with the evidence (when C is opened at random by you, you discover that the price is not there!). So:
\begin{align}
P( \text{prize behind } A|C\text{ open}) & = \frac{P(C\text{ open}| \text{prize behind } A)P( \text{prize behind } A)}{P(C\text{ open})} \\ 
& = \frac{(1/2)(1/3)}{(1/3)} = \frac{1}{2}
\end{align}
as claimed. 

\end{enumerate}
%
\item 
 Let $S_T$ denote the proposition ``statement $S$ is true''. Let $A_T$ denote the statement ``politician $A$ tells the truth,  $A_L$ denote the statement ``politician $A$ lies'' and similarly for $B_T$ and $B_L$. Under your prior, $P(A_T) = 4/5$, $P(A_L) = 1/5$, $P(B_T) = 1/4$ and $P(B_L) = 3/4$. Let ``BST`` denote the statement ``Politician $B$ says $S$ is true``. Then using Bayes theorem:
\begin{equation}
P(S_T | BST) = \frac{P(BST | S_T) P(S_T)}{P(BST)}.
\end{equation}
In the above equation, $P(S_T) = P(A_T) = 4/5$, as you don't know anything else about statement $S$ except what you heard from politician $A$, whom you trust to be truthful with probabilty $P(A_T)$. Also, $P(BST | S_T) = P(B_T) = 1/4$, for politician $B$ will say that statement $S$ is true (if this is indeed the case) with probability $P(B_T)$. It remains to compute
\begin{equation}
\begin{aligned}
P(BST) & = P(BST|S_T)P(S_T) + P(BST|\text{not } S_T)P(\text{not } S_T)  \\
	& = P(B_T)P(A_T) + P(B_L)P(A_L) .
\end{aligned}
\end{equation}
So the posterior probability for $S$ to be true after you have heard both politicians is  
\begin{equation}
\begin{aligned}
P(S_T | BST) & = \frac{P(B_T)P(A_T)}{ P(B_T)P(A_T) + P(B_L)P(A_L)}  \\ 
&= \frac{1}{1 + \frac{P(B_L)}{P(B_T)}\frac{P(A_L)}{P(A_T)}} = \frac{1}{1 + \frac{3}{4}} = 4/7 \approx 57\%.
\end{aligned}
\end{equation}

%
\item 
Let us denote by $od = 1$ the statement ``Mr Dunlop died because of drugs overdose''; by $Dd=1$ the statement ``Mr Dunlop is dead'' and by $u=1$ the statement ``Mr Dunlop used drugs''. 

We are looking for the posterior probability that Fuzzy Dunlop died of overdose, given that he was a drug addict ($u=1$) and that he is dead ($Dd = 1$):
\begin{equation*}
\begin{aligned}
& P(od = 1 | Dd = 1, u = 1) \\ 
& = \frac{P(u=1 | od = 1, Dd = 1) P(od =1 | Dd = 1)}{ P(u=1 | od = 1, Dd = 1)P(od =1 | Dd = 1) + P(u=1 | od = 0, Dd = 1)P(od =0 | Dd = 1)}
\end{aligned}
\end{equation*}
From the problem, we have that the probability of being a drug user and having been murdered (assuming that people only die of either overdose or murder in Baltimore) is $P(u=1 | od = 0, Dd = 1) = 0.3$.  Also, the probability of the person having died of overdose (given that we have the body) is 50\%, hence $P(od = 1 | Dd = 1) = 50\%$ so $P(od = 0 | Dd = 1) = 50\%$.

Finally, we need to estimate the probability that Mr Dunlop was a drug user, given that he died of overdose, $ P(u=1 | od = 1, Dd = 1)$. It seems highly unlikely that somebody would die of overdose the first time they try drugs, so perhaps we can assign $ P(u=1 | od = 1, Dd = 1) = 0.9$. 

So we have that 
\begin{equation*}
\begin{aligned}
P(od = 1 | Dd = 1, u = 1) & = \frac{1}{1 + \frac{P(u=1 | od = 0, Dd = 1)P(od =0 | Dd = 1)}{P(u=1 | od = 1, Dd = 1) P(od =1 | Dd = 1)}} \\ 
& = \frac{1}{1+3/9} = 75\%.
\end{aligned}
\end{equation*}
How sensitive is this conclusion to our guess for $P(u=1 | od = 1, Dd = 1) $? Changing this to $P(u=1 | od = 1, Dd = 1)  = 0.5$ (we are agnostic as to whether a drug overdose is more likely for usual drugs consumers or for novices) gives a posterior $P(od = 1 | Dd = 1, u = 1) = 62\%$, while increasing it to $P(u=1 | od = 1, Dd = 1)  = 0.99$ (most people overdosing are drugs users) gives $P(od = 1 | Dd = 1, u = 1) = 77\%$. So even looking at the two extreme cases we can still bracket our conclusion to be in the range from 62\% to 77\%.

\subsubsection{Bayesian parameter inference}
\label{sec:bayes_ex}
\item 
\begin{enumerate}
\item The likelihood is given by 
 \be 
 {\mathcal L}(\theta) = \prod_{i=1}^{N} \frac{1}{\sqrt{2\pi}\sigma}  \exp\left(-\frac{1}{2}\frac{(\theta-\xhati)^2}{\sigma^2}\right).
 \ee
Consider now the exponential term:
 \be
 \begin{split}
 \frac{1}{2} \sum_i \frac{(\theta-\xhati)^2}{\sigma^2} & = \frac{1}{2\sigma^2}\left(N\theta^2 - 2\sum_i \xhati \theta + \sum_i \xhati^2\right) \\
&  = \frac{N}{2\sigma^2}\left(\theta^2 - 2 \theta \bar{x} +  \bar{x}^2 -  \bar{x}^2  + \frac{1}{N}\sum_i \xhati^2\right) \\
& = 
 \frac{N}{2\sigma^2} (\theta-\bar{x})^2  + \frac{N}{2\sigma^2} \left(\frac{1}{N} \sum_i \xhati^2 -  \bar{x}^2  \right) 
\end{split}
\ee
So the likelihood can be written as 
\be
L(\theta) = L_0 \exp\left(-\frac{1}{2}\frac{(\theta-\bar{x})^2}{\sigma^2/N}\right),
 \ee
where $L_0$ is a constant that does not depend on $\theta$. 
\item The posterior pdf for $\theta$ is proportional to the likelihood times the prior (dropping the normalization constant in Bayes' Theorem):
\be
p(\theta | \hat{x}) \propto {\mathcal L}(\theta)p(\theta) \propto \exp\left(-\frac{1}{2}\frac{(\theta-\bar{x})^2}{\sigma^2/N}\right) \exp\left(-\frac{1}{2}\frac{\theta^2}{\Sigma^2}\right), 
\ee
where we have dropped normalization constants which do not depend on $\theta$ and we have used the Gaussian form of the prior. Collecting terms that depend on $\theta$ in the exponent and completing the square we get 
\be \label{eq:post1}
p(\theta | \hat{x})  \propto  \exp\left(-\frac{1}{2}\frac{(\theta-\bar{x}\frac{\Sigma^2}{\Sigma^2 + \frac{\sigma^2}{N}})^2}{\left[\frac{1}{\Sigma^2} + \frac{N}{\sigma^2}\right]^{-1}}\right),
\ee
which shows that the posterior for $\theta$ is a Gaussian with the mean and variance as given in the question.

\item When $N \rightarrow \infty$, we have that the variance $\left[\frac{1}{\Sigma^2} + \frac{N}{\sigma^2}\right]^{-1} \rightarrow \sigma^2/N$ (as $\frac{N}{\sigma^2} \gg \frac{1}{\Sigma^2} $)  and the mean $\bar{x}\frac{\Sigma^2}{\Sigma^2 + \frac{\sigma^2}{N}} \rightarrow \bar{x}$ (as $\Sigma^2 \gg \frac{\sigma^2}{N}$ and the fraction goes to unity). Thus the posterior pdf becomes
\be \label{eq:post1}
p(\theta | \hat{x}) \rightarrow  \exp\left(-\frac{1}{2}\frac{(\theta-\bar{x})^2}{{\sigma^2}/N}\right),
\ee
which shows that the posterior converges to the likelihood and the prior dependence disappears.
\item From the above result, we can use the posterior pdf to compute the posterior mean of $\theta$:  
\be
\langle \theta \rangle  = \int \theta p(\theta | \hat{x}) d\theta = \bar{x}.
\ee
Therefore the posterior mean tends to the sample mean, $\bar{x}$, which as we know is also the MLE for the mean. 

\end{enumerate}

\end{enumerate}

\section{Bayesian model selection}
\label{sec:modelselection}

\subsection{The three levels of inference}

For the purpose of this discussion, it is convenient to divide Bayesian inference in three different levels: 
 \begin{enumerate}
  \item {Level 1:} We have chosen a model $\mdl_0$, assumed true, and we want to learn about its parameters, $\theta_0$. E.g.: we assume $\Lambda$CDM to be the true model for the Universe and try to constrain its parameters. This is the usual parameter inference step.
  \item {Level 2:} We have a series of alternative models being considered ($\mdl_1, \mdl_2, \dots$) and we want to determine which of those is in best agreement with the data. This is a problem of model selection, or model criticism. For example, we might want to decide whether a dark energy equation of state $w=-1$ is a sufficient description of the available observations or whether we need an evolving dark energy model, $w=w(z)$. 
  \item {Level 3:} Of the $N$ models considered in Level 2, there is no clear ``best'' model. We want to report inferences on parameters that account for this model uncertainty. This is the subject of Bayesian model averaging. For example, we want to determine $\Omega_m$ independently of the assumed dark energy model. 
 \end{enumerate}

The Frequentist approach to model criticism is in the form of hypothesis testing (e.g., ``chi-squared-per-degree-of-freedom`` type of tests). One ends up rejecting (or not) a null hypothesis $H_0$ based on the p-value, i.e., the probability of getting data as extreme or more extreme than what has been observed if one assumes that $H_0$ is true. Notice that this is {\em not} the probability for the hypothesis! Classical hypothesis testing assumes the hypothesis to be true and determines how unlikely are our observations given this assumption. This is arguably not the quantity we are actually interested in, namely, the probability of the hypothesis itself given the observations in hand. Ref.~\cite{Sellke:2001} is a highly recommended read on this topic.
  
The Bayesian approach takes the view that there is no point in rejecting a model unless there are specific alternatives available: it takes therefore the form of model {\em comparison}. The key quantity for model comparison is the Bayesian evidence. Bayesian model comparison automatically implements a quantitative version of Occam's razor, i.e., the notion that simpler models ought to be preferred if they can explain the data sufficiently well. 

\subsection{The Bayesian evidence}

\subsubsection{Definition}

The evaluation of a model's performance in the light of the data
is based on the {\em Bayesian evidence}. This is the normalization integral on the
right--hand--side of Bayes' theorem, Eq.~\eqref{eq:evidence_def},
which we rewrite here conditioning explicitly on the model under
consideration, $\mdl$, with parameter space
$\Omega_\mdl$:
 \be
\label{eq:evidence_def_mdl}
 p(\data | \mdl) \equiv {\int_{\Omega_\mdl} p(\data | \params, \mdl)
p(\params | \mdl)\dr\params} \quad({\rm Bayesian~evidence }).
 \ee

The Bayesian evidence is the average of the likelihood under
the prior for a specific model choice. From the evidence, the
model posterior probability given the data is obtained by using
Bayes' Theorem to invert the order of conditioning:
 \be
 p(\mdl|\data) \propto p(\mdl)p(\data|\mdl),
 \ee
where we have dropped an irrelevant normalization constant that
depends only on the data and $p(\mdl)$ is the prior probability
assigned to the model itself. Usually this is taken to be
non--committal and equal to $1/N_m$ if one considers $N_m$
different models. 

When comparing two models, $\mdl_0$ versus
$\mdl_1$, one is interested in the ratio of the posterior
probabilities, or {\em posterior odds}, given by
 \be \label{eq:posterior_odds}
 \frac{p(\mdl_0|\data)}{p(\mdl_1|\data)} = B_{01}
 \frac{p(\mdl_0)}{p(\mdl_1)}.
\ee

\begin{definition}
The {\em Bayes factor} $B_{01}$ is the ratio of the
models' evidences:
\begin{equation}
 \label{eq:Bayes_factor}
 B_{01} \equiv \frac{p(\data | \mdl_0)}{p(\data | \mdl_1)} \quad({\rm Bayes~factor}).
\end{equation}
\end{definition}
A value $B_{01} > (<)~1$ represents an increase (decrease) of the
support in favour of model 0 versus model 1 given the observed
data (see \cite{Kass:1995} for more details on Bayes factors).

Bayes factors are usually interpreted against the Jeffreys'
scale~\cite{Jeffreys:1961} for the strength of evidence, given in
Table~\ref{Tab:Jeff}. This is an empirically calibrated scale,
with thresholds at values of the odds of about $3:1$, $12:1$ and
$150:1$, representing weak, moderate and strong evidence,
respectively. 

\begin{table}
 \centering
 \begin{tabular}{l l l l} 
  $|\ln B_{01}|$ & Odds & Probability & Strength of evidence \\\hline
 $<1.0$ & $\lsim 3:1$ & $<0.750$ & Inconclusive \\
 $1.0$ & $\sim 3:1$ & $0.750$ & Weak evidence \\
 $2.5$ & $\sim 12:1$ & $0.923$ & Moderate evidence \\
 $5.0$ & $\sim 150:1$ & $0.993$ & Strong evidence \\
 \end{tabular}
\caption{Empirical scale for evaluating the strength of evidence when
comparing two models, $\mdl_0$ versus $\mdl_1$ (so--called
``Jeffreys' scale''). Threshold values are empirically set, and
they occur for values of the logarithm of the Bayes factor of
$|\ln B_{01}|=1.0$, 2.5 and 5.0. The right--most column gives our
convention for denoting the different levels of evidence above
these thresholds. The probability column refers to the posterior
probability of the favoured model, assuming non--committal priors
on the two competing models, i.e.,~$p(\mdl_0) = p(\mdl_1) = 1/2$ and
that the two models exhaust the model space, $p(\mdl_0|\data) +
p(\mdl_1|\data) = 1$.\label{Tab:Jeff} }
\end{table}

\subsubsection{The Occam's razor effect}

We begin by considering the example of two nested models.  Consider
two competing models: $\mdl_0$ predicting that a parameter $\theta
= 0$ with no free parameters, and $\mdl_1$ which assigns to it a Gaussian prior distribution with 0 mean and variance $\Sigma^2$.
Assume we perform a measurement of $\theta$ described by a normal
likelihood of standard deviation $\sigma$, and with the maximum
likelihood value lying $\lambda$ standard deviations away from 0,
i.e. $|\pml/\sigma| = \lambda$. Then the Bayes factor between the
two models is given by, from Eq.~\eqref{eq:Bayes_factor}
 \be \label{eq:B01_example}
 B_{01} = \sqrt{1 + (\sigma/\Sigma)^{-2}} \exp \left( -
 \frac{\lambda^2}{2(1 + (\sigma/\Sigma)^2)} \right).
 \ee
For $\lambda \gg 1$, corresponding to a detection of the new
parameter with high significance, the exponential term dominates and
$B_{01} \ll 1$, favouring the more complex model with a non--zero
extra parameter, in agreement with what one would get using Frequentist hypothesis testing. But if
$\lambda \lsim 1$ and $\sigma/\Sigma \ll 1$ (i.e., the likelihood
is much more sharply peaked than the prior and in the vicinity of
0), then the prediction of the simpler model that $\theta = 0$ has
been confirmed. This leads to the Bayes factor being dominated by
the Occam's razor term, and $B_{01} \approx \Sigma/\sigma$, i.e.
evidence accumulates in favour of the simpler model proportionally
to the volume of ``wasted'' parameter space. If however
$\sigma/\Sigma \gg 1$ then the likelihood is less informative than
the prior and $B_{01} \rightarrow 1$, i.e. the data have not
changed our relative belief in the two models.

\begin{figure}
\centerline{\includegraphics[width=0.5\linewidth]{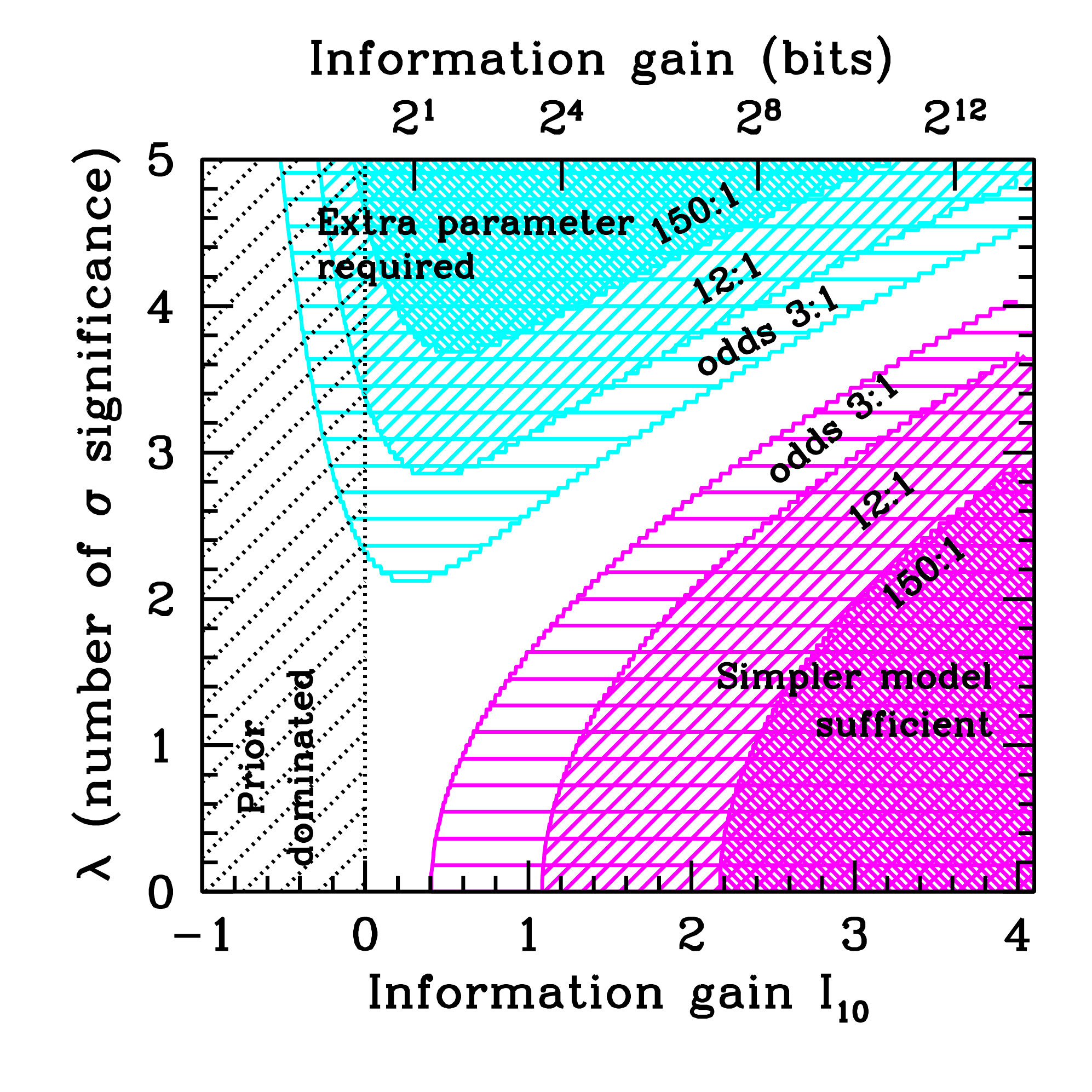}}
\caption{Illustration of Bayesian model comparison for two nested
models, where the more complex model has one extra parameter. The
outcome of the model comparison depends both on the information
content of the data with respect to the {\em a priori} available
parameter space, $\It$ (horizontal axis) and on the quality of fit
(vertical axis, $\lambda$, which gives the number of sigma
significance of the measurement for the extra parameter). Adapted
from~\cite{Trotta:2005ar}.} \label{fig:evidence_plan}
\end{figure}

In the above example, if the
data are informative with respect to the prior on the extra
parameter (i.e., for $\sigma/\Sigma \ll 1$) the logarithm of the
Bayes factor is given approximately by
 \begin{equation}
 \label{eq:B01_info}
 \ln B_{01} \approx \ln \left( \Sigma/\sigma \right) -
 \lambda^2/2,
 \end{equation}
where as before $\lambda$ gives the number of sigma away from a
null result (the ``significance'' of the measurement). The first
term on the right--hand--side is approximately the logarithm of
the ratio of the prior to posterior volume. We can interpret it as
the information content of the data, as it gives the factor by
which the parameter space has been reduced in going from the prior
to the posterior. This term is positive for informative data, i.e.
if the likelihood is more sharply peaked than the prior. The
second term is always negative, and it favours the more complex
model if the measurement gives a result many sigma away from the
prediction of the simpler model (i.e., for $\lambda \gg 0$). We
are free to measure the information content in base--10 logarithm
(as this quantity is closer to our intuition, being the order of
magnitude of our information increase), and we define the quantity
$\It \equiv \log_{10}\left(\Sigma/\sigma \right)$.
Figure~\ref{fig:evidence_plan} shows contours of $\vert \ln B_{01}
\vert = $~const for const~$= 1.0,2.5,5.0$ in the $(\It, \lambda)$
plane, as computed from Eq.~\eqref{eq:B01_info}. The contours
delimit significative levels for the strength of evidence,
according to the Jeffreys' scale (Table~\ref{Tab:Jeff}). For
moderately informative data ($\It \approx 1 - 2$) the measured
mean has to lie at least about $4\sigma$ away from 0 in order to
robustly disfavor the simpler model (i.e., $\lambda \gsim 4$).
Conversely, for $\lambda \lsim 3$ highly informative data ($\It
\gsim 2$) do favor the conclusion that the extra parameter is
indeed 0. In general, a large information content favors the
simpler model, because Occam's razor penalizes the large volume of
``wasted'' parameter space of the extended model.

An useful properties of Figure~\ref{fig:evidence_plan} is that the
impact of a change of prior can be easily quantified. A different
choice of prior width (i.e., $\Sigma$) amounts to a {\em
horizontal shift} across Figure~\ref{fig:evidence_plan}, at least
as long as $\It>0$ (i.e., the posterior is dominated by the
likelihood). Picking more restrictive priors (reflecting more
predictive theoretical models) corresponds to shifting the result
of the model comparison to the left of
Figure~\ref{fig:evidence_plan}, returning an inconclusive result
(white region) or a prior--dominated outcome (hatched region).
Notice that results in the 2--3 sigma range, which are fairly
typical in cosmology, can only support the more complex model in a
very mild way at best (odds of $3:1$ at best), while actually
being most of the time either inconclusive or in favour of the
simpler hypothesis (pink shaded region in the bottom right
corner).

Notice that Bayesian model comparison is usually {\em conservative} when it
comes to admitting a new quantity in our model, even in the case
when the prior width is chosen ``incorrectly'' (whatever that means!).
In general the result of the model comparison will eventually
override the ``wrong'' prior choice (although it might take a long
time to do so), exactly as it happens for parameter inference.

Bayesian model selection does not penalize parameters which are unconstrained by the data. This is easily seen from Eq.~\eqref{eq:B01_info}: if a parameter is unconstrained, its posterior width $\sigma$ is approximately equal to the prior width, $\Sigma$, and the Occam's razor penalty term goes to zero. In such a case, consideration of the Bayesian model complexity might help in judging model performance, see~\cite{Kunz:2006mc} for details. 

\subsubsection{Comparison with p-values}

Classical hypothesis testing relies on comparing the observed value of some test statistics, $T(X)$ (where $X$ is a RV with density $p(X|\theta)$) with its the expected distribution under a null hypothesis (usually denoted by $H_0$). The hypothesis test is to compare $H_0: \theta = \theta_0$ vs an alternative $H_1: \theta \neq \theta_0$.  The test statistics is so chosen that more extreme values denote a stronger disagreement with the null. 

\begin{definition}
The p-value (or observed significance level) is given by the probability under the null that $T$ achieves values as extremes or more extremes that have been observed (assuming here that the larger the value of $T$, the stronger the disagreement):
\be
\pval = p(T(X) \geq T^\obs |H_0). 
\ee
\end{definition} 

As an example, consider the case where under $H_0$, $x_i \sim \norm (\theta_0,\sigma)$ for
fixed $\theta_{0}$ (the null hypothesis), while under the alternative
$H_1$, $x\sim\norm(\theta,\sigma)$ and $n$ data samples are available
(with $\sigma$ known). The usual test statistics is then given by 
\be
T(X) = \sqrt{n}\frac{|\bar{X} - \theta_0|}{\sigma}.
\ee
The p-value is then given by 
\be \label{eq:computepval}
\pval = 2(1-\text{erf}(T^\obs))
\ee
where the observed value of the test statistics is
\be
T^\obs = \sqrt{n}\frac{|\bar{x} - \theta_0|}{\sigma}
\ee
and $\bar{x}$ is the sample mean. 

The classical procedure of reporting the observed $\pval$ leads to a gross misrepresentation of the evidence against the null (this is in contrast with the Neyman-Person procedure of setting a threshold p-value before the experiment is performed, and then only reporting whether or not that threshold has been exceeded). 
This is because it {\em does
not} obey the frequentist principle: in repeated use of a
statistical procedure, the long--run average actual error should
not be greater than the long--run average reported error
\cite{Berger:2003}. This means that, for example, of all reported
95\% confidence results, on average many more than 5\% turn out to
be wrong, and typically more than 50\% are wrong.

Jeffreys famously criticised the use of p-values thus (\cite{Jeffreys1980} cited in~\cite{Berger:1987a}): 
\begin{quotation}
I have always considered the arguments for the use of [p-values] absurd. They amount to saying that a hypothesis that may or may not be true is rejected because a greater departure from the trial value was improbable; that is, that it has not predicted something that has not happened.
\end{quotation}

An interesting illustration is given in~\cite{Berger:1987a}. Consider the case described above, and let us generate data from a
random sequence of null hypothesis ($H_0$) and alternatives
($H_1$), with $\theta_0=0$, $\sigma=1$ and $\theta \sim \norm(0,1)$.
Suppose that the proportion of nulls and alternatives is equal. We
then compute the p-value using Eq.~\eqref{eq:computepval} and we
select all the tests that give $\pval  \in
[\alpha-\epsilon,\alpha+\epsilon]$, for a certain value of
$\alpha$ and $\epsilon\ll\alpha$ (the exact value of $\epsilon$ is unimportant).
Among such results, which rejected the null hypothesis at the
$1-\alpha$ level, we then determine the proportion that actually
came from the null, i.e. the percentage of wrongly rejected nulls.
The
results are shown in Table~\ref{tab:nulls}. We notice that among
all the ``significant'' effects at the $95\%$ level  about 50\%
are wrong, and in general when there is only a single alternative
at least 29\% of the 95\% confidence level results will be wrong.

\begin{table}
\begin{center}
\begin{tabular}{l|l l|l|}
p-value         &  sigma &  fraction of true nulls    &  lower
bound
                \\
\hline
0.05            &   1.96 & $0.51$ &  0.29\\
0.01            &   2.58 & $0.20$ &  0.11   \\
0.001           &   3.29 & $0.024 $ & 0.018        \\
\end{tabular}
\end{center}
\caption{\label{tab:nulls} Proportion of wrongly rejected nulls
among all results reporting a certain p-value (simulation
results). The ''lower bound'' column gives the minimum fraction of true nulls (derived in~\cite{Berger:1987a}). This illustrates that the reported p-value is
 not equal to the fraction of wrongly rejected true nulls, which can
be considerably worse.}
\end{table}

\begin{svgraybox}
The root of this striking disagreement with a common
misinterpretation of the p-value (namely, that the p-value gives
the fraction of wrongly rejected nulls in the long run) is
twofold. While the p-value gives the probability of obtaining
data that are as extreme or more extreme than what has actually
been observed {\em assuming the null hypothesis is true}, one is
not allowed to interpret this as the probability of the null
hypothesis to be true, which is actually the quantity one is
interested in assessing. The latter step requires using Bayes
theorem and is therefore not defined for a frequentist. Also,
quantifying how rare the observed data are under the null is not
meaningful unless we can compare this number with their rareness
under an alternative hypothesis. 
\end{svgraybox}

A useful rule of thumb is obtained by \cite{Berger:1987a}: it is recommended to think of a $n\sigma$ result as of a $(n-1)\sigma$ result. Reducing the number of sigma significance brings the naive p-value interpretation in better alignment with the above results. 
All these points are discussed
in greater detail in \cite{Berger:1987a,Sellke:2001,Berger:2003,Lyons:2013ch,2014arXiv1408.6123D}.

\subsection{Computation of the evidence}

\subsubsection{Nested sampling} 

\begin{figure}
\begin{center}
\includegraphics[width=0.45\linewidth]{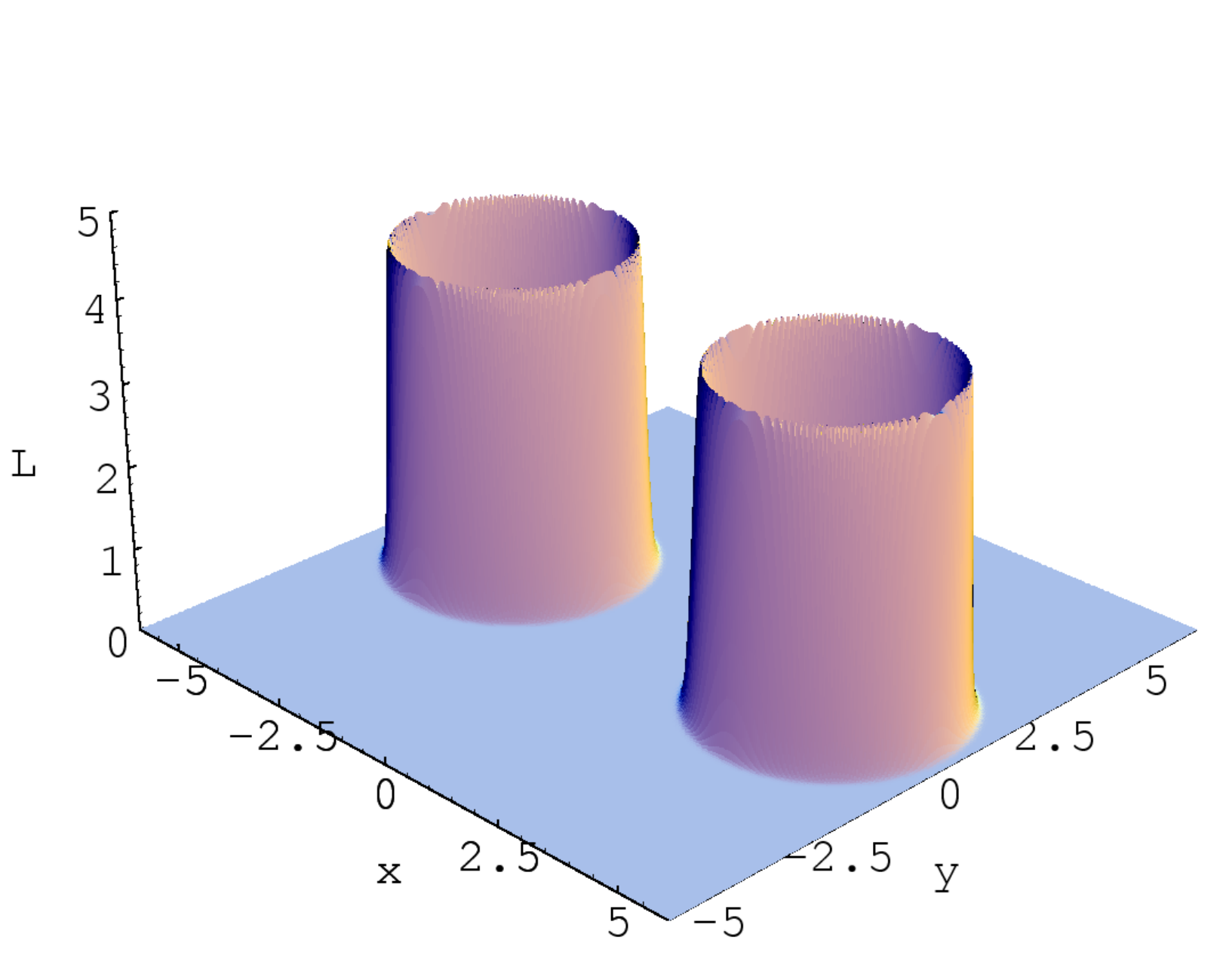}
\includegraphics[width=0.45\linewidth]{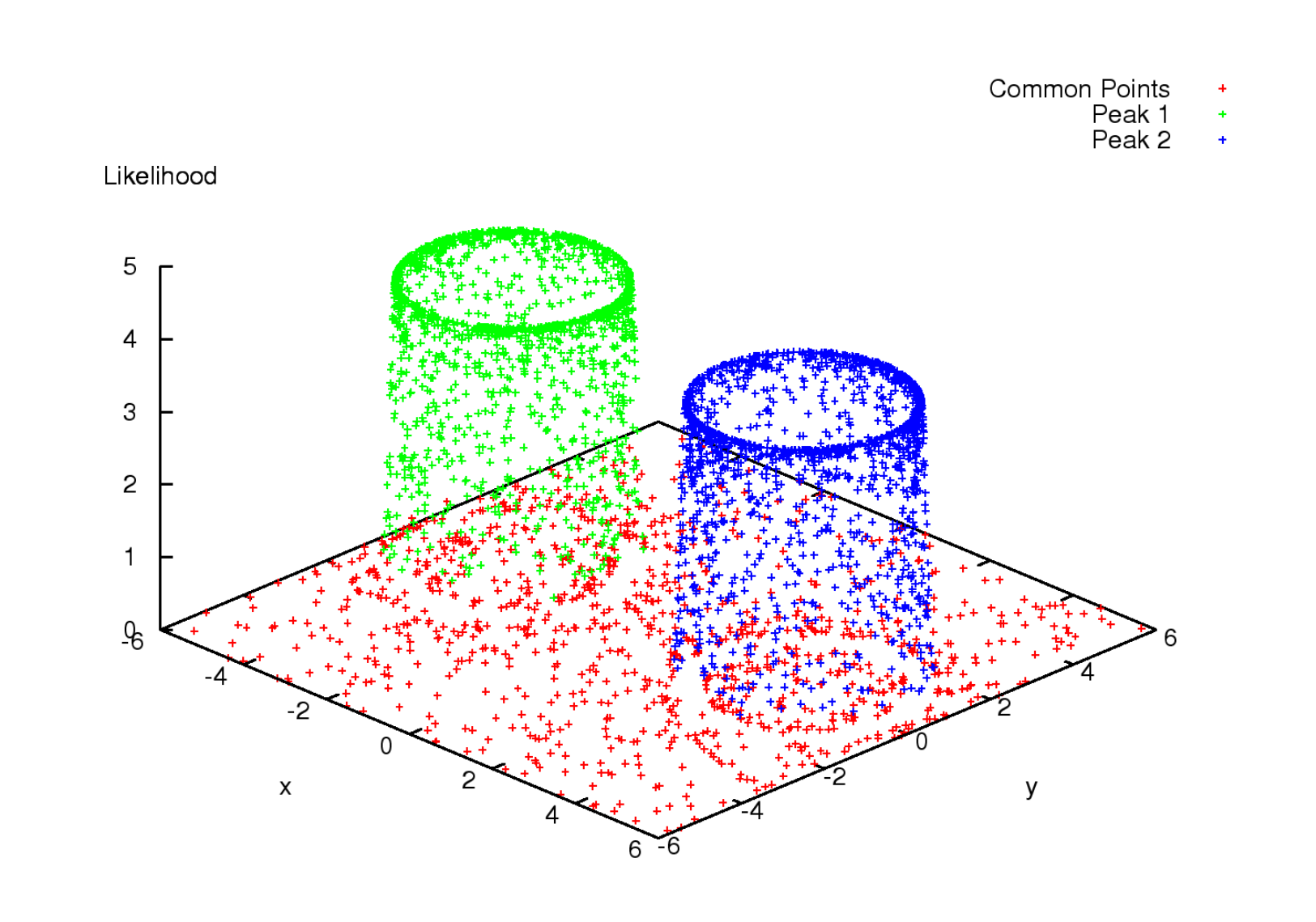}\\
\caption{Example of posterior reconstruction using Nested Sampling. Left panel: target likelihood in a 2D parameter space $(x,y)$. Right panel: reconstructed posterior (with flat priors) using Nested Sampling. From Ref.~\cite{Feroz:2007kg}.}
\label{fig:NS_samples}
\end{center}
\end{figure}

A powerful and efficient alternative to classical MCMC methods has emerged in the last few years in the form of the so--called ``nested sampling'' algorithm, out forward by John Skilling~\cite{SkillingNS}. Although the original motivation for nested sampling was to compute the evidence integral of Eq.~\eqref{eq:evidence_def_mdl}, the development of the multi--modal nested sampling technique~\cite{Feroz:2007kg} (and more recently the PolyChord algorithm~\cite{Handley:2015fda}, as well as diffusive nested sampling, implemented in the DNest4 code~\cite{Brewer:2016scw}) provides a powerful and versatile algorithm that can sample efficiently from complex, multi-modal likelihood surfaces, see Fig.~\ref{fig:NS_samples}.

The gist of nested sampling is that the
multi--dimensional evidence integral is recast into a
one--dimensional integral, by defining the
prior volume $X$ as $\dr X \equiv p(\params|\mdl) \dr
\params $ so that
 \begin{equation} \label{eq:def_prior_volume}
  X(\lambda) = \int_{\like(\params)>\lambda} p(\params|\mdl) \dr
  \params
 \end{equation}
where $\like(\params) \equiv p(\data|\params, \mdl)$ is the
likelihood function and the integral is over the parameter space
enclosed by the iso--likelihood contour $\like(\params) =
\lambda$. So $X(\lambda)$ gives the volume of parameter space
above a certain level $\lambda$ of the likelihood. 

The
Bayesian evidence, Eq.~\eqref{eq:evidence_def_mdl}, can be written
as
 \begin{equation} \label{eq:nested_integral}
 p(\data | \mdl) = \int_0^1 \like(X) \dr X,
 \end{equation}
where $\like(X)$ is the inverse of
Eq.~\eqref{eq:def_prior_volume}. Samples from $\like(X)$ can be
obtained by drawing uniformly samples from the likelihood volume
within the iso--contour surface defined by $\lambda$. This is the difficult part of the algorithm. 

Finally, the 
1--dimensional integral of Eq.~\eqref{eq:nested_integral} can be
obtained by simple quadrature, thus
 \begin{equation}
 p(\data | \mdl) \approx \sum_i \like(X_i) W_i ,
 \end{equation}
where the weights are $W_i = \frac{1}{2}(X_{i-1} - X_{i+1})$, see~\cite{SkillingNS,Mukherjee:2005wg} for details\footnote{Publicly available software
implementing nested sampling can be found at
\texttt{http://www.mrao.cam.ac.uk/software/cosmoclust/} (MultiNest) and \texttt{https://github.com/eggplantbren/DNest4/} (DNest4) (accessed Jan 5th 2017).}.

\subsubsection{Thermodynamic integration} 

Thermodynamic integration computes the evidence integral by defining
\be
 E(\mu) \equiv {\int_{\Omega_\mdl} \like(\params)^\mu
p(\params | \mdl)\dr\params},
\ee
where $\mu$ is an annealing parameter and $\like(\params) \equiv p(\data | \params, \mdl)$. Obviously the desired evidence corresponds to $E(1)$. One starts by performing a standard MCMC sampling with $\mu = 0$ (i.e., sampling from the prior), then gradually increases $\mu$ to 1 according to some annealing schedule. The log of the evidence is then given by
\be
\ln E(1) = \ln E(0) + \int_0^1 \frac{d\ln E}{d\mu} \dr \mu =   \int_0^1 \langle \ln \like  \rangle_\mu \dr \mu,
\ee
where the average log-likelihood is taken over the posterior with annealing parameter $\mu$, i.e.
\be
 \langle \ln \like  \rangle_\mu  = \frac{{\int_{\Omega_\mdl}(\ln \like) \like(\params)^\mu
p(\params | \mdl)\dr\params}}{{\int_{\Omega_\mdl} \like(\params)^\mu
p(\params | \mdl)\dr\params}}.
\ee
The drawback is that the end result might depend on the annealing schedule used and that typically this methods takes 10 times as many likelihood evaluations as parameter estimation. For an overview of so-called ``population Monte Carlo'' algorithms and annealed importance sampling, see~\cite{2001TJSAI..16..279I,10.2307/27594084}.

\subsubsection{Laplace approximation} 

An approximation to the Bayesian evidence can be
obtained when the likelihood function is unimodal and
approximately Gaussian in the parameters. Expanding the likelihood
around its peak to second order one obtains the Laplace
approximation
 \be \label{eq:like_Gaussian}
 p(\data | \params, \mdl) \approx
 \lmax
 \exp\left[-\frac{1}{2}(\params-\pml)^t L
(\params-\pml)\right],
 \ee
where $\pml$ is the maximum--likelihood point, $\lmax$ the maximum
likelihood value and $L$ the likelihood Fisher matrix (which is
the inverse of the covariance matrix for the parameters). Assuming
as a prior a multinormal Gaussian distribution with zero mean and
Fisher information matrix $P$ one obtains for the evidence,
Eq.~\eqref{eq:evidence_def_mdl}
 \be
 \label{eq:evidence_Gaussian}
 p(\data | \mdl) = \lmax \frac{|F|^{-1/2}}{|P|^{-1/2}}
 \exp\left[-\frac{1}{2}({\pml}^t L \pml - \overline{\params}^tF\overline{\params})
 \right],
 \ee
where the posterior Fisher matrix is $F = L + P$ and the posterior
mean is given by $\overline{\params} = F^{-1}L \pml$.

\subsubsection{The Savage-Dickey density ratio} 
A useful approximation to the Bayes factor, Eq.~\eqref{eq:Bayes_factor},
is available for situations in which the models being compared
are {\em nested} into each other, i.e. the more complex model
($\mdl_1$) reduces to the original model ($\mdl_0$) for specific
values of the new parameters. This is a fairly common scenario when one wishes to evaluate whether the inclusion of
the new parameters is supported by the data (e.g., do we need isocurvature contributions to the
initial conditions for cosmological perturbations, or whether a
curvature term in Einstein's equation is needed, or whether a
non--scale invariant distribution of the primordial fluctuation is
preferred). 

Writing for the extended model parameters $\params =
(\pzero, \pone)$, where the simpler model $\mdl_0$ is obtained by
setting $\pone = 0$, and assuming further that the prior is
separable (which is usually the case), i.e.\ that
 \begin{equation}
 p(\pzero, \pone| \mdl_1) = p(\pone | \mdl_1) p(\pzero | \mdl_0),
 \end{equation}
the Bayes factor can be written in all generality as
 \begin{equation} \label{eq:savagedickey}
 B_{01} = \left.\frac{p(\pone \vert \data, \mdl_1)}{p(\pone |
 \mdl_1)}\right|_{\pone = 0}.
 \end{equation}
This expression is known as the Savage--Dickey density ratio (see~\cite{Trotta:2005ar} and references therein). The
numerator is simply the marginal posterior under the more complex
model evaluated at the simpler model's parameter value, while the
denominator is the prior density of the more complex model
evaluated at the same point. This technique is particularly useful
when testing for one extra parameter at the time, because then the
marginal posterior $p(\pone \vert \data, \mdl_1)$ is a
1--dimensional function and normalizing it to unity probability
content only requires a 1--dimensional integral, which is simple
to do using for example the trapezoidal rule.

\subsubsection{Information criteria for approximate model selection} 

Sometimes it might be useful to employ methods that aim at an
approximate model selection under some simplifying assumptions
that give a default penalty term for more complex models, which
replaces the Occam's razor term coming from the different prior
volumes in the Bayesian evidence~\cite{Liddle:2004nh}. 

\runinhead{Akaike Information Criterion (AIC):} the AIC is an essentially frequentist
criterion that sets the penalty term equal to twice the number of
free parameters in the model, $k$:
  \begin{equation} \label{eq:def_AIC}
  {\rm AIC} \equiv - 2 \ln \lmax + 2k
  \end{equation}
where $\lmax \equiv p(\data | \pml, \mdl)$ is the maximum
likelihood value. 
\runinhead{Bayesian Information Criterion (BIC):}
 the BIC follows from a
Gaussian approximation to the Bayesian evidence in the limit of
large sample size:
  \begin{equation} \label{eq:def_BIC}
  {\rm BIC} \equiv - 2 \ln \lmax + k\ln N
  \end{equation}
where $k$ is the number of fitted parameters as before and $N$ is
the number of data points. The best model is again the one that
minimizes the BIC.
\runinhead{Deviance Information Criterion (DIC):}
the DIC can be written as
  \begin{equation} \label{eq:def_DIC}
  {\rm DIC} \equiv - 2 {D_\text{KL}} + 2 {\mathcal C}_b .
  \end{equation}
In this form, the DIC is reminiscent of the AIC, with the $\ln
\lmax$ term replaced by the estimated KL divergence $ {D_\text{KL}}$ and the number
of free parameters by the effective number of parameters, ${\mathcal C}_b$ (see \cite{Trotta:2008qt} for definitions).

The information criteria ought to be interpreted with care when
applied to real situations. Comparison of Eq.~\eqref{eq:def_BIC}
with Eq.~\eqref{eq:def_AIC} shows that for $N>7$ the BIC penalizes
models with more free parameters more harshly than the AIC.
Furthermore, both criteria penalize extra parameters regardless of
whether they are constrained by the data or not, unlike the
Bayesian evidence. In conclusion,
what makes the information criteria attractive, namely the absence
of an explicit prior specification, represents also their
intrinsic limitation.

\subsection{Example: model selection for the inflationary landscape}

The inflationary model comparison carried out in Ref.~\cite{Martin:2013nzq,2014PhRvD..90f3501M} is an example of the application of the above formalism to the problem of deciding which theoretical model is the best description of the available observations. Although the technical details are fairly involved, the underlying idea can be sketched as follows.

The term ``inflation'' describes a period of exponential expansion of the Universe in the very first instants of its life, some $10^{-32}$ seconds after the Big Bang, during which the size of the Universe increased by at least 25 orders of magnitude. This huge and extremely fast expansion is required to explain the observed isotropy of the cosmic microwave background on large scales. It is believed that inflation was powered by one or more scalar fields. The behaviour of the scalar field during inflation is determined by the shape of its potential, which is a real-valued function $V(\phi)$ (where $\phi$ denotes the value of the scalar field). The detailed shape of $V(\phi)$ controls the duration of inflation, but also the spatial distribution of inhomogeneities (perturbations) in the distribution of matter and radiation emerging from inflation. It is from those perturbations that galaxies and cluster form out of gravitational collapse. Hence the shape of the scalar field can be constrained by observations of the large scale structures of the Universe and of the CMB anisotropies. 

Theories of physics beyond the Standard Model motivate certain functional forms of $V(\phi)$, which however typically have a number of free parameters, $\theta$. The fundamental model selection question is to use cosmological observations to discriminate between alternative models for $V(\phi)$ (and hence alternative fundamental theories). The major obstacle to this programme is that very little if anything at all is known {\em a priori} about the free parameters $\theta$ describing the inflationary potential. What is worse, such parameters can assume values across several orders of magnitude, according to the theory. Hence the Occam's razor effect of Bayesian model comparison can vary in a very significant way depending on the prior choices for $\Psi$. Furthermore, a non-linear reparameterization of the problem (which leaves the physics invariant) does in general change the Occam's razor factor, and hence the model comparison result.

\begin{figure}
\begin{center}
\includegraphics[width=\linewidth, angle=90]{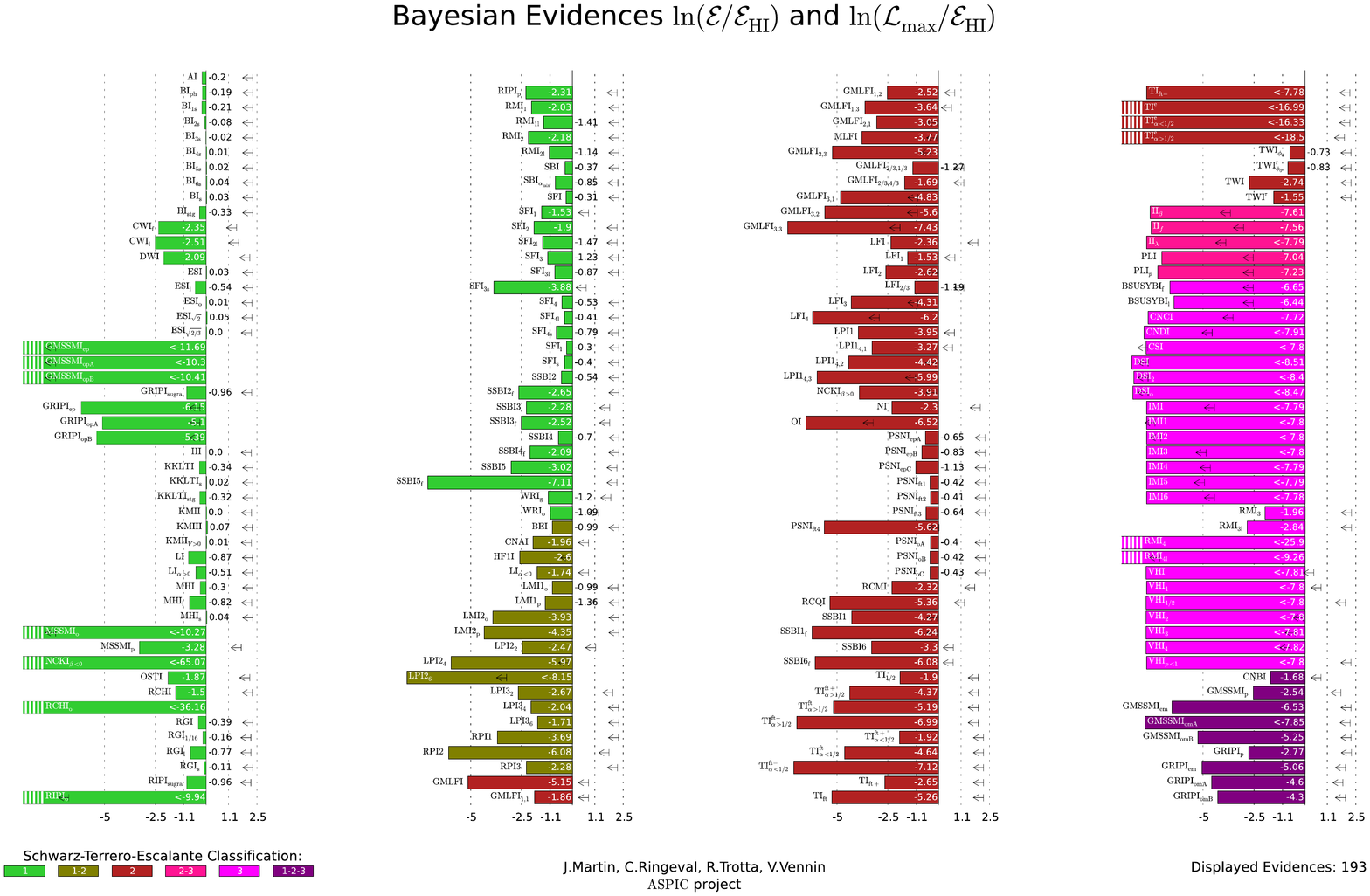}
\caption{Bayes factors (bars) and absolute upper bound to the Bayes
  factors (arrows) for inflationary scenarios, with Higgs
  inflation as the reference model (see~\cite{Martin:2013nzq} for further details). \label{fig:evid}}
\end{center}
\end{figure}

In Ref.~\cite{Martin:2013nzq} inflationary model selection was considered from a principled point of view.  The Bayesian evidence and complexity of 198 
  slow-roll single-field models of inflation was computed, using the Planck 2013
  Cosmic Microwave Background data. The models considered
  represented an almost complete and systematic scan of the entire landscape of
  inflationary scenarios proposed so far (More recently, this works has been extended to more complex scenarios with more than one scalar field~\cite{Vennin:2015vfa}). The analysis singled out the
  most probable models (from an Occam's razor point of view) that are
  compatible with Planck data. The resulting Bayes factors (normalised to the case of Higgs Inflation) are displayed in Fig.~\ref{fig:evid}. 
\subsection{Open challenges} 

I conclude by listing what I think are some of the open questions and outstanding challenges in the application of Bayesian model selection to cosmological model building. 

\begin{itemize}

\item Is Bayesian model selection always applicable?  The Bayesian model comparison approach as applied to cosmological and particle physics problems has been strongly criticized by some authors. E.g., George Efstathiou~\cite{Efstathiou:2008ed} and Bob Cousins~\cite{Cousins:2008gf,2013arXiv1310.3791C}  pointed out (in different contexts) that often insufficient attention is given to the selection of models and of priors, and that this might lead to posterior model probabilities which are largely a function of one's unjustified assumptions. This draws attention to the difficult question of how to choose priors on phenomenological parameters, for which theoretical reasoning offers poor or no guidance (as in the inflationary model comparison example above).   

\item How do we deal with Lindley's paradox? It is simple to construct examples of situations where Bayesian model comparison and classical hypothesis testing disagree (Lindley's paradox~\cite{Lindley:1957}). This is not surprising, as frequentist hypothesis testing and Bayesian model selection really ask different questions of the data~\cite{Sellke:2001}. Furthermore, as the scaling with the number of data points is different, there isn't even a guarantee that the two approaches will agree in the asymptotic regime of large data sample size. As Louis Lyons aptly put it:
\begin{quotation}
Bayesians address the question everyone is interested in by
using assumptions no--one believes, while frequentists use
impeccable logic to deal with an issue of no interest to
anyone~\cite{Lyons:2006}. 
\end{quotation}
However, such a disagreement is more likely to occur in situations where the signal is weak, which are precisely the kind of ``frontier science'' cases which are the most interesting ones (e.g., discovery claims). Is there a way to evaluate e.g. the loss function from making the ``wrong'' decision about rejecting/accepting a model? In this context, perhaps a decision theoretical approach would be beneficial: the loss function of making the wrong decision has to be explicitly formulated, thus helping putting the user's subjective biases and values in the open.

\item How do we assess the completeness of the set of known models? Bayesian model selection always returns a best model among the ones being compared, even though that model might be a poor explanation for the available data. Is there a principled way of constructing an {\em absolute} scale for model performance in a Bayesian context? (for example, along the lines of the notion of Bayesian doubt, introduced in~\cite{March:2010ex}).

\item Is Bayesian model averaging useful?  Bayesian model averaging can be used to obtain final inferences on parameters which take into account the residual model uncertainty (examples of applications in cosmology can be found in~\cite{Liddle:2006kn,Parkinson:2010zr,Vardanyan:2011in,Hee:2015eba}). However, it also propagates the prior sensitivity of model selection to the level of model-averaged parameter constraints. Is it useful to produce model-averaged parameter constraints, or should this task be left to the user, by providing model-specific posteriors and Bayes factors instead? 

\item Is there such a thing as a ``correct'' prior? In fundamental physics, models and parameters (and their priors) are supposed to represent (albeit in an idealized way) the real world, i.e., they are not simply useful representation of the data (as they are in other statistical problems, e.g.~as applied to social sciences). In this sense, one could imagine that there exist a ``correct'' prior for e.g. the parameters $\params$ of our cosmological model, which could in principle be derived from fundamental theories such as string theory (e.g., the distribution of values of cosmological parameters across the landscape of string theory~\cite{Tegmark:2004qd}). This raises interesting statistical questions about the relationship between physics, reality and probability. 

\end{itemize}

\subsection{Exercises}

\begin{enumerate}
\item  A coin is tossed $N=250$ times and it returns $H=140$ heads. Evaluate the evidence that the coin is biased using Bayesian model comparison and contrast your findings with the usual (frequentist) hypothesis testing procedure (i.e.\, testing the null hypothesis that $p_H = 0.5$). Discuss the dependency on the choice of priors.  
\item In 1919 two expeditions sailed from Britain to measure the light deflection from stars behind the Sun's rim during the solar eclipse of May 29th. Einstein's General Relativity predicts a deflection angle 
$$
\alpha = \frac{4GM}{c^2 R},
$$
where $G$ is Newton's constant, $c$ is the speed of light, $M$ is the mass of the gravitational lens and $R$ is the impact parameter. It is well known that this result it exaclty twice the value obtained using Newtonian gravity. For $M=M_\odot$ and $R=R_\odot$ one gets from Einstein's theory that $\alpha=1.74$ arc seconds.

The team led by Eddington reported $1.61 \pm 0.40$ arc seconds (based on the position of 5 stars), while the team headed by Crommelin reported $1.98\pm0.16$ arc seconds (based on 7 stars).  

What is the Bayes factor between Einstein and Newton gravity from those data? Comment on the strength of evidence. \\

%
\item  Assume that the combined constraints from CMB, BAO and SNIa on the density parameter for the cosmological constant can be expressed as a Gaussian posterior distribution on $\OL$ with mean 0.7 and standard deviation 0.05. Use the Savage-Dickey density ratio to estimate the Bayes factor between a model with $\OL = 0$ (i.e., no cosmological constant) and the $\Lambda$CDM model, with a flat prior on $\OL$ in the range $0 \leq \OL \leq 2$. Comment on the strength of evidence in favour of $\Lambda$CDM. %
\item If the cosmological constant is a manifestation of quantum fluctuations of the vacuum, QFT arguments lead to the result that the vacuum energy density $\rho_\Lambda$ scales as
\begin{equation}
\rho_\Lambda \sim \frac{c\hbar}{16\pi} k_\text{max}^4
\end{equation}
where $k_\text{max}$ is a cutoff scale for the maximum wavenumber contributing to the energy density (see e.g. ~\cite{Carroll:1991mt}). Adopting the Planck mass as a plausible cutoff scale (i.e., $k_\text{max} = c/\hbar M_\text{Pl} $) leads to ``the cosmological constant problem'', i.e., the fact that the predicted energy density
\begin{equation}
\rho_\Lambda \sim 10^{76} \text{GeV}^4
\end{equation}
is about 120 orders of magnitude larger than the observed value, $\rho_\text{obs} \sim 10^{-48} \text{GeV}^4$.

\begin{enumerate}
\item Repeat the above estimation of the evidence in favour of a non-zero cosmological constant, adopting this time a flat prior in the range $0 \leq \Omega_\Lambda/\Omega_\Lambda^\text{obs} < 10^{120}$. What is the meaning of this result? What is the required observational accuracy (as measured by the posterior standard deviation) required to override the Occam's razor penalty in this case?
\item It seems that it would be very difficult to create structure in a universe with $\Omega_\Lambda\gg100$, and so life (at least life like our own) would be unlikely to evolve. How can you translate this ``anthropic'' argument into a quantitative statement, and how would it affect our estimate of $\Omega_\Lambda$ and the  model selection problem?
\end{enumerate}

\item This problem follows up the cosmological parameter estimation problem from supernovae type Ia (for a more thorough treatment, see \cite{Vardanyan:2009ft,Vardanyan:2011in}).

\begin{enumerate}
\item Adopt uniform priors $\Omega_\text{m} \sim U(0,2)$ and $\Omega_\Lambda \sim U(0,2)$. Produce a 2D marginalised posterior pdf in the $(\Omega_\text{m},\Omega_\Lambda)$ plane.
\item Produce a 1D marginalised posterior pdf for the curvature parameter, $\Omega_\kappa = 1 - \Omega_\Lambda - \Omega_\text{m}$, paying attention to normalising it to unity probability content. 
What is the shape of the prior on $\Omega_\kappa$ implied by your choice of a uniform prior on $\Omega_\text{m}, \Omega_\Lambda$? 
\item Use the Savage-Dickey density ratio formula to estimate from the above 1D posterior the evidence in favour of a flat Universe, $\Omega_\kappa = 0$, compared with a non-flat Universe, $\Omega_\kappa \neq 0$, with prior $P(\Omega_\kappa) = U(-1,1)$. 

Discuss the dependency of your result on the choice of the above prior range.
\end{enumerate}
\end{enumerate}

\subsection{Solutions to selected exercises}
 
\begin{enumerate}
\item  This is a model comparison problem, where we are comparing model $\mdl_0$ that the coin is fair (i.e., $p_H = 0.5$) with a model $\mdl_1$ where the probability of heads is $\neq 0.5$. We begin by assigning under model 1 a flat prior to $p_H$ between 0 and 1. 

The Bayes factor (or ratio of the two models' evidences) is given by
\be \label{eq:Bayes_fac}
B = \frac{P(H=140 | \mdl_1)}{P(H=140 | \mdl_0)} = \frac{\frac{H!(N-H)!}{(H+1)!}}{(1/2)^{N}}{\Big \vert}_{N=250, H = 140} = \frac{\frac{140!110!}{251!}}{(1/2)^{250}} \approx 0.48 \sim 2:1
\ee
(notice that we have cancelled the ``choose'' terms in the numerator and denominators above). So there is not even weak evidence in favour of the model that the coin is biased. The log of the Bayes factor is plotted as a function of $H$ in Fig.~\ref{fig:evidencecoin}.  By inspection it is apparent that values $107 \leq H \leq 143$ favour the fair coin model ($\ln B < 0$). In order to obtain ``strong evidence'' in favour of the biased coin model ($\ln B > 5$), it is necessary that either $H < 94$ or $H>31$.

The usual Frequentist hypothesis testing procedure would be to compute the tail probability of obtaining data as extreme or more extreme than have been observed under the null hypothesis, i.e., that the coin is fair. This gives the p-value:
\be
\text{p-value} =  \left(\frac{1}{2}\right)^{N} \sum_{H = H_\text{obs}}^N  {N\choose{H}}  \approx 0.033 
\ee 
 
 \begin{figure}
\centering
\includegraphics[width = 0.5\linewidth]{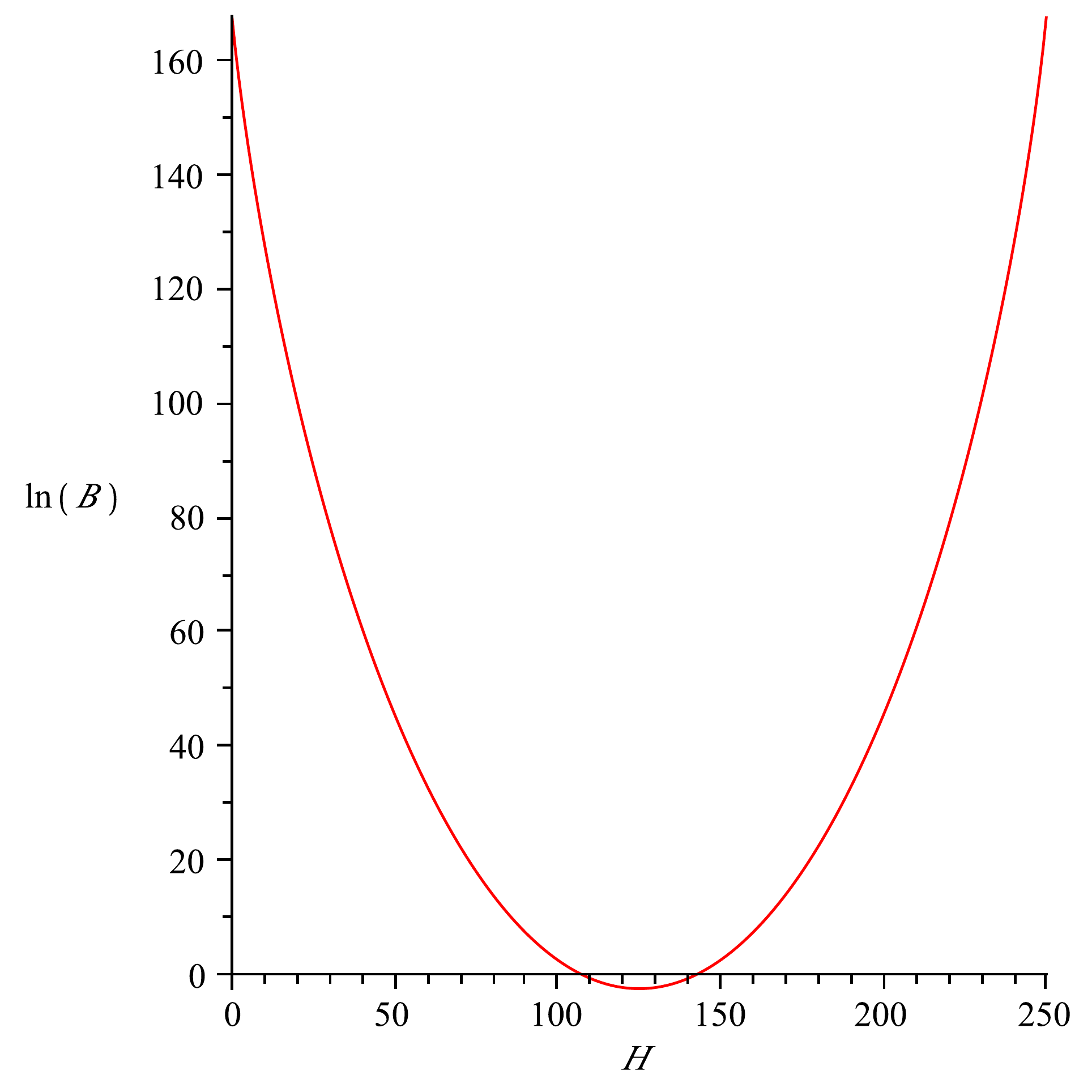}
\caption{Natural log of the Bayes factor between the model ``the coin is biased'' (with flat prior) and the model ''the coin is fair'', as a function of the number of heads ($H$) in 250 flips, see Eq.~\eqref{eq:Bayes_fac}. Values $\ln B > 0$ favour the biased coin model. The Jeffreys' threshold for ``strong evidence'' is at $\ln B = 0$ \label{fig:evidencecoin}}
\end{figure}

So for a Frequentist, the data would exclude the null hypothesis that the coin is fair at more than the 95\% CL. 

How does the Bayesian result depend on the choice of prior for the alternative hypothesis? Above we have given to $p_H$ a flat prior between 0 and 1. If we wanted to give the maximum possible advantage to a model where the coin is not fair, we could put all of its prior probability in a delta-function concentrated at the value of $p_H$ that maximizes the probability of what has been observed. So under this maximally advantageous model for the unfairness hypothesis (let's call this $\mdl_2$), we would select a ``prior'' (in quotation marks, for this prior is actually selected after the data have been gathered, so we are effectively using the data twice here!) of the form $P(p_H) = \delta(p_H - H/N)$. In this case the odds in favour of this new model are 
\be 
B =  \frac{P(H=140 | \mdl_2)}{P(H=140 | \mdl_0)} = \frac{(H/N)^H(1-H/N)^{N-H}}{(1/2)^{250}}{\Big \vert}_{H=140, H=250} \approx 6.1. 
\ee
Even in this most favourable setup for the hypothesis that the coin is biased, we find only weak evidence (odds of 6 to 1) against the model of a fair coin. Therefore we can safely conclude that the data do not warrant to conclude that the coin is unfair.

%
\item 
We are comparing here two models which both make exact predictions for the deflection angle, with no free parameters. If you prefer, you might consider the prior on $\alpha$ under each theory to be a delta-function centered at the predicted value. This of course neglects the uncertainty associated with $M_\odot$ and $R_\odot$. 

In this case, the evidence is thus simply the likelihood function for the observed data under each theory (you can convince yourself that this is correct by explicitly computing the evidence for each model assuming the delta-function prior above). This gives for the Bayes factor in favour of Einstein gravity vs Newton (assuming Gaussian likelihoods)
\be
B = \frac{\like_0 \exp\left(-\frac{1}{2} \frac{(\hat{\alpha} - \alpha_E)^2}{\sigma^2} \right)}{\like_0 \exp\left(-\frac{1}{2} \frac{(\hat{\alpha} - \alpha_N)^2}{\sigma^2} \right)}
\ee
where $\alpha_E = 1.74''$,  $\alpha_N = 87''$, $\hat{\alpha}$ is maximum likelihood value of the experiment and $\sigma$ is the standard deviation. 

Using the supplied data from Eddington, one obtains $B \sim 5$, so ``weak evidence'' in favour of Einstein theory according to the Jeffreys' scale for the strength of evidence. The Crommelin data instead give $B\sim 10^{10}$, so very strong evidence for Einstein. Notice that this comes about because the measurement from Crommelin is on the high side (i.e., higher than Einstein prediction, even), and therefore the assumed Gaussian tail becomes tiny for $\alpha = \alpha_N$. It is worth noticing that, although the above calculation is formally correct, it is likely to overestimate the evidence against Newton, because the Gaussian approximation made here is certain to break down that far into the tails (i.e, $\alpha_N$ is $\sim 11 \sigma$ away from the value measured by Crommelin. No distribution is exactly valid that far into the tails!).

%
\item 
Here we are comparing two nested model, $\mdl_0$ with $\OL = 0$ and a more complicated model, $\mdl_1$,where $\OL \leq 0$ and a flat prior $P(\OL | \mdl_1) = 1/2$ for $0\leq \OL \leq 2$ and 0 elsewhere (notice that the prior needs to be normalized, hence the factor $1/2$). We can therefore use the Savage-Dickey density ratio to compute the Bayes factor between $\mdl_0$ and $\mdl_1$:
\be \label{eq:Bayes_Lambda_1}
B_{01} = \frac{P(\OL = 0 | \text{CMB+BAO+SN}, \mdl_1)}{P(\OL = 0 | \mdl_1)} = 
\frac{ 
\frac{1}
{\sqrt{2\pi} \sigma} \exp \left( -\frac{1}{2} \frac{(0-\hat{\Omega}_\Lambda)^2 }{\sigma^2} \right)
}{1/2},
\ee 
where we have assumed thart the posterior under $\mdl_1$ can be approximated as a Gaussian of mean $\hat{\Omega}_\Lambda = 0.7$ and standard deviation $\sigma=0.05$. Numerical evaluation gives $B_{01} \sim 10^{-42}$, so with this prior the model that $\OL = 0$ can be ruled out with very strong evidence. Another way of looking at this  result is the following: if, after having seen the data, you remain unconvinced that indeed $\OL > 0$, this means that the ratio in your relative degree of prior belief in the two models should exceed $P(\mdl_0)/P(\mdl_1) > 10^{42}$. 

%
\item 
\begin{enumerate}
\item The calculation of the Bayes factor proceeds as above, but this time with a much larger prior range for the alternative model, $\OL >0$. This means that the prior height, $P(\OL = 0 | \mdl_1)$ , appearing in the denominator of Eq.~\eqref{eq:Bayes_Lambda_1} is very small, i.e. $P(\OL = 0 | \mdl_1) = 10^{-120}$, as the prior needs to be normalized. Repeating the above calculation, we get for the Bayes factor in favour of $\mdl_0$ (i.e., that $\OL = 0$)
\begin{equation}
B_{01} \sim \frac{10^{-42}}{10^{-120}} \sim 10^{88}.
\end{equation}
Now the Bayes factor is positive (and huge), a reflection of the enormous amount of prior range wasted by $\mdl_1$. Therefore under this new prior, the Bayesian model comparison favour the hypothesis that there is no cosmological constant despite the fact that the likelihood peaks about $0.7/0.05 \sim 14\sigma$ away from $\OL = 0$. This is an extreme example of Occam's penalty. 

In order for the Occam's factor to be overruled by the likelihood, we require that $B_{01} = 1$ (i.e., equal odds for the two models). This translates in the approximate condition for the number of sigma detection, $\lambda$: 
\be
\exp\left( -\frac{1}{2} \lambda^2 \right) \sim 10^{-120},
\ee
where we have dropped the term $1/\sigma$ in front of the likelihood for simplicity (as the likelihood is going to be dominated by the exponential anyhow). Solving for $\lambda$ gives
\be
\lambda \sim \sqrt{240\ln 10} \approx 23.
\ee
So we would need a $\sim 23\sigma$ detection of $\OL > 0$ to override completely the Occam's razor penalty.

\item The outcome of the model comparison changes dramatically if one is willing to impose a much more stringent upper cutoff to the prior range of $\OL$, based e.g. on anthropic arguments. The observations that structures cannot form if $\OL \gg 100$ (and therefore there would be no observers to measure dark energy, see e.g. the original argument by Weinberg~\cite{Weinberg:1987dv}) can be approximately translated in a prior range extending perhaps to $\OL \sim 10^3$. With this choice of range, the Bayes factor becomes 
\be
B_{01} \sim \frac{10^{-42}}{10^{-3}} \sim 10^{-39},
\ee
thus swinging back to support $\mdl_1$ with enormous odds. This illustrate that Bayesian model comparison can be difficult (and strongly dependent on the theoretical prior range adopted) in cases where there is no compelling (or unique) argument to define the prior. 

\end{enumerate}

\end{enumerate}

\begin{acknowledgement}
I would like to thank the many colleagues who provided invaluable input and discussions over the years: Bruce Bassett, Jim Berger, Bob Cousins, Eric Feigelson, Farhan Feroz, Alan Heavens, Mike Hobson, Andrew Jaffe, Martin Kunz, Andrew Liddle, Louis Lyons, Daniel Mortlock, John Peacock and David van Dyk. Many thanks to the Organizers of the 44th Saas Fee Advanced Course on Astronomy and Astrophysics, ``Cosmology with wide-field surveys''  (held in March 2014) for inviting me to present these lectures, and to the students for their piercing and stimulating questions. I am grateful to many cohorts of students, at Imperial College London and in various advanced schools, for their valuable feedback and comments on earlier versions of these notes. Any remaining mistake is of course fully my own.
\end{acknowledgement}
\section*{Appendix}
\addcontentsline{toc}{section}{Appendix}
\label{app:background}
\section{Introductory and background material}

\subsection{The uniform, binomial and Poisson distributions}

\begin{figure}
\begin{center}
\begin{tabular}{cc}
\includegraphics[width=0.45\linewidth]{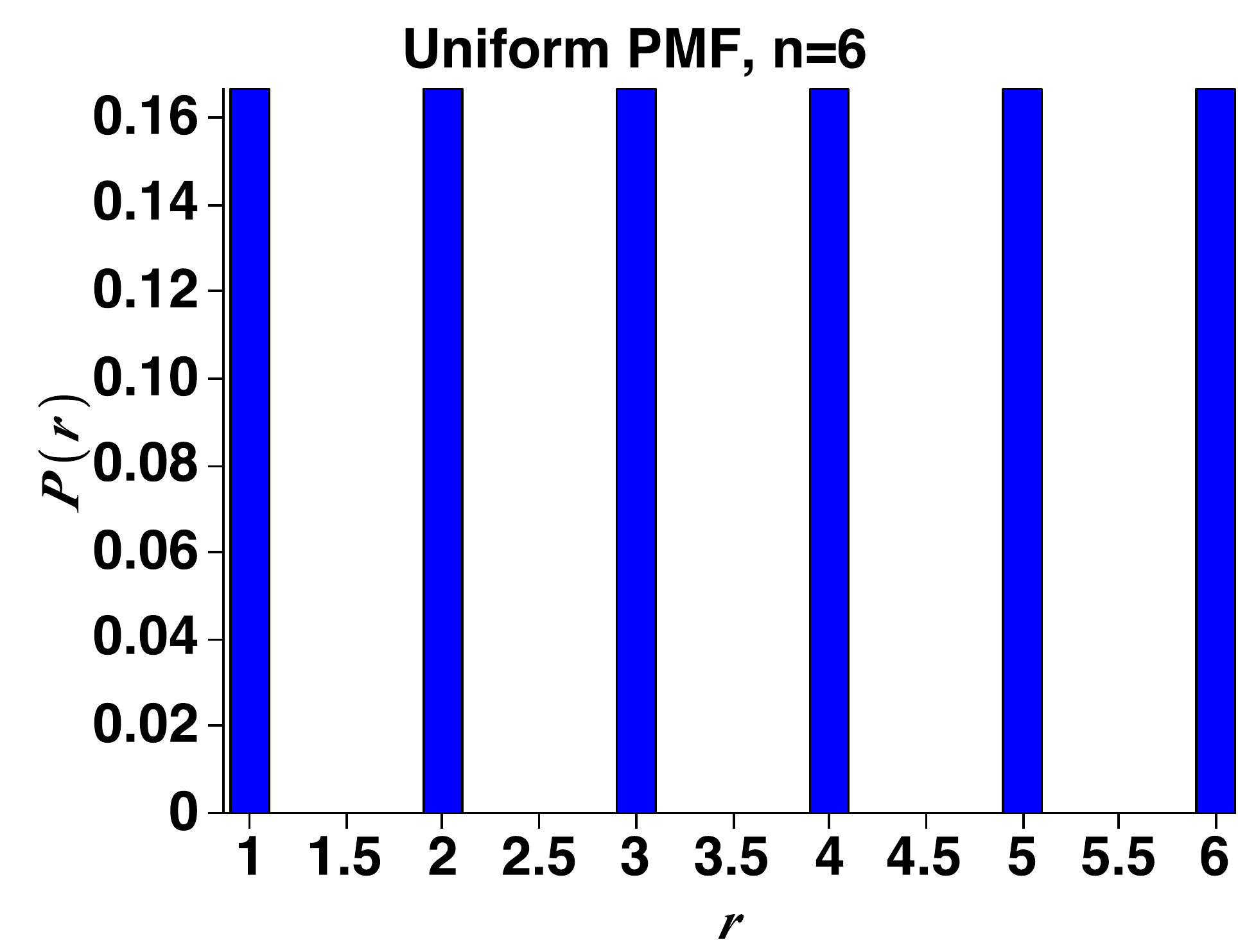} \hfill 
\includegraphics[width=0.45\linewidth]{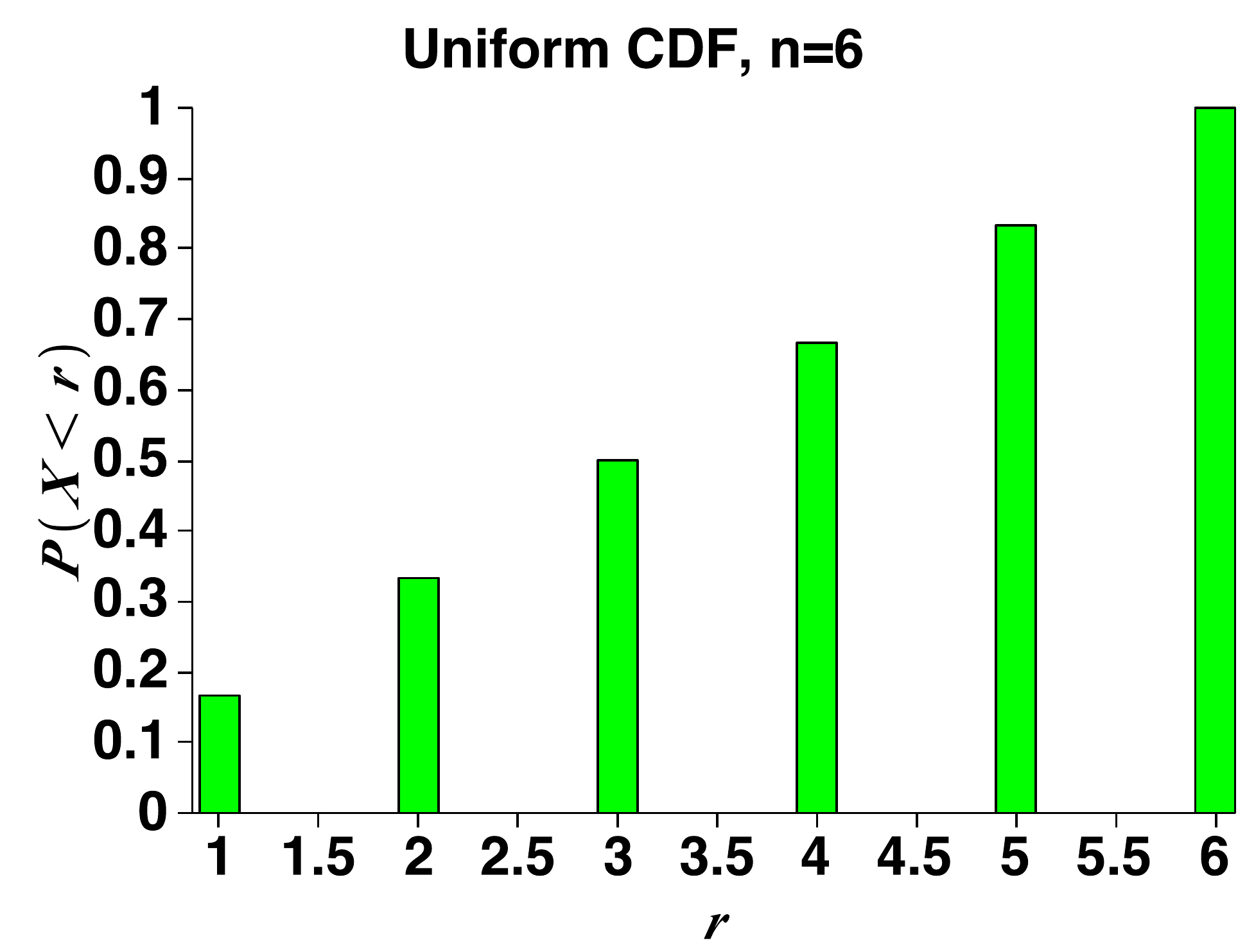} \hfill
\end{tabular}
\end{center}
\caption{Left panel: uniform discrete distribution for $n=6$. Right panel: the corresponding cdf. }
\label{fig:uniform}
\end{figure}

\begin{figure}
\begin{center}
\begin{tabular}{cc}
\includegraphics[width=0.45\linewidth]{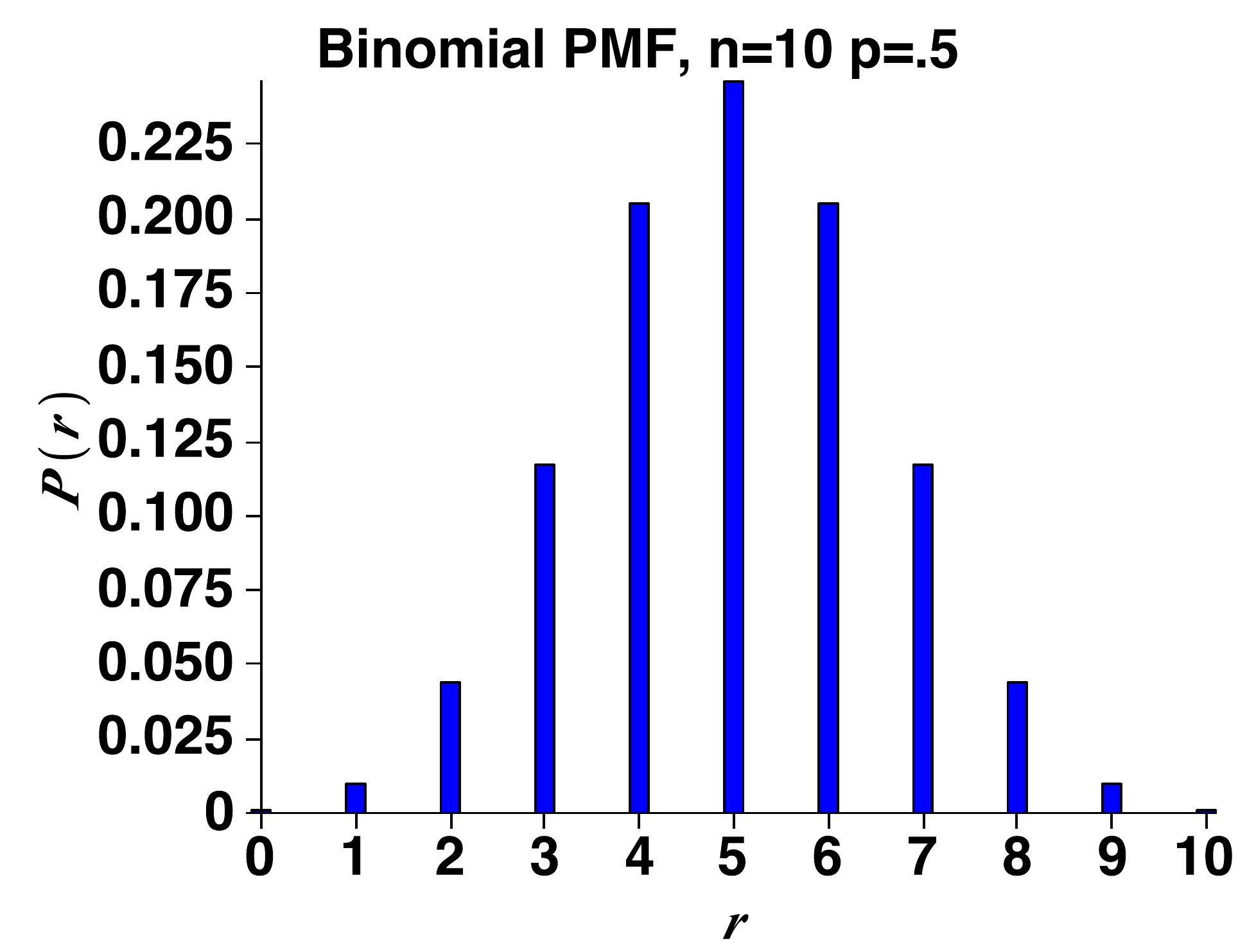} \hfill 
\includegraphics[width=0.45\linewidth]{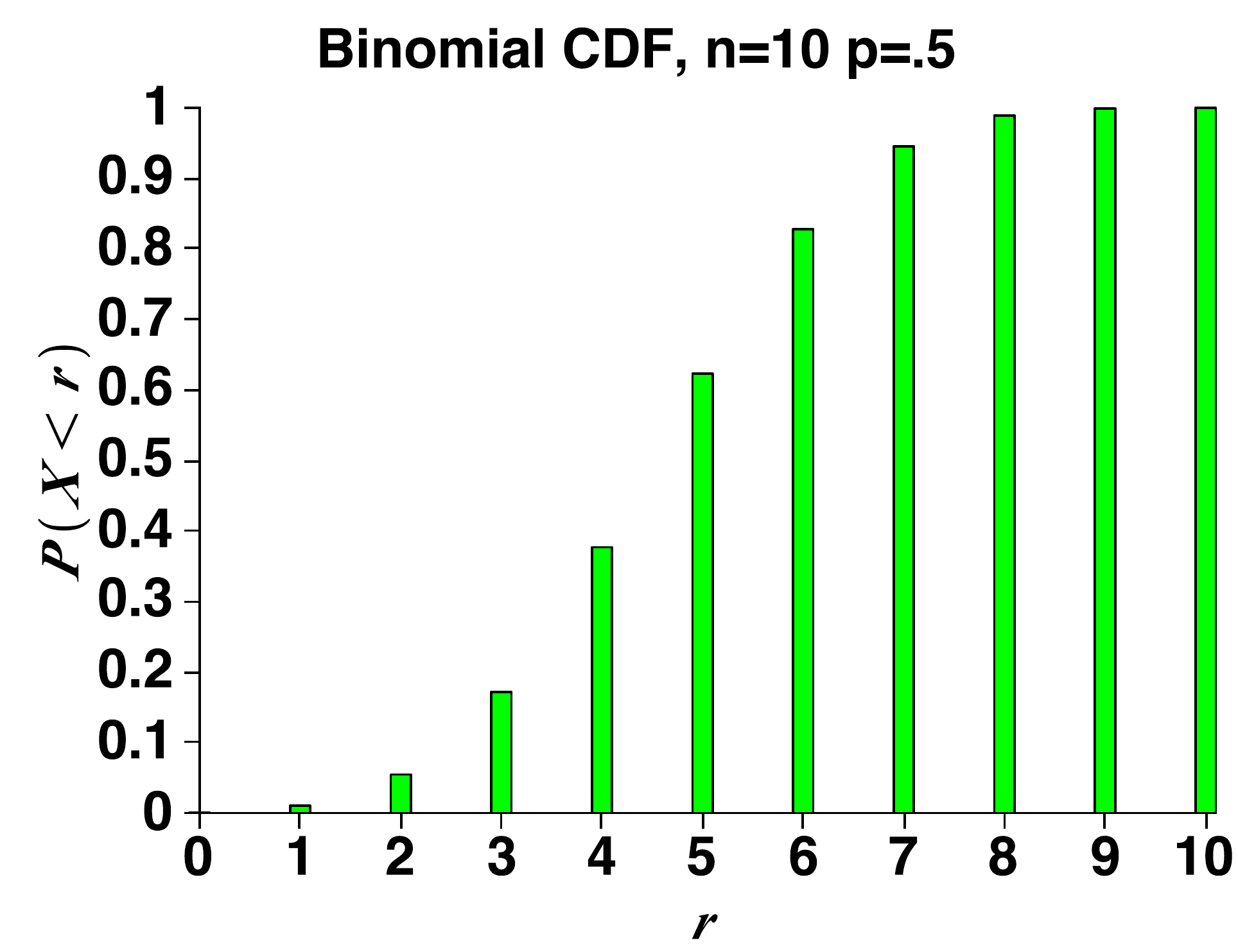} \hfill \\ 
\includegraphics[width=0.45\linewidth]{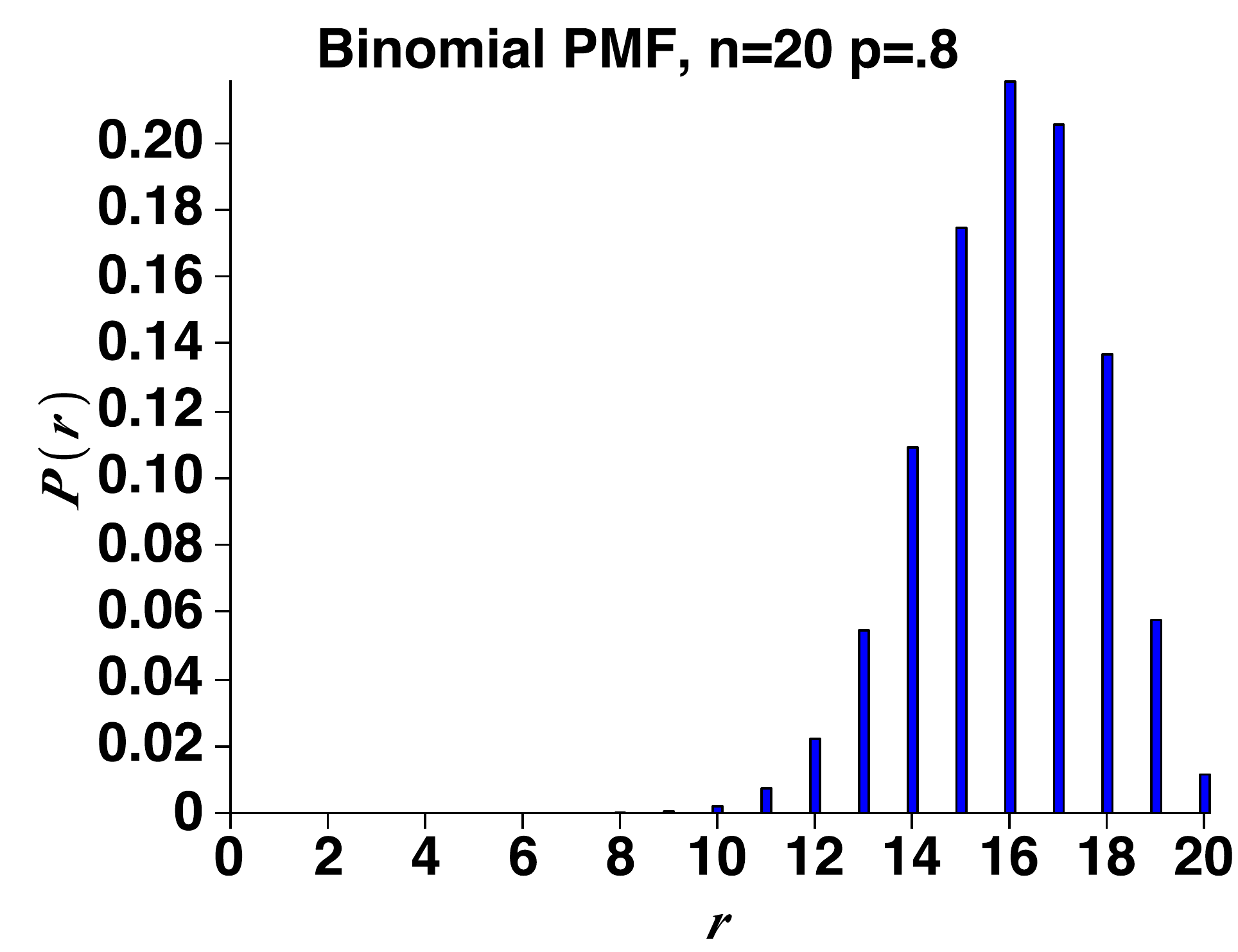} \hfill 
\includegraphics[width=0.45\linewidth]{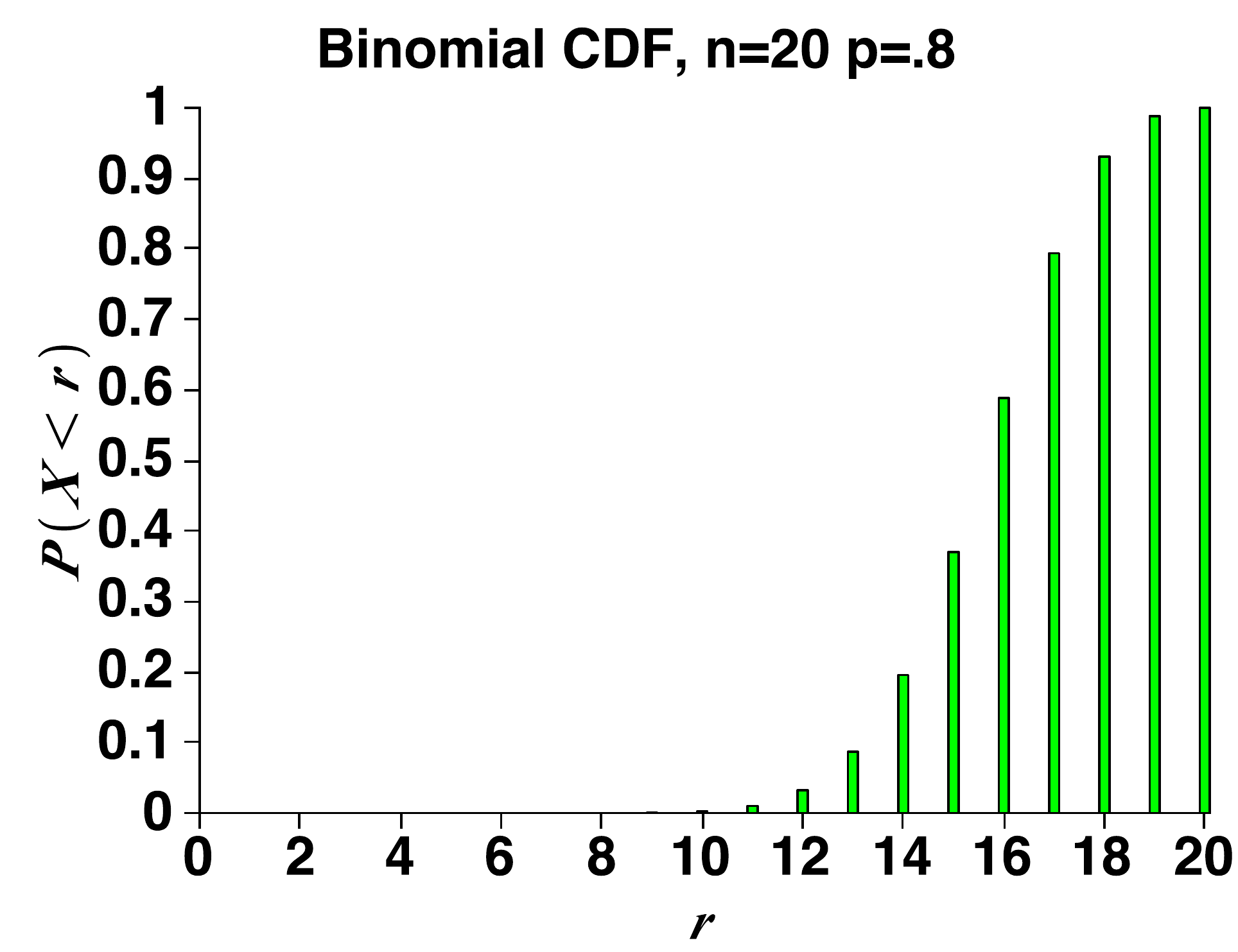} \hfill \\
\includegraphics[width=0.45\linewidth]{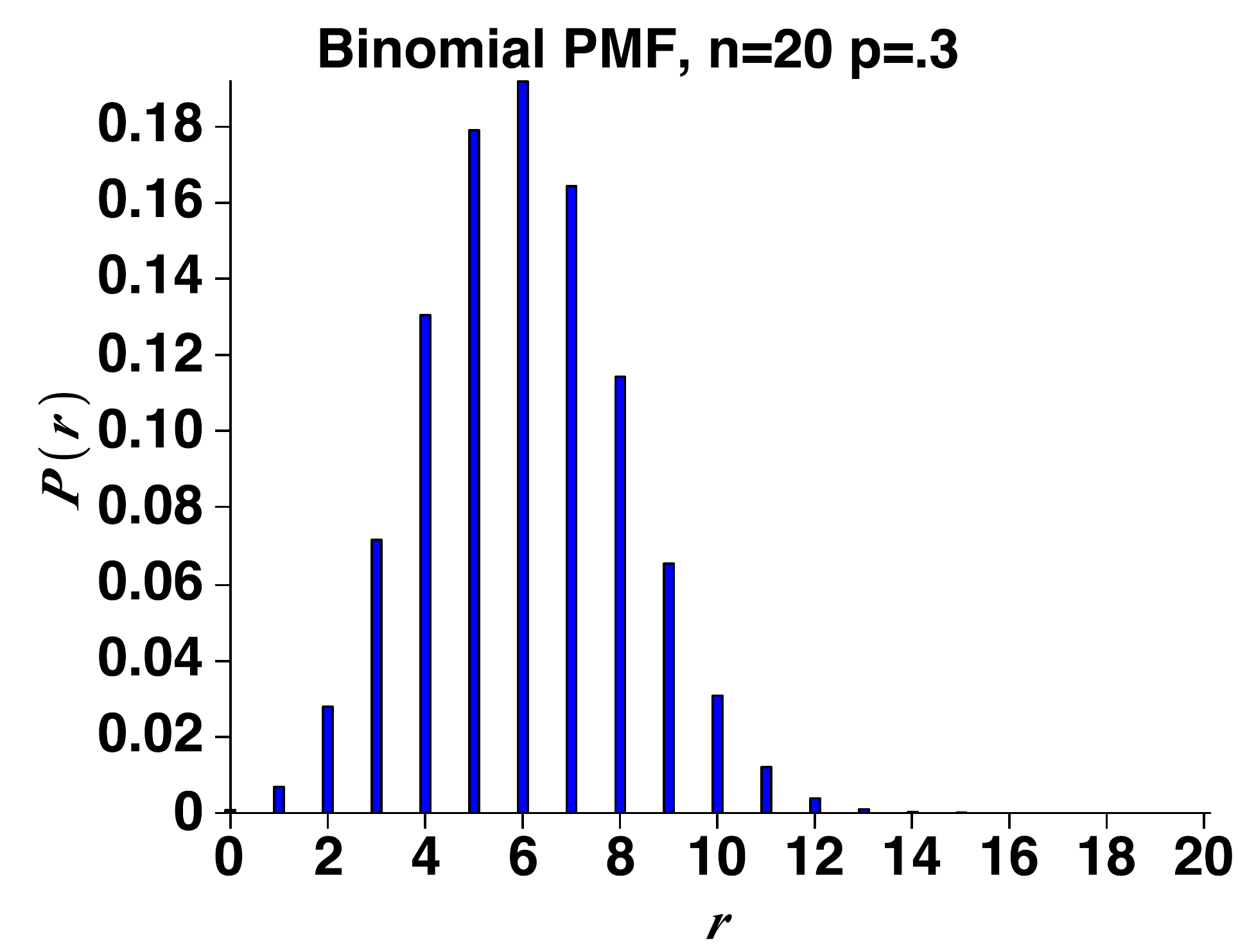} \hfill 
\includegraphics[width=0.45\linewidth]{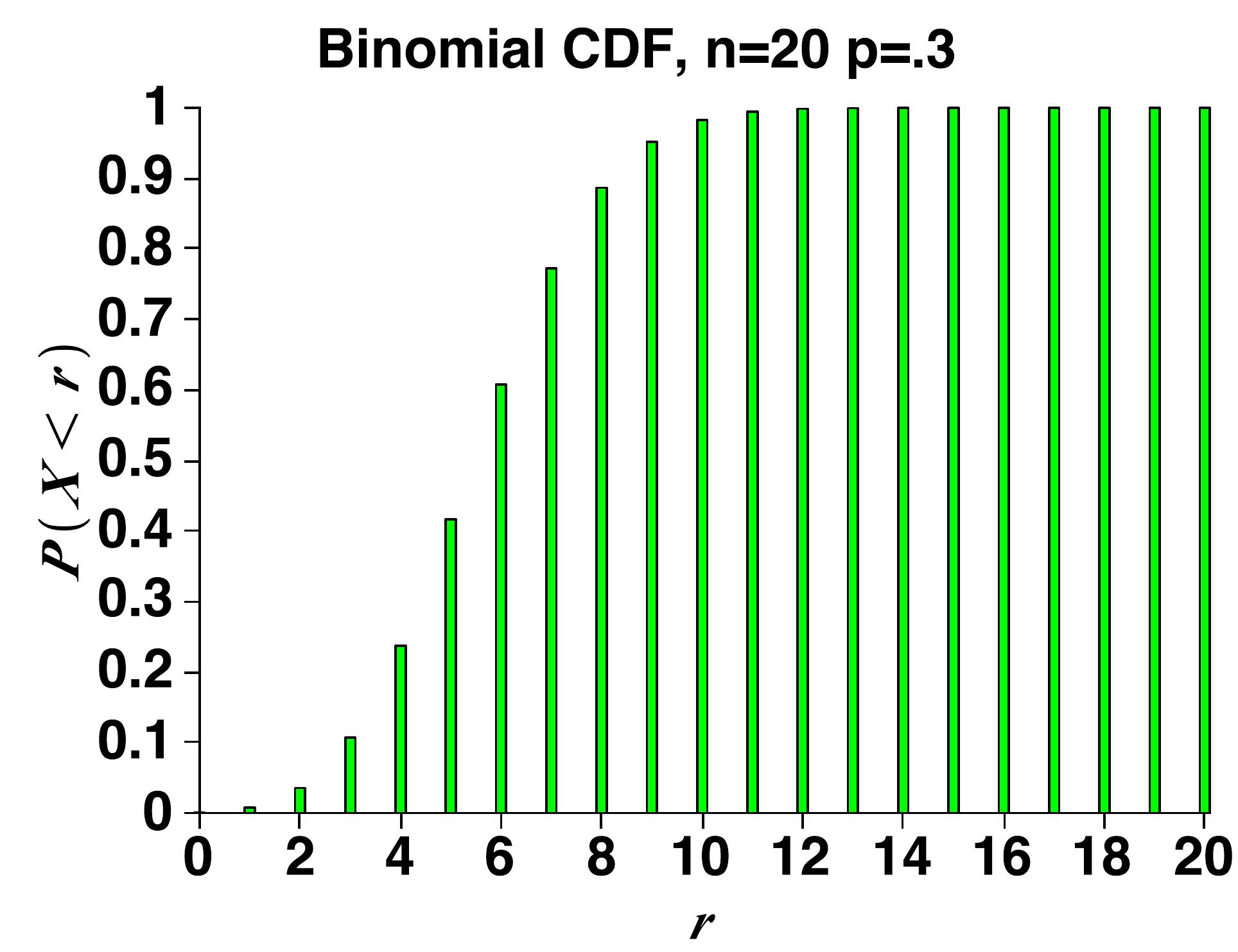} \hfill \\
\end{tabular}
\end{center}
\caption{Some examples of the binomial distribution, Eq.~\eqref{eq:binomialdef}, for different choices of $n,p$, and its corresponding cdf. }
\label{fig:binomial}
\end{figure}

\runinhead{The uniform distribution:} for $n$ equiprobable outcomes between 1 and $n$, the {uniform discrete distribution} is given by
\be \label{eq:uniform}
P(r) = \left\{
\begin{array}{c l}
  1/n & \mbox{for } 1 \leq r \leq n  \\
  0 &\mbox{otherwise}
\end{array}
\right.
\ee
It is plotted in Fig.~\ref{fig:uniform} alongside with its cdf for the case of the tossing of a fair die ($n = 6$). 
\runinhead{The binomial distribution:} the binomial describes the probability of obtaining $r$ ``successes'' in a sequence of $n$ trials, each of which has probability $p$ of success. 
Here, ``success'' can be defined as one specific outcome in a binary process (e.g., H/T, blue/red, 1/0, etc). The binomial distribution $B(n,p)$ is given by:
\be \label{eq:binomialdef}
 P(r|n,p) \equiv B(n,p) = {n \choose r} p^r (1-p)^{n-r},
\ee
where the ``choose'' symbol is defined as 
\be {n \choose r} \equiv \frac{n!}{(n-r)!r!}
\ee
for $0 \leq r \leq n$ (remember, $0! = 1$). Some examples of the binomial for different choices of $n,p$ are plotted in Fig.~\ref{fig:binomial}.

The derivation of the binomial distribution proceeds from considering the probability of obtaining $r$ successes in $n$ trials ($p^r$), while at the same time obtaining $n-r$ failures ($(1-p)^{n-r}$). The combinatorial factor in front is derived from considerations of the number of permutations that leads to the same total number of successes.  
\runinhead{The Poisson distribution:} the Poisson distribution describes the probability of obtaining a certain number of events in a process where events occur with a fixed average rate and independently of each other. The process can occur in time (e.g.,  number of planes landing at Heathrow, number of photons arriving at a photomultiplier, number of murders in London, number of electrons at a detector, etc \dots in a certain time interval) or in space (e.g., number of galaxies in a patch on the sky).  

Let's assume that $\lambda$ is the average number of events occuring per unit time or per unit length (depending on the problem being considered). Furthermore, $\lambda =$ constant in time or space. 

\example{For example, $\lambda = 3.5$ busses/hour is the {\it average} number of busses passing by a particular bus stop every hour; or $\lambda = 10.3$ droplets/m$^2$ is the {\it average} number of drops of water hitting a square meter of the surface of an outdoor swimming pool in a certain day. Notice that of course at every given hour an integer number of busses actually passes by (i.e., we never observe 3 busses and one half passing by in an hour!), but that the {average} number can be non-integer (for example, you might have counted 7 busses in 2 hours, giving an average of 3.5 busses per hour). The same holds for the droplets of water.} 

 For problems involving the time domain (e.g., busses/hour), the probability of $r$ events happening in a time $t$ is given by the {Poisson distribution}:   \attention 
\be \label{eq:poisson} 
P(r|\lambda, t) \equiv  {\rm Poisson}(\lambda) = \frac{(\lambda t)^r}{r!}e^{-\lambda t}.
\ee
If the problem is about the spatial domain (e.g., droplets/m$^2$), the probability of $r$ events happening in an area $A$ is given by:  
\be \label{eq:poisson_space}
P(r|\lambda, A) \equiv  {\rm Poisson}(\lambda) = \frac{(\lambda A)^r}{r!}e^{-\lambda A}.
\ee

Notice that this is a discrete pmf in the number of events $r$, and {not} a continuous pdf in $t$ or $A$. The probability of getting $r$ events in a unit time interval is obtained by setting $t=1$ in Eq.~\eqref{eq:poisson}; similarly, the probability of getting $r$ events in a unit area is obtained by setting $A=1$ in Eq.~\eqref{eq:poisson_space}

\example{A particle detector measures protons which are emitted with an average rate $\lambda = 4.5$/s. What is the probability of measuring 6 protons in 2 seconds? \\ {Answer:} 
\be
P(6 |\lambda = 4.5 \text{s}^{-1}, t=2 \text{s}) = \frac{(4.5 \cdot 2)^6}{6!}e^{-4.5 \cdot 2} = 0.09.
\ee
So the probability is about 9\%.}

\begin{figure}
\begin{center}
\begin{tabular}{cc}
\includegraphics[width=0.45\linewidth]{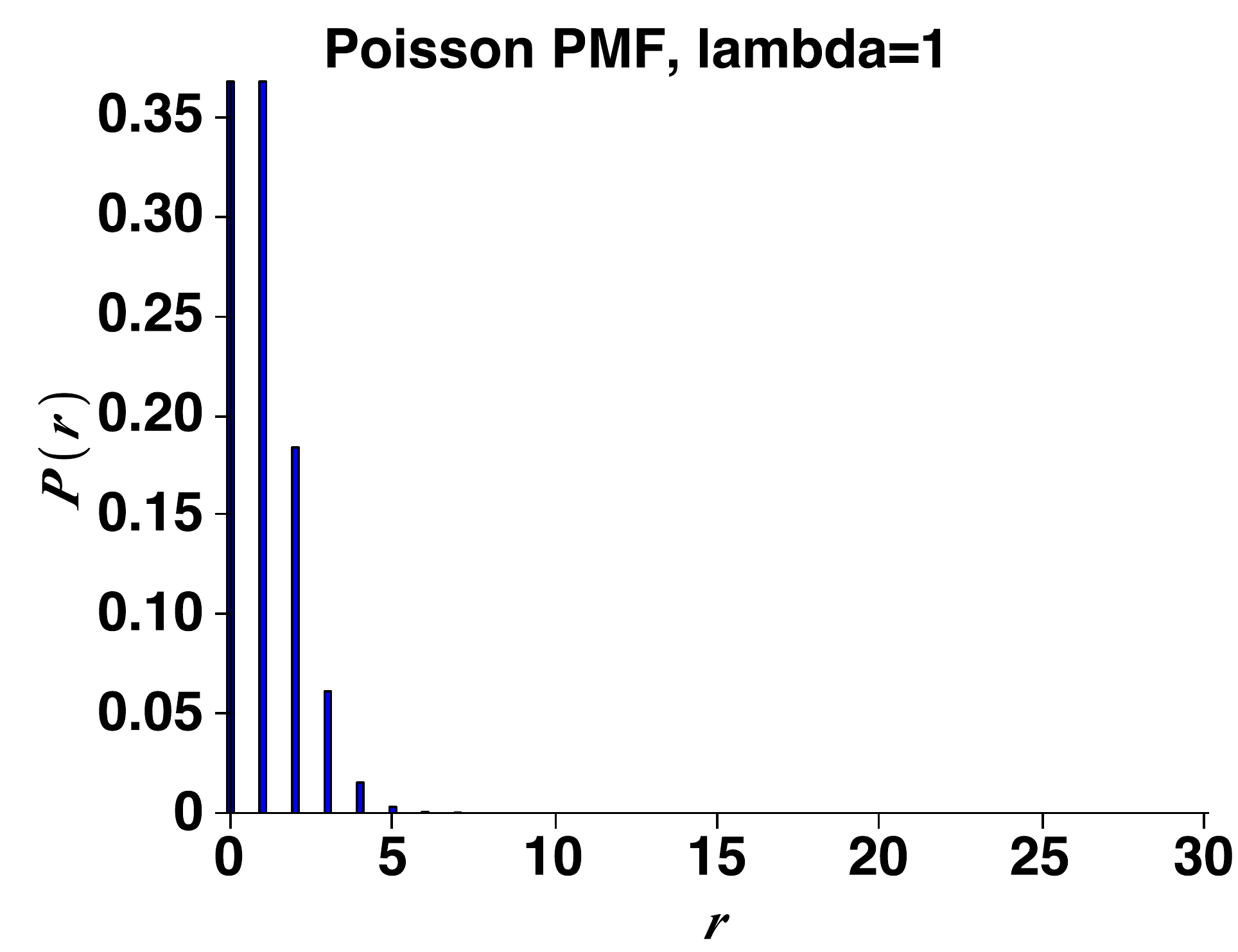} \hfill 
\includegraphics[width=0.45\linewidth]{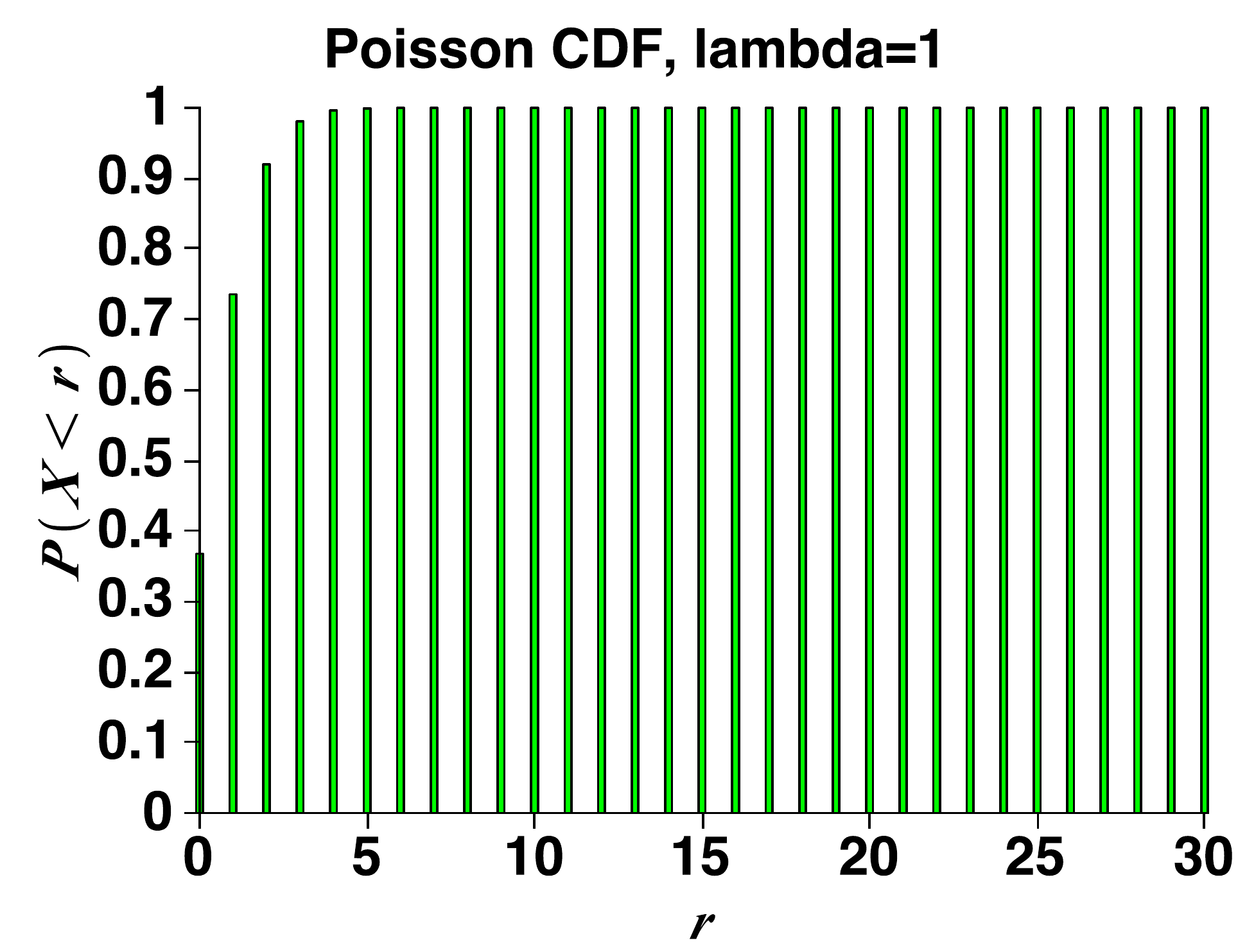} \hfill \\ 
\includegraphics[width=0.45\linewidth]{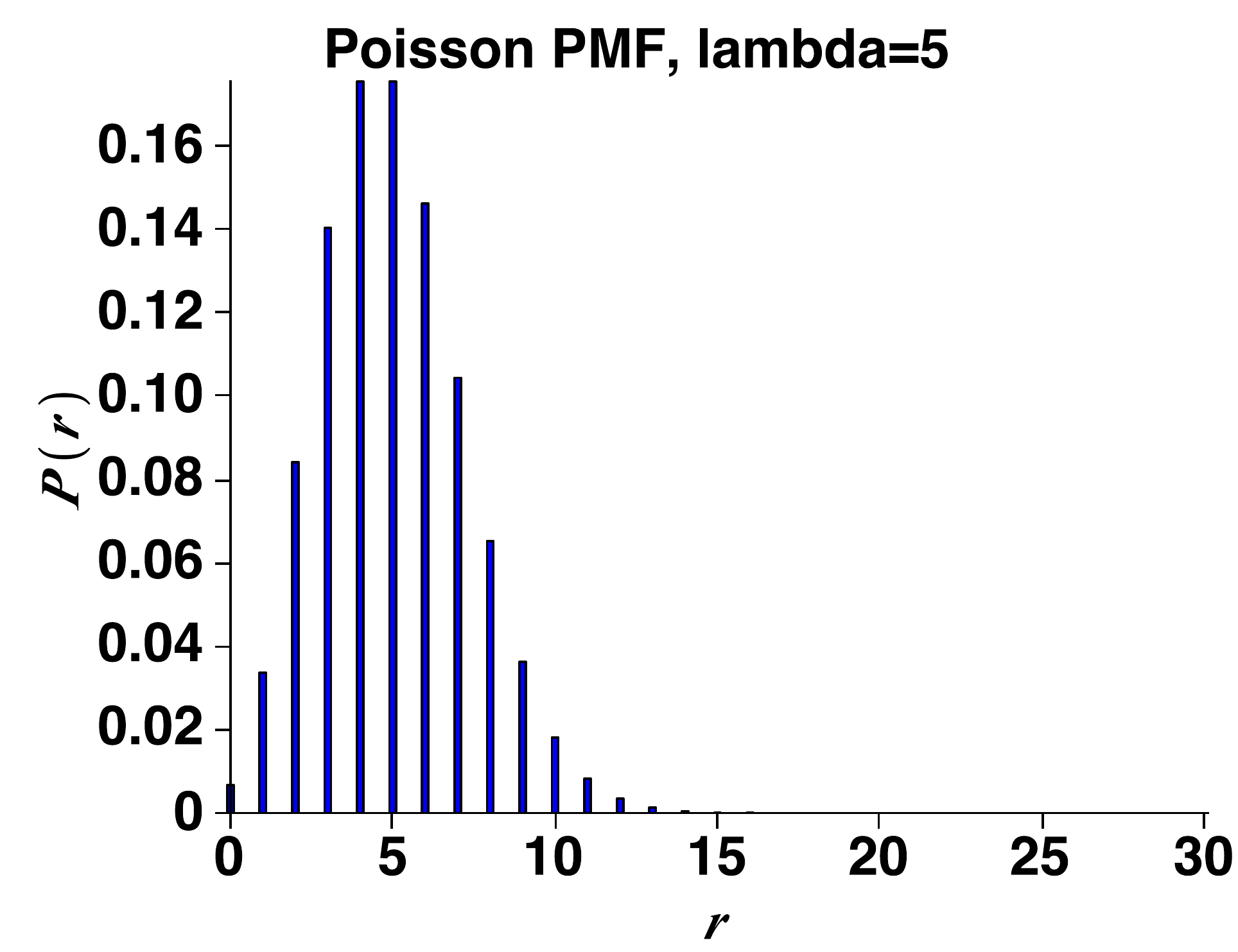} \hfill 
\includegraphics[width=0.45\linewidth]{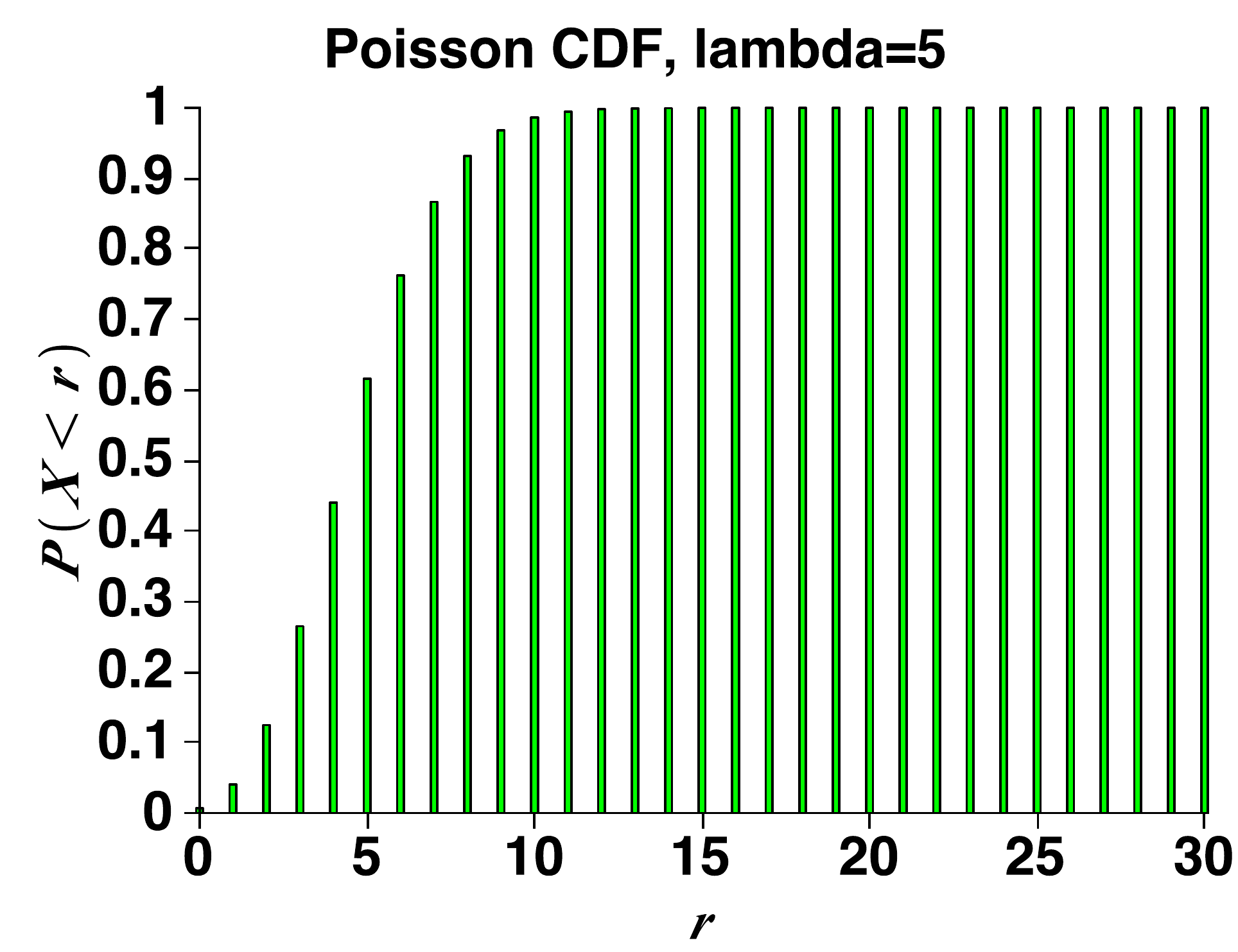} \hfill \\
\includegraphics[width=0.45\linewidth]{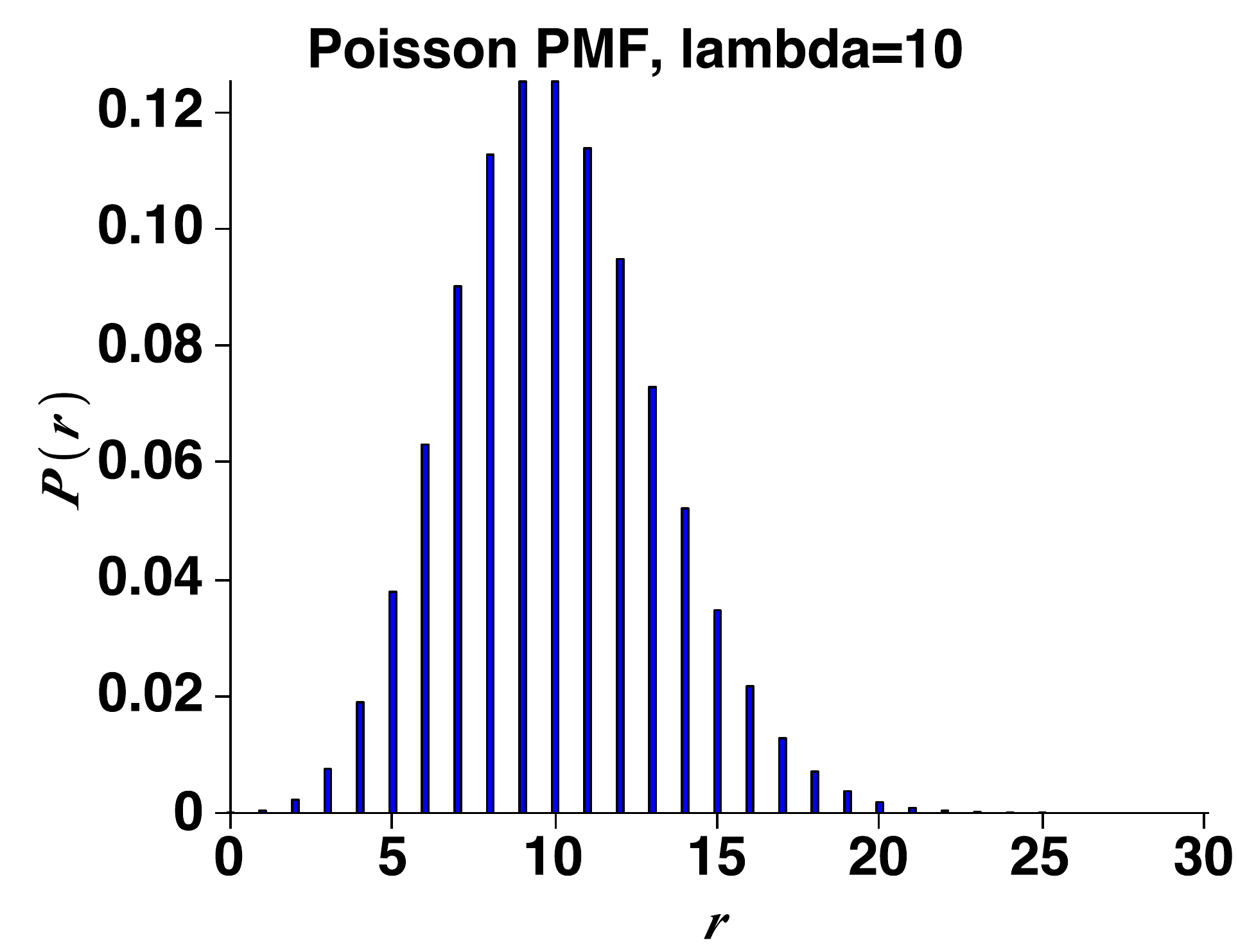} \hfill 
\includegraphics[width=0.45\linewidth]{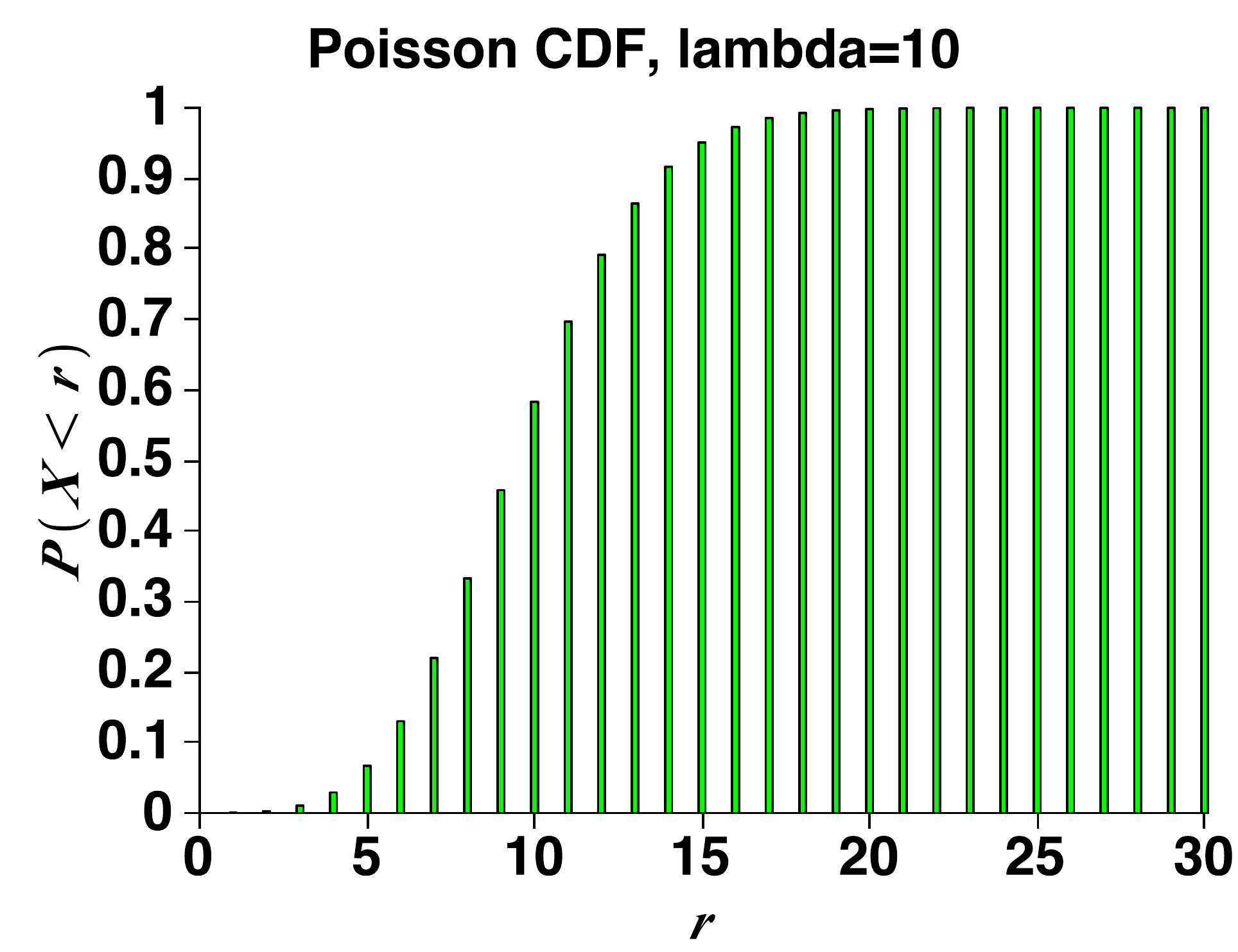} \hfill \\
\end{tabular}
\end{center}
\caption{Some examples of the Poisson distribution, Eq.~\eqref{eq:poisson}, for different choices of $\lambda$, and its corresponding cdf. }
\label{fig:poisson}
\end{figure}

The Poisson distribution of Eq.~\eqref{eq:poisson} is plotted in Fig.~\ref{fig:poisson} as a function of $r$ for a few choices of $\lambda$ (notice that in the figure $t=1$ has been assumed, in the appropriate units).  
The derivation of the Poisson distribution follows from considering the probability of 1 event taking place in a small time interval $\Delta t$, then taking the limit $\Delta t \rightarrow dt \rightarrow 0$. It can also be shown that the Poisson distribution arises from the binomial in the limit $p n \rightarrow \lambda$ for ${n \rightarrow \infty}$, assuming $t=1$ in the appropriate units (see lecture).  

\example{In a post office, people arrive at the counter at an average rate of 3 customers per minute. What is the probability of 6 people arriving in a minute? }
\\ {\em  Answer:} The number of people arriving follows a Poisson distribution with average $\lambda = 3$ (people/min). The probability of 6 people arriving in a minute is given by
\be
P(n=6 | \lambda, t=1\, {\rm min} ) = \frac{(\lambda t)^n}{n!} e^{-\lambda t} \approx 0.015
\ee
So the probability is about 1.5\%. 

The discrete distributions above depend on parameters (such as $p$ for the binomial, $\lambda$ for Poisson), which control the shape of the distribution. If we know the value of the parameters, we can compute the probability of an observation (as done it the examples above). This is the subject of {probability theory}, which concerns itself with the theoretical properties of the distributions. The inverse problem of making inferences about the parameters from the observed samples (i.e., learning about the parameters from the observations made) is the subject of statistical inference, addressed later. 

\subsection{Expectation value and variance}

Two important properties of distributions are the {expectation value} (which controls the location of the distribution) and the {variance or dispersion} (which controls how much the distribution is spread out). Expectation value and variance are functions of a RV. 
\begin{definition} The expectation value $E[X]$ (often called ``mean'', or ``expected value''\footnote{We prefer not to use the term ``mean'' to avoid confusion with the {sample mean}.}) of the discrete RV $X$ is defined as
\be
 E[X] = \langle X \rangle \equiv \sum_i x_i P_i.
 \ee 
\end{definition}

\example{You toss a fair die, which follows the uniform discrete distribution, Eq.~\eqref{eq:uniform}. What is the expectation value of the outcome?} \\
{Answer:} the expectation value is given by $E[X] = \sum_i i \cdot \frac{1}{6} = 21/6$.

\begin{definition}
The {variance or dispersion} $\Var(X)$ of the discrete RV $X$ is defined as
\be
\Var(X) \equiv E[(X - E[X])^2] = E(X^2) - E[X]^2.
\ee
The square root of the variance is often called ``standard deviation'' and is usually denoted by the symbol $\sigma$, so that $\Var(X) = \sigma^2$. 
\end{definition}

\example{For the case of tossing a fair die once, the variance is given by}
\be
\Var(X) = \sum_i(x_i - \langle X \rangle)^2 P_i = \sum_i x_i^2 P_i - \left(\sum_i x_i P_i\right)^2 = \sum_i i^2\frac{1}{6} - \left(\frac{21}{6}\right)^2 = \frac{105}{36}.
\ee  
\item For the binomial distribution of Eq.~\eqref{eq:binomialdef}, the expectation value and variance are given by:
\be \label{eq:properties_binomial}
E[X] = n p, \qquad \Var(X) = n p(1-p).
\ee 

\example{A fair coin is tossed $N$ times. What is the expectation value for the number of heads, $H$? What is its variance?}
For $N = 10$, evaluate the probability of obtaining 8 or more heads. \\
{Answer:} The expectation values and variance are given by Eq.~\eqref{eq:properties_binomial}, with $p=1/2$ (as the coin is fair), thus
\begin{equation}
E(H) = N p = N/2 \quad {\rm and} \quad {\rm{Var}}(H) = Np (1-p) = N/4.
\end{equation}
The probability of obtaining 8 or more heads is given by
\be
P(H=8 = \sum_{H=8}^{10} P(H \, {\rm heads} | N, p=1/2) = \frac{1}{2^{10}} \sum_{H=8}^{10} {10 \choose H} = \frac{56}{1024} \approx 0.055.
\ee
So the probability  of obtaining 8 or more heads is about 5.5\%.

For the Poisson distribution of Eq.~\eqref{eq:poisson}, the expectation value and variance are given by:
\be \label{eq:properties_poisson}
E[X] = \lambda t, \qquad \Var(X) = \lambda t,
\ee
while for the spatial version of the Poisson distribution, Eq.~\eqref{eq:poisson_space}, they are given by:
\be \label{eq:properties_poisson}
E[X] = \lambda A, \qquad \Var(X) = \lambda A.
\ee

As we did above for the discrete distribution, we now define the following properties for continuous distributions. 
\begin{definition}
The expectation value $E[X]$ of the continuous RV $X$ with pdf $p(X)$ is defined as
\be
 E[X] = \langle X \rangle \equiv \int x p(x) dx.
 \ee 
 \end{definition}
 \begin{definition}
 The variance or dispersion $\Var(X)$ of the continuous RV $X$ is defined as
\be
\Var(X) \equiv E[(X - E[X])^2] = E(X^2) - E[X]^2 = \int x^2 p(x) dx - \left(\int x p(x) dx\right)^2.
\ee
\end{definition}

\subsection{The exponential distribution} 

The exponential distribution describes the time one has to wait between two consecutive events in a Poisson process, e.g. the waiting time between two radioactive particles decays. If the Poisson process happens in the spatial domain, then the exponential distribution describes the distance between two events (e.g., the separation of galaxies in the sky). In the following, we will look at processes that happen in time (rather than in space). 

To derive the exponential distribution, one can consider the arrival time of Poisson distributed events with average rate $\lambda$ (for example, the arrival time particles in a detector).  The probability that the first particle arrives at time $t$ is obtained by considering the probability (which is Poisson distributed) that no particle arrives in the interval  $[0, t]$, given by $P(0 | \lambda, t) = \exp(-\lambda t)$ from Eq.~\eqref{eq:poisson}, times the probability that one particle arrives during the interval  $[t,t+\Delta t]$, given by $\lambda \Delta t$. Taking the limit $\Delta t \rightarrow 0$ it follows that the probability density (denoted by a symbol $p()$) for observing the first event happening at time $t$ is given by \attention 
\be \label{eq:exponential}
p(\text{1st event happens at time } t | \lambda ) = \lambda e^{-\lambda t},
\ee
where $\lambda$ is the mean number of events per unit time. This is the exponential distribution. 

\example{Let's assume that busses in London arrive according to a Poisson distribution, with average rate $\lambda = 5$ busses/hour. You arrive at the bus stop and a bus has just departed. What is  the probability that you will have to wait more than 15 minutes? \\
{Answer:} the probability that you'll have to wait for $t_0 = 15$ minutes or more is given by 
\be
\int_{t_0}^\infty p(\text{1st event happens at time } t | \lambda ) dt = \int_{t_0}^\infty\lambda e^{-\lambda t} dt = e^{-\lambda t_0} = 0.29,
\ee
where we have used $\lambda = 5\text{busses/hour} = 1/12$  busses/min.}

If we have already waited for a time $s$ for the first event to occur (and no event has occurred), then the probability that we have to wait for another time $t$ before the first event happens satisfies 
\begin{equation} \label{eq:lack_of_memory}
p(T>t+s | T>s) = p(T>t).
\end{equation} 
This means  that having waited for time $s$ without the event occuring, the time we can expect to have to wait has the same distribution as the time we have to wait from the beginning. The exponential distribution has no ``memory'' of the fact that a time $s$ has already elapsed.

For the exponential distribution of Eq.~\eqref{eq:exponential}, the expectation value and variance for  the time $t$ are given by 
\be \label{eq:properties_exponential}
E(t) = 1/\lambda, \qquad \Var(t) = 1/\lambda^2.
\ee

\subsection{The Gaussian (or Normal) distribution}

The Gaussian pdf (often called ``the Normal distribution'') is perhaps the most important distribution. It is used as default in many situations involving continuous RV, since it naturally flows from the the Central Limit Theorem, section~\ref{sec:clt}. 

\begin{figure}
\begin{center}
\includegraphics[width=0.45\linewidth]{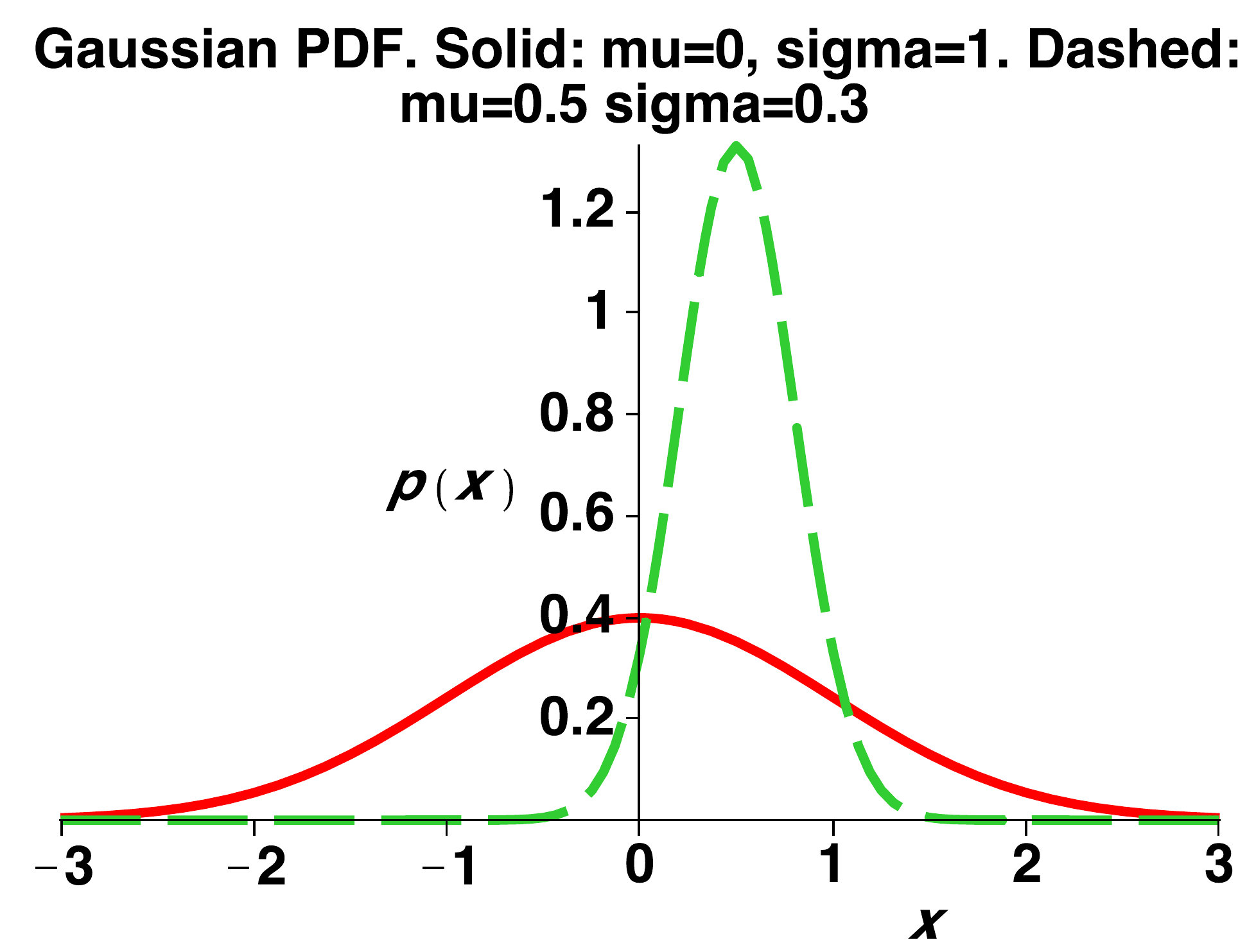} \hfill 
\includegraphics[width=0.45\linewidth]{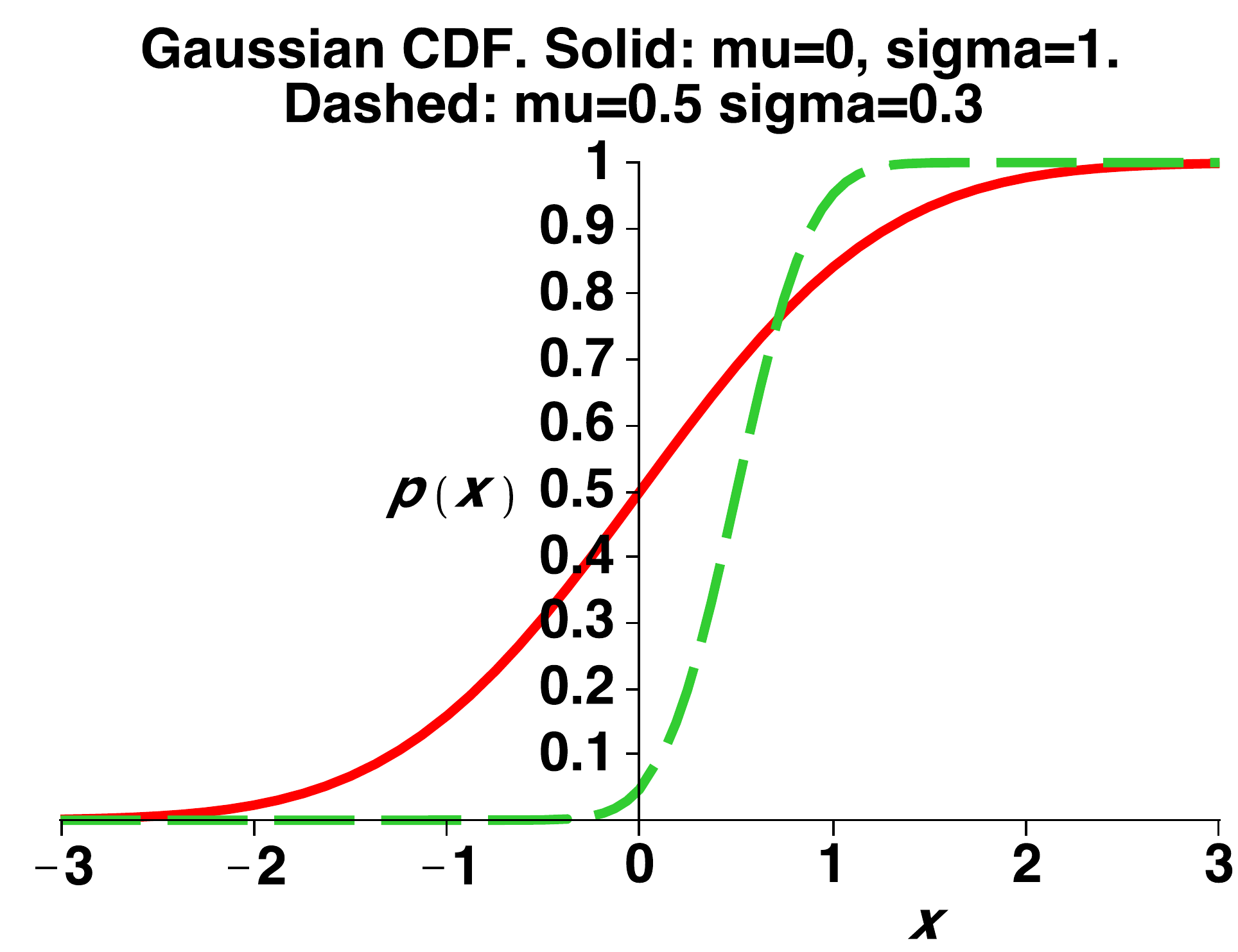} \hfill \\ 
\end{center}
\caption{Two examples of the Gaussian distribution, Eq.~\eqref{eq:gaussian}, for different choices of $\mu, \sigma$, and its corresponding cdf. The expectation value $\mu$ controls the location of the pdf (i.e., when changing $\mu$ the peak moves horizontally, without changing its shape), while the standard deviation $\sigma$ controls its width (i.e., when changing $\sigma$ the spread of the peak changes but not its location). }
\label{fig:gauss}
\end{figure}

The Gaussian pdf is a continuous distribution with mean $\mu$ and standard deviation $\sigma$ is given by
\be \label{eq:gaussian}
p(x|\mu,\si) = \frac{1}{\sqrt{2\pi} \si} \exp\left(-\frac{1}{2}\frac{(x-\mu)^2}{\si^2} \right),
\ee 
and it is plotted in Fig.~\ref{fig:gauss} for two different choices of $\{\mu, \si\}$. The Gaussian is the famous bell-shaped curve. 

For the Gaussian distribution of Eq.~\eqref{eq:gaussian}, the expectation value and variance are given by:
\be \label{eq:Gaussian_properties}
E[X] = \mu, \qquad \Var(X) = \si^2.
\ee

\begin{figure}
\begin{center}
\includegraphics[width=0.53\linewidth]{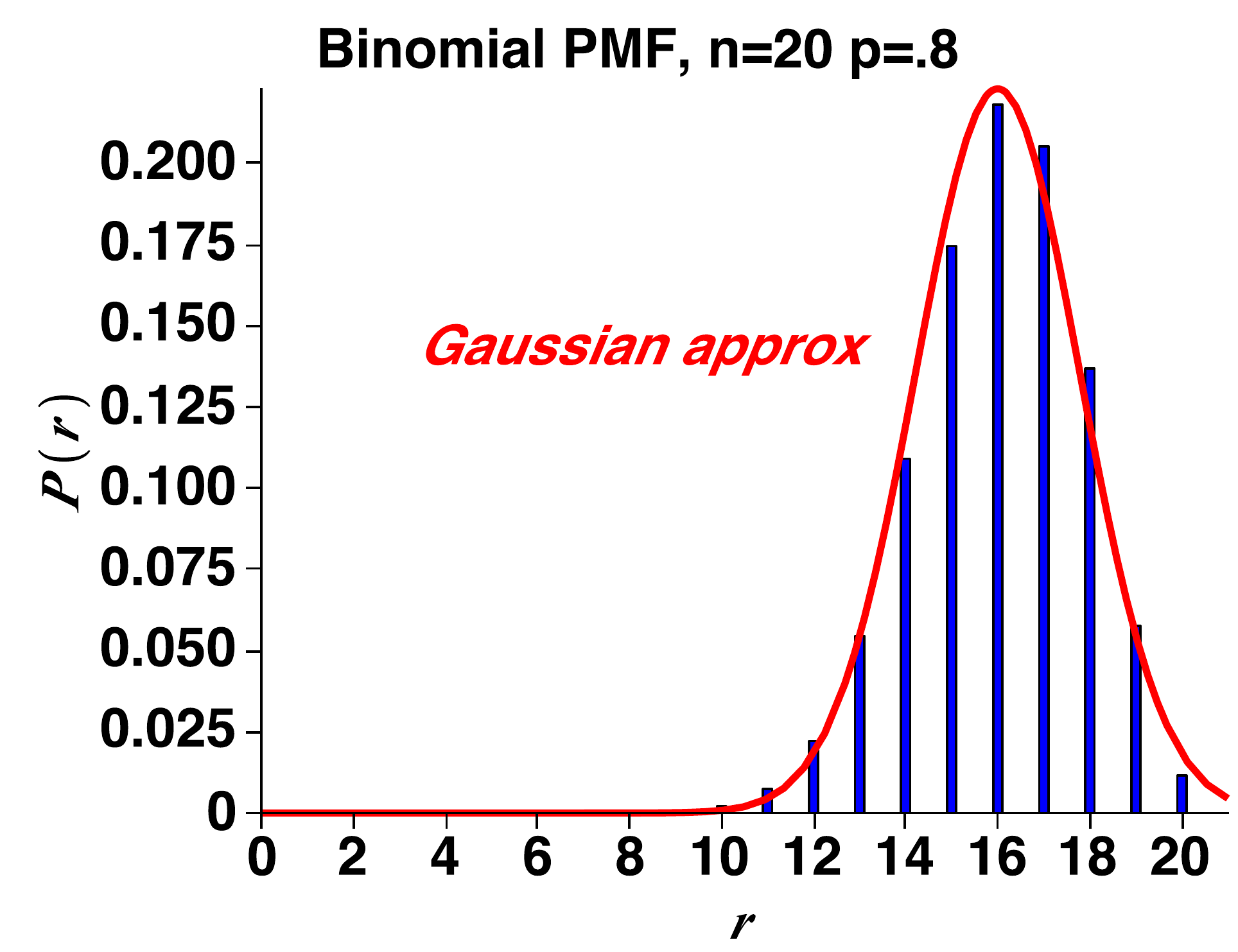} \hfill 
\includegraphics[width=0.42\linewidth]{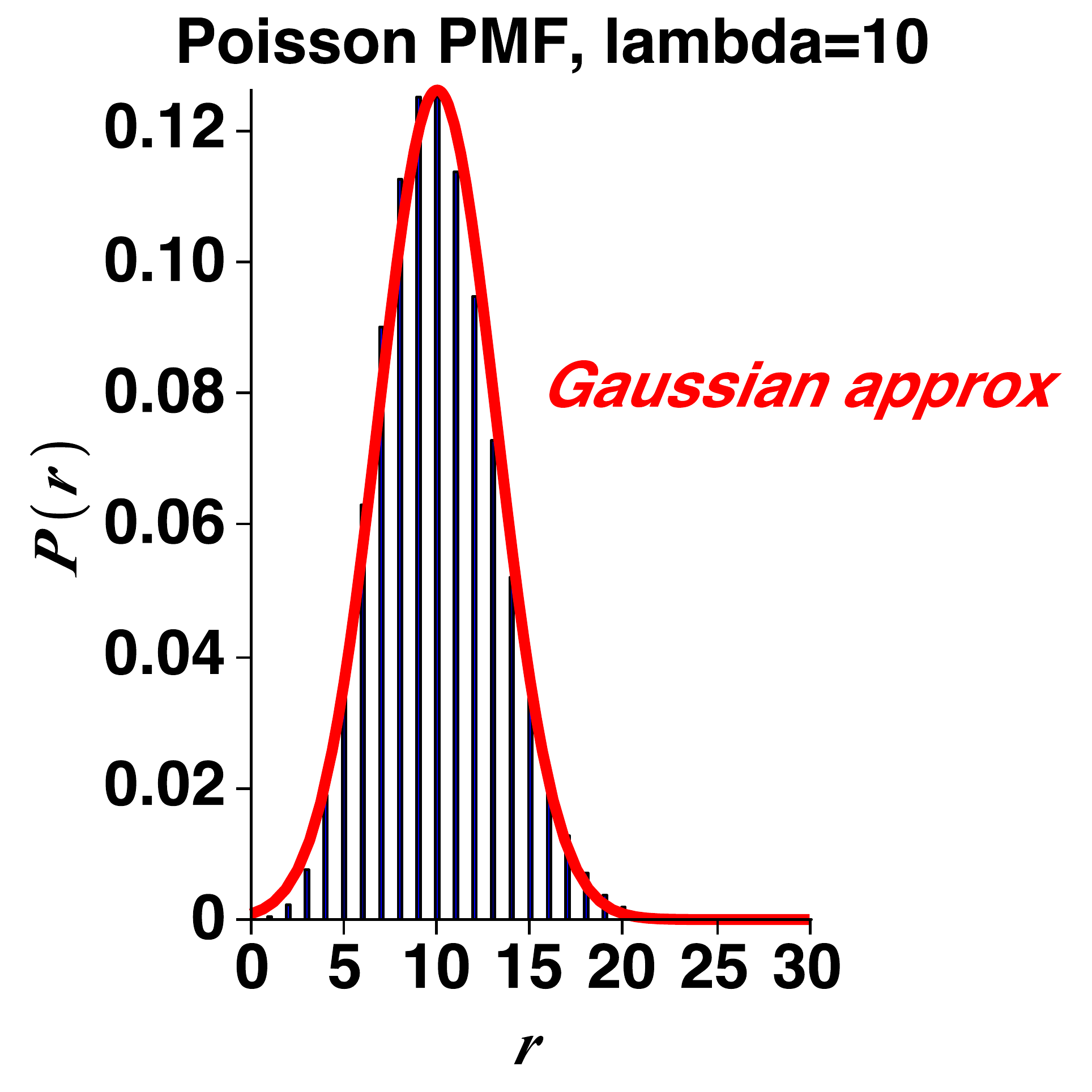} \hfill \\ 
\end{center}
\caption{Gaussian approximation to the binomial (left panel) and the Poisson distribution (right panel).  The solid curve gives in each case the Gaussian approximation to each pmf. }
\label{fig:gauss_approx}
\end{figure}

It can be shown that the Gaussian arises from the binomial in the limit $n\rightarrow\infty$ and from the Poisson distribution in the limit $\lambda \rightarrow \infty$.  As shown in Fig.~\ref{fig:gauss_approx}, the Gaussian approximation to either the binomial or the Poisson distribution is very good even for fairly moderate values of $n$ and $\lambda$. 

The probability content of a Gaussian of standard deviation $\sigma$ for a given symmetric interval around the mean of width $\kappa \sigma$ on each side is given by
\begin{align} \label{eq:prob_content_gaussian}
P(\mu - \kappa \sigma < x < \mu +\kappa \sigma) & = \int_{\mu-\kappa \sigma}^{\mu+\kappa \sigma}\frac{1}{\sqrt{2\pi}{\si}} \exp\left(-\frac{1}{2}\frac{(x-\mu)^2}{\si^2} \right) dx \\
& = \frac{2}{\sqrt{\pi}}\int_0^{\kappa/\sqrt{2}} \exp\left(-y^2 \right) dy \\
& = {\rm erf}(\kappa/\sqrt{2}),
\end{align}
where the {error function} erf is defined as 
\be
{\rm erf}(x) =  \frac{2}{\sqrt{\pi}}\int_0^{x} \exp\left(-y^2 \right) dy, 
\ee
and can be found by numerical integration (also often tabulated and available as a built-in function in most mathematical software). Also recall the useful integral:
\be
 \int_{-\infty}^{\infty} \exp\left(-\frac{1}{2}\frac{(x-\mu)^2}{\si^2} \right) dx = \sqrt{2\pi}{\si}.
 \ee

Eq.~\eqref{eq:prob_content_gaussian} allows to find the probability content of the Gaussian pdf for any symmetric interval around the mean. Some commonly used values are given in Table~\ref{tab:gauss_interval}. 
\begin{table}
\centering
\begin{tabular}{r r r }
\hline
$\kappa$  & $P(-\kappa < \frac{x-\mu}{\si} < \kappa)$ & Usually called \\
``number of sigma'' & Probability content & \\ 
\hline
1 & 0.683 & $1\si$ \\
2 & 0.954 & $2\si$ \\
3 & 0.997 & $3\si$ \\
4 & 0.9993 & $4\si$ \\
5 & $1 - 5.7\times10^{-7}$ & $5\si$ \\ \hline
1.64 & 0.90 & 90\% probability interval\\
1.96 & 0.95 & 95\% probability interval\\
2.57 & 0.99 & 99\% probability interval\\
3.29 & 0.999 & 99.9\% probability interval\\\hline
\end{tabular}
\caption{Relationship between the size of the interval around the mean and the probability content for a Gaussian distribution. \label{tab:gauss_interval}}
\end{table}

\example{Measurements are often reported with the notation $T = (100 \pm 1)$ K (in this case, we assume we have measured a temperature, $T$). If nothing else is specified, it is usually implied that the error follows a Gaussian distribution. In the example above, $\pm 1$ K is the so-called ``1$\si$ interval''. This means that 68.3\% of the probability is contained within the range $[ 99, 101 ]$ K. A ``2$\si$ interval'' would have a length of 2 K on either side, so 95.4\% of the probability is contained in the interval $[ 98, 102]$ K. If one wanted a 99\% interval, one would need a 2.57$\sigma$ range (see Table~\ref{tab:gauss_interval}). Since in this case the 1$\sigma$ error is 1 K, the 2.57$\sigma$ error is 2.57 K and the 99\% interval is $[ 97.43, 102.57]$ K. }

A heuristic derivation of how the Gaussian arises follows from this example involving darts throwing.
Suppose we are throwing darts towards a target (located at the center of the coordinate system, at the position $x=0, y=0$), with the following rules:
\begin{enumerate}
\item Throws are independent.
\item Errors in the $x$ and $y$ directions are independent.
\item Large errors are less probable than small ones.
\end{enumerate}
The probability of a dart landing in an infinitesimal square located at coordinates $(x,y)$ and of size $(\Delta x, \Delta y)$ (i.e., the dart landing in the interval $[x, x+\Delta x]$ and $[y,y+\Delta y]$) is given by:
\be
p(x)\Delta x \cdot p(y) \Delta y = f(r) \Delta x \Delta y,
\ee
where $p(x)$ is the probability density of landing at position $x$ (and similarly for $p(y)$), which is what we are trying to determine. On the l.h.s.~of this equation, we can multiply the probabilities of landing in the $x$ and $y$ direction because of rule number (1) and (2). On the l.h.s., $f(r)$ is a function that only depends on the radial distance from the center, because of rule (2). 

We now differentiate the above equation w.r.t. the polar coordinate $\phi$:
\be \label{eq:d1}
\left( p(x) \frac{dp(x)}{d\phi} + p(y) \frac{dp(y)}{d\phi} \right)\Delta x \Delta y  = 0.
\ee
(Note that the r.h.s. becomes 0 as it does not depend on $\phi$). In polar coordinates, $x = r\cos\phi, y = r\sin\phi$, hence 
\begin{align}
\frac{dp(x)}{d\phi} & = \frac{\partial p}{\partial x}\frac{\partial x}{\partial \phi} = - \frac{\partial p}{\partial x} y, \\
\frac{dp(y)}{d\phi} & = \frac{\partial p}{\partial y}\frac{\partial y}{\partial \phi} =  \frac{\partial p}{\partial y} x.
\end{align}
Eq.~\eqref{eq:d1} becomes
\be
\left( - p(x) \frac{\partial p}{\partial x} y +  p(y)\frac{\partial p}{\partial y} x \right) \Delta x \Delta y  = 0,
\ee
which implies
\be
 \frac{p(x)}{x} \frac{\partial p}{\partial x}  =   \frac{p(y)}{y}\frac{\partial p}{\partial y}. 
 \ee
Since each side only depends on one of the variables, they must both equal a constant $C$, and we obtain the differential equation:
\be
\frac{\partial p}{\partial x}  = C x p(x)
\ee
(and similarly for $y$). Integration gives the solution
\be
p(x) = A e^{\frac{C}{2}x^2}
\ee
and $C<0$ because of rule (3). We thus define $C = -1/\sigma^2$. Requiring that the distribution is normalized gives $A = \frac{1}{\sqrt{2\pi}\sigma}$, and therefore $p(x)$ has the shape of a Gaussian (similarly for $p(y)$).

\subsection{The Chi-Square distribution}

We define the RV $\chi^2$ as the sum of the squares of $n$ standardised independent identically distributed Gaussian RV, $x_1, \dots, x_n$, where $x_i \sim \norm(\mu, \sigma)$:
\be 
\chi^2 = \sum_i^n \left( \frac{x_i - \nu}{\sigma} \right)^2  
\ee
The the RV $\chi^2$ is distributed according to the Chi-Square distribution with $n$ degrees of freedom, 
\be \label{eq:chisquare}
p(\chi^2) = \frac{1}{\Gamma(n/2)2^{n/2}} (\chi^2)^{\frac{n}{2}-1}\exp(-\frac{1}{2}\chi^2).
\ee
For the Chi-Square distribution of Eq.~\eqref{eq:chisquare}, the expectation value and variance are given by:
\be \label{eq:chisquare_properties}
E[X] = n \qquad \Var(X) = 2n.
\ee

\bibliography{SaasFe_biblio}{}
\bibliographystyle{spmpsci}

\end{document}